\documentclass[twoside,11pt]{article}
\usepackage{jair, theapa, rawfonts}
\usepackage{subfiles}
\usepackage[disable]{todonotes}
\usepackage{amssymb}
\usepackage[cmex10]{amsmath}
\usepackage{breqn}
\usepackage{enumitem}
\usepackage{subfig}
\usepackage[ruled,linesnumbered]{algorithm2e}
\usepackage[noend]{algpseudocode}
\usepackage{multicol}

\DeclareMathOperator*{\argmax}{arg\,max}

\newcommand\xqed[1]{%
	\leavevmode\unskip\penalty9999 \hbox{}\nobreak\hfill
	\quad\hbox{#1}}
\newcommand\demo{\xqed{$\square$}}

\newtheorem{lemma}{Lemma}
\newtheorem{proposition}{Proposition}
\newtheorem{example}{Example}
\newtheorem{definition}{Definition}

\begin{document}
	
	\title{Optimal Auction Design for the Gradual Procurement of Strategic Service Provider Agents}
	
	\author{\name Farzaneh Farhadi \email f.farhadi@imperial.ac.uk \\
		\addr Department of Computing, Imperial College London, UK
		\AND
		\name Maria Chli \email m.chli@aston.ac.uk \\
		\addr School of Engineering and Applied Science, Aston University, UK
		\AND
		\name Nicholas R. Jennings \email n.jennings@imperial.ac.uk \\
		\addr Department of Computing, Imperial College London, UK}

	\maketitle

	\begin{abstract}
		We consider an outsourcing problem where a software agent procures multiple services from providers with uncertain reliabilities to complete a computational task before a strict deadline. The service consumer requires a procurement strategy that achieves the optimal balance between success probability and invocation cost. However, the service providers are self-interested and may misrepresent their private cost information if it benefits them. For such settings, we design a novel procurement auction that provides the consumer with the highest possible revenue, while giving sufficient incentives to providers to tell the truth about their costs. This auction creates a contingent plan for gradual service procurement that suggests recruiting a new provider only when the success probability of the already hired providers drops below a time-dependent threshold. To make this auction incentive compatible, we propose a novel weighted threshold payment scheme which pays the minimum among all truthful mechanisms. Using the weighted payment scheme, we also design a low-complexity near-optimal auction that reduces the computational complexity of the optimal mechanism by $99\%$ with only marginal performance loss (less than $1\%$). We demonstrate the effectiveness and strength of our proposed auctions through both game theoretical and numerical analysis. The experiment results confirm that the proposed auctions exhibit $59\%$ improvement in performance over the current state-of-the-art, by increasing success probability up to $79\%$ and reducing invocation cost by up to $11\%$.

	\end{abstract}

	\section{Introduction} \label{sec:intro}
	
	There are many reasons why a business may choose to outsource a particular task, job or process. Some of the recognized benefits of outsourcing include an improved focus on core business activities, increased efficiency due to the task being performed by specialists, and reduced costs. Outsourcing has gained wide-spread popularity in a wide range of application areas, including finance \cite{Reddy2008}, IT \cite{LACITY2009}, cloud computing \cite{Motahari2009}, supply chain management \cite{Chli-old,Chli2015}, and marketing \cite{Kotabe2011}. The focus of this paper is on outsourcing applications where both the outsourcer and the service providers are automated software agents and the task can be procured and delivered by an automated computer system.
	
	Although there are many benefits, outsourcing presents a number of challenges that must be addressed. First, when a task is outsourced, the outsourcing agent, the consumer, has no direct control over the process. Therefore, it may face some uncertainties in the task's delivery time. Uncertain delivery times require a consumer with a strict deadline for task execution to deal with a tradeoff between the probability of completing the task on time and the overall cost. Recruiting faster and more reliable providers improves the task's success probability but at the expense of an increased cost. The balance of this tradeoff varies from situation to situation and is greatly influenced by parameters such as the deadline for task execution, the task's value for the consumer, and the diversity of service providers' costs and speeds. For example, the higher the value of the task, the more money the consumer is willing to spend to reduce the risk of failure. On the other hand, when the task is more urgent, the consumer may place more weight on the providers' speeds rather than their costs, when evaluating providers. 
	
	Some of the information needed to find the optimal point of the success-cost tradeoff, such as the providers' execution costs, is often unavailable to the consumer and must be extracted from the providers. In real-world applications, however, service providers are often self-interested and may misrepresent their private information upon request, if this promises to increase their profits. For instance, a self-interested provider, which only aims to maximize its own profit, may have an incentive to inflate its cost to earn higher revenue. Eliciting truthful information from self-interested providers is the second challenge faced by an outsourcing agent.

	The main goal of this paper is to design an outsourcing mechanism that optimally addresses the two challenges mentioned above. To address the first challenge, we introduce the following two outsourcing techniques that give the consumer sufficient degrees of freedom to adapt its strategy to situations: \textit{Redundant allocation} and \textit{Gradual recruitment}. Redundant allocation, which means outsourcing the same task to multiple service providers in parallel, is a powerful approach for improving the success probability \cite{Ha2006}. The second technique, gradual recruitment, is an addition to redundant allocation, which enables the consumer to not recruit all providers at once, but gradually over time. Gradual recruitment allows the consumer to use the information revealed over time to adjust the success-cost tradeoff. Using this technique, the consumer can reduce its costs by first hiring some cheaper providers and then hiring more expensive and more reliable ones, if the deadline is approaching and the task is not completed yet. 

	To address the challenge of information elicitation from selfish providers, we need to incorporate strong performance-based incentives into the outsourcing mechanism such that the providers find truthful information disclosure to be optimal.
	Existing incentive mechanisms that aim to balance the success-cost tradeoff in the presence of selfish providers restrict their attention to simultaneous deterministic outsourcing techniques, where a set of providers which are selected deterministically based on the reported costs are hired simultaneously to attempt the task in parallel \cite{Zhang2018,Feldman2020}. 
	This limited subset of outsourcing techniques does not include gradual recruitment where the recruitments happen not simultaneously but sequentially and are stochastic instead of deterministic, as each provider's invocation depends on the uncertain behaviors of others hired earlier. 
	In this paper, we show that sequential stochastic techniques such as gradual recruitment can help the consumer to significantly increase its revenue. However, to enable consumer to use such techniques in dealing with self-interested providers, the need for a novel incentive mechanism is heavily felt.
	
	To address this gap, we design a novel incentive mechanism, in the form of a procurement auction, that is flexible and adapts to the consumer's needs by employing redundant allocation and gradual recruitment techniques. This auction, which we call WGPA, possesses the following desired properties at the same time: 1) \textit{Incentive compatibility}: The auction provides sufficient incentives to providers to report their costs truthfully; 2) \textit{Interim- and ex-post individual rationality}: It encourages providers to voluntarily participate in the auction (interim-individual rationality) and guarantees that they do not regret the decision they have made (ex-post individual rationality); and 3) \textit{Revenue-optimality}: It provides the consumer with the highest possible revenue by achieving the optimal balance between the success probability and the total cost. The auction also guarantees that the consumer will never run into a deficit. 
	
	An important point about our proposed revenue-optimal auction is that its payment function is different from that of the well-known threshold payment scheme which is 
	proven to be optimal in auctions with deterministic allocation functions \cite{Myerson1981}. We prove in this paper that the threshold payment scheme, which is also known as the critical payment scheme \cite{Parkes2007}, is not optimal for auctions with sequential stochastic allocation functions. For such auctions, we propose a ``weighted'' threshold payment scheme that guarantees incentive compatibility and individual rationality at the minimum cost.
	This payment scheme determines the price of recruiting each provider based on a weighted integral over the provider's possible bid values.

	Using the weighted threshold payment scheme, we decompose the optimal auction design problem into several subproblems, each of which is a mixture of continuous and combinatorial optimization problems. We then give a branch-and-bound algorithm to solve each subproblem individually and accurately.  We also present a low-complexity heuristic algorithm that can achieve performance very close to the optimal branch-and-bound algorithm. To test the effectiveness of the proposed auction, several benchmarks are adopted to demonstrate the performance in terms of revenue and social-welfare. The results show that employing the redundant allocation and gradual recruitment techniques can improve the revenue and social-welfare by up to $120\%$ and $123\%$, respectively.
	
	Against this background, we advance the state-of-the-art in the following ways.
	\begin{itemize}
		\item We propose the first payment scheme that can be used to design revenue-maximizing auctions in settings with stochastic and/or dynamic allocation.
		\item We are the first to develop an incentive compatible and ex-post individually rational outsourcing mechanism that achieves the optimal tradeoff between success probability and overall cost. It achieves this by employing redundant allocation and gradual recruitment techniques.
		\item Using the weighted threshold payment scheme, we present a low-complexity suboptimal incentive mechanism that retains incentive compatibility and individual rationality and achieves at least $0.99$ of the optimal expected revenue.
	\end{itemize}
	
	The rest of the paper is organized as follows. After a review of the main relevant literature (Section \ref{sec:literature}), a specification of our outsourcing problem is given in Section \ref{sec:model}. In Section \ref{sec:define-ass-strategy}, we formally define redundant allocation and gradual recruitment as the two main outsourcing techniques used in this paper. In Section \ref{sec:auct-design}, we formulate the outsourcing problem with self-interested agents as an auction design problem and derive conditions that guarantee our desirable properties. Sections \ref{sec:payment-function} and \ref{sec:allocation-function} are devoted to solving the optimal auction design problem introduced in Section \ref{sec:auct-design}. We first derive the payment function of the optimal auction in Section \ref{sec:payment-function}. We use this result in Section \ref{sec:allocation-function} to derive the optimal allocation function. The theoretical properties of the designed optimal auction are discussed in Section \ref{sec:properties}. In Section \ref{sec:numerical}, we evaluate our proposed auction by simulations compared to several benchmarks. We conclude our paper in Section 10.

	\section{Related Work} \label{sec:literature}

Reaching a desired outcome when agents have private information can be achieved through some form of negotiation process \cite{Sierra2001,Fatima2004,Jennings2005,Zheng2016}. In \cite{Yao2010,Wang2021}, the authors propose protocols that can be used for negotiation between the consumer and the service providers in an outsourcing application. These protocols provide a desired outcome when agents are truthful in information sharing. However, their performance significantly degrades when providers are selfish and may misreport their private information if it benefits them.

Determining appropriate rules to achieve a desired outcome when providers are selfish is the subject of mechanism design \cite{Myerson1988}. In a special case, when the consumer faces a single selfish provider, the goal could be achieved by designing a menu of contracts from which the provider can make a choice \cite{Rogerson2003,Iyer2005,Chao2009}. However, when multiple providers exist and the outcome depends on all providers' private information, a more general form of incentive mechanism is required \cite{Borgers2015,Farhadi-faithful,Farhadi-interdependent}. Incentive mechanisms can be very diverse. However, according to the direct revelation principle \cite{Myerson1979}, the consumer can restrict attention to direct mechanisms where the providers are asked to reveal their private information in terms of bids, and the recruitment strategy and payments are determined based on these bids. Such mechanisms are often known as auctions and these are the basis for our work.

Auctions are categorized into two groups based on the auction's primary goal: \textit{efficient auctions} or \textit{optimal auctions}. The former are designed to maximize the social-welfare, while the latter aim to maximize the auctioneer's expected revenue. Because optimality and efficiency usually cannot be achieved simultaneously, the auctioneer has to make the choice before starting. A cooperative auctioneer may prefer the efficient auctions, whereas a self-interested auctioneer may prefer the optimal auctions to gain more revenue. Most of the early works in the field have dealt with efficient auction design \cite{Vickrey1961,Clarke1971,Groves1973}. The first insights into the design of optimal auctions have started with the works of Myerson \cite{Myerson1981} and Riley and Samuelson \cite{Riley1981} in 1981.
Since then, much effort has been devoted to both issues in auction theory. In the following, we review the related works in each of these fields.

\subsection{Social-Welfare Maximizing (Efficient) Auctions}

The efficiency problem is theoretically solved by the Vickrey-Clarke-Groves (VCG) mechanism \cite{Clarke1971,Groves1973} and its variants, including execution–contingent VCG \cite{Ramchurn2009,Gerding2010}, dynamic VCG \cite{Bergemann2006}, multi-stage VCG \cite{Zhang2012}, ad/position dependent cascade VCG (APDC-VCG) \cite{Farina2017}, and online VCG \cite{Parkes2003}. Nevertheless,  VCG-based mechanisms suffer from two drawbacks: (i) lack of ex-post individual rationality; and (ii) high computational complexity \cite{Nisan2007-2}.

To avoid such drawbacks, researchers started to develop some heuristic low-complexity mechanisms that approximate the optimal social-welfare while satisfying incentive compatibility and individual rationality. Examples of such approximations can be found in \cite{Babaioff2009,Huang2012,Kraft2014,Jin2015,Stein-conference,STEIN2011}. In \cite{Stein-conference,STEIN2011}, the authors propose several service procurement mechanisms that vary in their information requirement. A pairing mechanism is the most general mechanism presented in \cite{STEIN2011} that does not need tuning of the parameters. This mechanism first pairs providers randomly. Then, for each pair, it puts the provider with the lowest bid into a candidate set $\mathcal{K}$ and assigns it a virtual cost equal to its pair's bid. When this procedure ends, the mechanism restricts itself to providers in $\mathcal{K}$ and computes the social-welfare maximizing recruitment strategy by assuming that the providers' costs are equal to their virtual costs. Each provider will receive a payment equal to its virtual cost upon recruitment. The pairing mechanism is proved to be incentive compatible and ex-post individually rational and is currently the state-of-the-art in the class of approximate social-welfare maximizing mechanisms.

The focus of this paper is not on designing efficient auctions, but rather on designing optimal auctions (whose related literature will be discussed in Section \ref{sec:literature-optimal}). However, we will show in Section \ref{sec:numerical} that our proposed auction is so well-designed that it also outperforms the current state-of-the-art (i.e. the pairing mechanism) in terms of social-welfare by up to $122.7\%$.

\subsection{Revenue Maximizing (Optimal) Auctions} \label{sec:literature-optimal}

The problem of designing an auction that maximizes the auctioneer's revenue (optimal auction) is usually more challenging, as the ultimate goal depends not only on the allocations but also on the payments. Our work in this paper falls within the framework of optimal auction design, which is less studied due to its complexity. 

The study of optimal auctions started with the seminal work of Myerson in 1981 \cite{Myerson1981}. In this work, Myerson introduced the first optimal auction for single-object problems. 
Since then, there has been some progress in extending Myerson's fundamental result to multi-object environments, however, a general and analytical optimal auction in this framework has yet to be found \cite{Nedelec2021}. 

The available literature on multi-object auction design can be categorized based on whether the objects are identical or not (homogeneous \cite{Maskin1989,Malakhov2009,Pycia2021} vs. heterogeneous auctions \cite{Vries2003,Ledyard2007,Xu2020}).  The problem we investigate in this paper is a homogeneous multi-object auction 
design, as the auctioneer can create multiple copies of the task and assign them to different providers. However, there are two fundamental differences between our problem and the standard multi-object optimal auction design studied in the literature. 
\begin{enumerate}
	\item In available multi-object auctions, the decision is on the number or the set of objects (goods or tasks) assigned to each bidder. There is no time element in such auctions and it is often assumed that all assignments are made simultaneously. These auctions are often called \textit{simultaneous auctions} \cite{Zhang2018,Feldman2020}. There is a set of auctions called sequential auctions where the decisions about distinct sets of objects are made separately and sequentially in time \cite{Tardos2012,Hosseinalipour2017,Donna2018,Narayan2019,Kong2021}. The goal of this approach is to simplify the optimal auction design problem by restricting attention to a subset of objects at each round. However, none of these works consider \textit{time} as a decision factor. Our paper is the first work that considers \textit{allocation time} as a deciding factor in the optimal auction design. Time adds a continuous aspect to the allocation design part of the problem and hence significantly increases its complexity.
	
	\item Almost all of the revenue-maximizing auctions available in the literature, either single-object or multi-object auctions, utilize a deterministic allocation function. A deterministic allocation function determines a winning set of bidders for each object, based on the received bids, where bidders within the winning set receive and bidders out of the winning set do not receive the object for sure. In our problem, however, the recruitment process continues until the task is completed by at least one provider. Therefore, due to the uncertainty in delivery times, the recruitment of providers who are not first in the line is always conditional and probabilistic. The work presented in this paper is the first auction design that employs a non-deterministic allocation function. 
	
\end{enumerate} 

The features mentioned above differentiate our work from the existing literature, making it the first work to address designing revenue-maximizing auctions when recruitments can take place at arbitrary points in time.

\subsection{Robust Service Procurement} 
Our work is also related to the area of robust service procurement. There is a body of work that suggests the use of redundancy to overcome uncertainty. This is based on techniques that duplicate the critical components of a system in order to increase its reliability \cite{Tillman1977,Coit1996}. The tools in this area are mainly focused on either maximizing the success probability or minimizing the cost. In this paper, instead, we concentrate on maximizing the service consumer's revenue, which implicitly balances the success probability and cost.

Redundancy can be achieved with either a parallel or a serial configuration. The parallel redundancy, where the providers attempt the task concurrently, has seen a large amount of research \cite{Huhns2003,KOIDE2009,Zhang2009}. The serial redundancy, where a new provider is recruited when the previous service fails or takes too long, has also been studied in \cite{Friese2005,Oinn2006,Erradi2006}. The serial redundancy is also known as gradual recruitment. Protocols that employ serial redundancy often use pre-defined deadlines to determine when to switch to an alternative provider \cite{Oinn2006,Erradi2006}.

A major drawback of the works within this area is that they rely on heuristic techniques to choose the number of redundant services and deadlines. The works of \cite{Lukose2000} and \cite{Glatard2007} tackle this drawback by studying when the current service should be timed out to invoke a new one. However, these studies are under the assumption that only one service provider could be active at any time. This shortcoming is addressed by \cite{STEIN2011}, where an algorithm for deriving the efficient combination of parallel and serial redundancy has been proposed. This algorithm gives a contingent plan for how different services should be invoked over time to maximize the social-welfare, when the providers' cost information is publicly known. The authors also propose an incentive compatible and individually rational heuristic mechanism, called pairing, for settings with unknown cost information and selfish providers. However, as we show in Section \ref{sec:numerical} this heuristic mechanism is far from efficiency.

Against the existing literature, our proposed approach is not heuristic, but rather systematic and guaranteed to produce the revenue-optimal redundancy-based procurement auction. Our proposed optimal auction is different from the first-past-the-post auctions, where the providers work on the task in parallel and then the first one that delivers the task is the only one that gets paid \cite{Pandichi2011}. The main reason is that such redundancy-based auctions are not ex-post individually rational, as they cause regret to those that attempt the task but not win.

	\section{Problem Specification} \label{sec:model}

A consumer $C$ would like a task to be completed before a deadline $D$. The task has a value $V$ for the consumer if it is executed before the deadline. The consumer is no longer interested in the task when the deadline is passed. The consumer cannot accomplish the task itself and hence needs to outsource it to other agents that are capable of performing it. There are $n$ service providers (SPs) that can perform the task for the consumer. We denote the set of available SPs by $\mathcal{N}=\{1, \ldots, n \}$. The consumer may be aware of this set in advance or obtain this information when it designs an auction and calls for bids. 

The time that each SP needs for executing the task is uncertain to the consumer, due to concurrent orders from other consumers, hardware or network problems and the provider's scheduling policies. Therefore, the consumer may prefer to increase the chance of completing the task before the deadline, by procuring multiple SPs to attempt the task in parallel. However, executing the task has a cost for each provider, and the consumer must pay recruited providers a fee that covers at least their costs. Therefore, the consumer is looking for the optimal balance between the chance of success and the service invocation cost. This problem would be a standard optimization problem, if the costs are known to the consumer (as discussed in Section \ref{sec:intro}).
However, each provider's cost is its own private information and cannot be observed by either the consumer or other providers. The consumer cannot extract the providers' costs by simply asking them, since the providers are strategic and have incentive to misreport their costs.

\begin{figure}[t]
	\centering
	\includegraphics[height=0.35\textwidth]{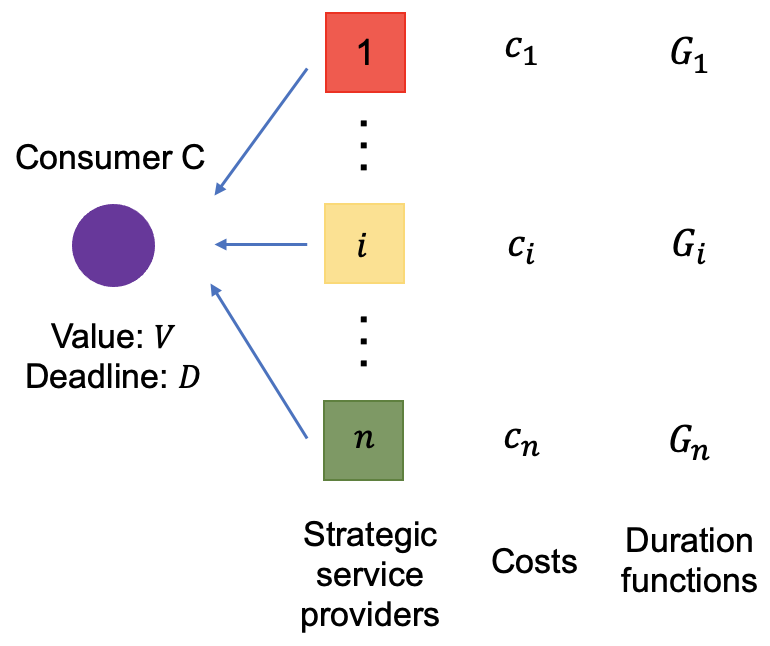}
	\caption{Schematic of the problem}\label{fig:scenario}
\end{figure}

Dealing with intelligent and strategic providers holding private information adds a new dimension to the problem, as the consumer needs to pay an extra cost to incentivize the providers to reveal truthful information. Our goal in this paper is to propose a flexible service procurement mechanism that can help a consumer to maximize its revenue when information is distributed among selfish providers. In the rest of this section, we model the main components of our problem precisely. A schematic of the problem is depicted in Fig. \ref{fig:scenario}. We start the design process from Section \ref{sec:define-ass-strategy}.

\subsection{Uncertainty in Service Provisioning}
Providers may be different from each other in terms of the speed of performing the task. The time $X_i$ that each SP $i$ requires to perform the task is uncertain. This uncertainty can be due to different reasons, e.g., the service lines of a SP might be less or more congested at different times; the resources required for performing the task could also be either available or unavailable at a given moment. We model this uncertainty by assuming that the time $X_i$ required by provider $i$ to complete the task is not a deterministic, but a random variable obtained from a distribution with cumulative density function $G_i(t)=Prob(X_i \leq t)$. We refer to function $G_i(.)$ as provider $i$'s duration function and assume that $G_i(0)=0$ for all $i \in \mathcal{N}$. The providers' duration functions are assumed to be independent and commonly known.

\subsection{Providers' Private Costs}\label{sec:private-costs}
Executing the task has a cost $c_i$ for each SP $i$. The cost of each provider $i$ is drawn from an arbitrary distribution with support on the interval $\mathcal{C}=[0,c_{max}]$. We denote the cumulative distribution function of the provider $i$'s cost by $F_i(.)$ and its corresponding density function by $f_i(.)$. The cost distribution of each SP $i$ is assumed to be regular, i.e., $c_i+\frac{F_i(c_i)}{f_i(c_i)}$ is increasing in $c_i$. The regularity condition is satisfied by many distributions such as the uniform, normal, and Pareto distributions and is a common assumption in the auction literature \cite{Salek2008,Feng2019}. The cost of each provider $i$ is only known to itself and is considered as provider $i$'s private information. The providers' private information is not observed by either the other providers or the consumer. However, the consumer knows the distribution $F_i$ from which each $c_i$ is drawn.

We assume that the costs of $n$ providers are stochastically independent random variables. Thus, the joint density function for the cost vector $\boldsymbol{c}=(c_1,\ldots,c_n)$ is $f(\boldsymbol{c})=\prod_{i \in \mathcal{N}}{f_i(c_i)}$. Of course, each provider $i$ considers its own cost $c_i$ to be a known quantity, not a random variable. However, we assume that SP $i$ assesses the probability distribution for the other providers' costs in the same way as the consumer does. That is, both the consumer and provider $i$ assess the joint distribution function for the vector $\boldsymbol{c}_{-i}=(c_1,\ldots,c_{i-1},c_{i+1},\ldots,c_n)$ of costs of all providers other than $i$ to be $f_{-i}(\boldsymbol{c}_{-i})=\prod_{j \neq i}{f_j(c_j)}$.

\subsection{Providers' Strategic Behavior}\label{sec:strategic}
In order to invoke a service, the consumer has to make a payment to the provider at least equal to its cost. However, we discussed in Section \ref{sec:private-costs}, the consumer is not aware of the providers' costs. 
In this paper, the providers are assumed to be self-interested and will act selfishly to maximize their own utilities. The utility of each provider $i$ is the difference between the payment $t_i$ it receives from the consumer and the cost $c_i$ it incurs for performing the task. If a self-interested provider is asked to declare a cost, it may have an incentive to either ``mark up" the price to earn more profit or ``mark down" the price to attract the consumer.

In this situation, the consumer should design its recruitment strategy and the payments such that the providers cannot benefit by misreporting their costs. This can be done by designing an incentive compatible auction\footnote{We define the concepts of auctions and incentive compatibility formally in Section \ref{sec:auct-design}.}\cite{Brubaker1980}. The consumer announces the auction's rules in advance and then each provider seeks an action that maximizes its own utility. The auction must be designed such that truth telling is the best strategy for each provider.

\subsection{Consumer's Objective}
In most practical settings, the consumer is also self-interested and would like to maximize its own revenue. The consumer's revenue is the difference between the value it gets from the task (if any) and all the payments it should make. The consumer gets a value $V$ from the task if it is executed successfully before the deadline. Thus, a self-interested consumer often faces a tradeoff between the success probability and the cost, as faster and more reliable SPs, which can increase the chance of getting the task done before the deadline, are often more expensive to hire. The goal of the consumer is to design auction rules such that it can lead the outcome to the optimum point of this tradeoff.

In the next section, we present two outsourcing techniques that allow the consumer to better navigate this tradeoff. We incorporate these techniques into our auction design problem in Section \ref{sec:auct-design}.

	\section{Flexible Outsourcing Techniques}\label{sec:define-ass-strategy}

When the task's delivery time is uncertain, a revenue-maximizing consumer with a strict deadline should look for effective ways to increase the success probability at a reasonable cost. In this section, we introduce two outsourcing techniques that can help the consumer to achieve this goal and lead the outcome to the optimal point of the success-cost tradeoff. Since the tipping point of this tradeoff depends on various factors, such as the task's deadline and value along with the providers' diversity in terms of reliability and cost, the outsourcing techniques introduced below provide the consumer with a high degree of freedom to adapt its strategy to situations.

\subsection{Technique 1: Redundant Allocation}

In this technique, the consumer increases the chance of task success by procuring multiple SPs to attempt the task in parallel. In this case, the consumer obtains value $V$ if at least one of the providers completes the task before the deadline $D$. Completing the task by more than one SP does not provide any additional value to the consumer. However, the consumer cannot stop hired providers from performing the task when the task is completed by one SP or the deadline is reached. This means that the task is assumed to be non-interruptible and needs to run to completion once it started.

In the redundant allocation technique, the consumer should make a payment to each hired provider to incentivize it to perform the task upon recruitment. This makes it apparent that recruiting more providers increases the consumer's cost and could hurt the revenue if it is not managed properly. To avoid this negative effect, the consumer should decide wisely which set of providers leads to a better balance between the success probability and the total cost.

\subsection{Technique 2: Gradual Recruitment} \label{sec:gradual-recruitment}

This technique is an extension to redundant allocation which aims to lower the cost while still retaining high success probability. Specifically, instead of hiring a set of providers all at the beginning, gradual recruitment lets the consumer make its decisions gradually over time. For example, the consumer has an option to hire some SPs at time $0$ and wait for a specific period of time (which is less than the deadline) to see whether any of the hired providers can complete the task. If they couldn't, then the consumer may prefer to spend some money and recruit some new providers to increase the success likelihood. 

The gradual recruitment technique gives the consumer the opportunity to use the information revealed over time to better adjust the success-cost tradeoff. This ability can be illustrated by the following simple example.

\begin{example}\label{Ex-gradual-recruitment}
	Suppose that a high-valued task can be performed by two SPs. Provider $1$ is cheap to hire, but its delivery time varies based on its workload. Under a low workload, provider $1$ delivers the task after $D/2$ units of time. However, it needs at least $2D$ time-units to deliver the task when working at a high workload. The second provider is expensive to hire, but it is totally reliable and always delivers the task by time $D/2$. 
	
	In this situation, if the consumer knew provider $1$'s workload, the optimal decision would be to recruit provider $1$ when the workload is low and to recruit provider $2$ when the workload is high. This information is not available to the consumer beforehand. However, the consumer can build a cheap experiment based on the gradual recruitment technique to first find out provider $1$'s workload and then make the decision accordingly. Specifically, the consumer can first recruit provider $1$ at a low cost and give it $D/2$ units of time to perform the task. Whether or not provider $1$ delivers the task by $D/2$ serves as a signal to the consumer to discover provider $1$'s workload. That is, provider $1$'s workload is low if and only if the job is delivered by time $D/2$. If based on this information, the workload is revealed to be high, the consumer can spend more money to guarantee success by recruiting provider $2$. This strategy provides the same value to the consumer as the one that recruits provider $2$ from the beginning (i.e., the probability of success is $1$ in both cases), however, the former incurred a much lower expected cost to the consumer than the latter.
	\demo
\end{example}

Example \ref{Ex-gradual-recruitment} shows the ability of gradual recruitment technique to find and utilize new information over time. However, efficiently exploiting this potential can be challenging, as it presents the consumer with some new challenges. In addition to selecting a set of SPs to be hired, now the consumer also has to decide on the order of hiring sellers and the time gaps between them.

We start facing this challenge in the next subsection, by embedding the concepts of redundant allocation and gradual recruitment into the consumer's decision making problem.

\subsection{Modeling Redundant Allocation and Gradual Recruitment}
We define the consumer's recruitment strategy as a contingent plan that specifies the providers that should be hired at specific times if the task is not completed yet.

\begin{definition}[Recruitment strategy] 
	A recruitment strategy is a vector $\rho = ((s_1\allowbreak,\allowbreak\tau_1),\allowbreak \ldots,\allowbreak (s_m\allowbreak,\allowbreak \tau_m))$, where each element represents the time $\tau_i \in [0,D]$ that provider $s_i \in \mathcal{N}$ is recruited and $m$ is an arbitrary integer number between $0$ and $n$. The elements of this vector are all different in their first component (i.e., $s_i \neq s_j$ if $i \neq j$) and ordered increasingly based on their second component $\tau_i$, i.e., $\tau_{i+1} \geq \tau_i$, for all $i \in \{1, \ldots,m-1\}$.
	Provider $s_i$ is recruited at time $\tau_i$ if and only if the task has not been completed so far. 
\end{definition}

A recruitment strategy $\rho$ specifies an ordered set of SPs  $\rho_s=(s_1,\ldots,s_m)$ that are candidates to be invoked, as well as the time of their invocation $\rho_{\tau}=(\tau_1,\ldots,\tau_m)$. The consumer starts by recruiting the provider with the first turn, i.e., $s_1$, at time $\tau_1$ and waits up to time $\tau_2$ to see if the task is completed. If the task is not completed, the consumer increases the success probability by invoking provider $s_2$, which is second in the line, to attempt the task in parallel with $s_1$. This gradual recruitment continues until either the task is completed or all the candidate providers have been invoked.

We denote the set of all recruitment strategies by $\mathcal{H}$. This set is completely general, it includes the extremes of ``single strategies'', where the consumer recruits only one single provider (i.e., $\rho = ((s_1,\tau_1))$), ``simultaneous strategies'', where a set of providers are recruited simultaneously at time $\tau$ (i.e. $\rho = ((s_1,\tau),\ldots,(s_m,\tau))$), and all conceivable intermediate strategies. 

The consumer would like to choose a recruitment strategy that maximizes its expected revenue. We denote the consumer's expected revenue when it employs the recruitment strategy $\rho$ and makes payments $\boldsymbol{t}=(t_i)_{i\in \mathcal{N}}$ to providers upon their recruitment, by
\begin{equation}\label{eq:exp-util-cons1}
	\mathbb{E}\left[ U(\rho,\boldsymbol{t})\right]=V(\rho,D)-C(\rho,\boldsymbol{t}),
\end{equation}
where $V(\rho,D)$ is the expected value the consumer gets from task completion and $C(\rho,\boldsymbol{t})$ is the expected cost of hiring providers.
The effect of requirement strategy $\rho$ on the expected value and the expected cost is quite different and opposite. Recruiting more providers early on can raise the revenue by increasing the success probability and hence the expected value $V(\rho,D)$, but it can also reduce the revenue by increasing the total cost $C(\rho,\boldsymbol{t})$. Our goal in this paper is to propose an auction to find the optimal balance between these two conflicting factors.

In the rest of this subsection, we quantify the expected value and the expected cost for any arbitrary recruitment strategy $\rho_s=(s_1,\ldots,s_n)$. Then, in Section \ref{sec:auct-design}, we design an auction to find the tipping point of the above-mentioned tradeoff.

\subsubsection{Expected Value of a Recruitment Strategy}\label{sec:value-strategy}

The consumer gets a value $V$ if the recruitment strategy $\rho$ is successful in delivering the task before deadline $D$. A recruitment strategy $\rho$ is successful if at least one of the recruited SPs completes the task before the deadline. We define the completion time of the task when recruitment strategy $\rho$ is employed as the first completion time for providers $\rho_s$ and denote it by $X_{\rho} = \min_{i \in \{1, \ldots, m \}}{(\tau_i + X_{s_i})}$. The task completion time $X_{\rho}$ is a random variable whose distribution depends on the duration functions of the recruited providers. Since each $X_{s_i}$, for $s_i \in \rho_s$, is drawn independently from distribution $G_{s_i}$, the cumulative distribution function of $X_{\rho}$ can be derived as follows:
\begin{equation}\label{eq:CDF-complete}
	G_{\rho}(t)=Prob(X_{\rho} \leq t)=1-Prob(X_{\rho} > t)=1-\prod_{i=1}^m{(1-G_{s_i}(t-\tau_i))}.
\end{equation}
Based on \eqref{eq:CDF-complete}, the success probability of strategy $\rho$ for a task with deadline $D$ can be formulated as 
\begin{equation}\label{eq:psucc}
	P_{succ}(\rho,D)=Prob(X_{\rho} \leq D)=1-\prod_{i=1}^m{(1-G_{s_i}(D-\tau_i))}.
\end{equation}
We can use \eqref{eq:psucc} to derive the expected value of a recruitment strategy $\rho$ as 
\begin{equation}\label{eq:exp-value}
	V(\rho,D)=V P_{succ}(\rho,D)=V (1-\prod_{i=1}^m{(1-G_{s_i}(D-\tau_i))}).
\end{equation}

\subsubsection{Expected Cost of a Recruitment Strategy}\label{sec:cost-strategy}

The consumer makes a payment $t_i$ to each provider $i$ upon recruitment. Therefore, the expected cost of a recruitment strategy $\rho$ when employed along with the payment strategy $\boldsymbol{t}$ can be written as
\begin{equation}\label{eq:cost-strategy1}
	C(\rho,\boldsymbol{t})=\sum_{i=1}^n{P_i(\rho) \hspace{0.05cm} t_i},
\end{equation}
where $P_i(\rho)$ is the probability that provider $i$ will be invoked in the recruitment strategy $\rho$. Providers out of the set of candidate providers $\rho_s$ have no chance for being hired, i.e., $P_i(\rho)=0$, where $i \notin \rho_s$. However, each provider $i \in \rho_s$ will be recruited at a specific time if all the providers with earlier turns fail to complete the task by then. 

To formalize this, we define the order of provider $i$'s invocation in the recruitment strategy $\rho$ as $o_i(\rho)$, where
\begin{align}\label{eq:position}
	o_i(\rho)=\hspace{-0.05cm}
	\begin{cases}
		k, & \text{if} \quad s_k=i,\\
		0,			& \text{if} \quad i \notin \rho_s.\\
	\end{cases}
\end{align}
Any provider $i \in \rho_s$ will be invoked at time $\tau_{o_i(\rho)}$ if all the providers $j$ with $o_j(\rho)<o_i(\rho)$ fail to complete the task by $\tau_{o_i(\rho)}$. Since the providers' duration functions are independent random variables, we can derive provider $i$'s invocation probability as
\begin{equation}\label{eq:prob_selection}
	P_i(\rho)=\prod_{j: o_j(\rho)<o_i(\rho)}{\hspace{-0.5cm}Prob(\tau_{o_j(\rho)}+X_{j} > \tau_{o_i(\rho)})}=\hspace{-0.5cm}\prod_{j: o_j(\rho)<o_i(\rho)}{\hspace{-0.5cm}(1-G_{j}(\tau_{o_i(\rho)}-\tau_{o_j(\rho)}))},
\end{equation}
where $i \in \rho_s$. It can be seen from \eqref{eq:prob_selection} that the providers' invocation probabilities are inversely related to their invocation orders; providers with earlier turns have higher chances of being recruited.

By substituting \eqref{eq:prob_selection} in \eqref{eq:cost-strategy1}, we can derive the consumer's expected cost as
\begin{equation}\label{eq:cost-strategy}
	C(\rho,\boldsymbol{t})=\sum_{i \in \rho_s}{t_i \hspace{0.05cm} (\prod_{j: o_j(\rho)<o_i(\rho)}{(1-G_{j}(\tau_{o_i(\rho)}-\tau_{o_j(\rho)}))})}.
\end{equation}

\subsubsection{Expected Revenue of a Recruitment Strategy}
Combining the results of Subsections \ref{sec:value-strategy} and \ref{sec:cost-strategy}, we can obtain a closed form expression for the consumer's expected revenue, as follows:
\begin{align}\label{eq:exp-util-cons}
	&\mathbb{E}\left[ U(\rho,\boldsymbol{t})\right]=V P_{succ}(\rho,D)-\sum_{i=1}^n{P_i(\rho) \hspace{0.05cm} t_i}\nonumber\\
	&=V (1-\prod_{i=1}^m{(1-G_{s_i}(D\hspace{-0.05cm}-\hspace{-0.05cm}\tau_i))})-\sum_{i \in \rho_s}{t_i \hspace{0.05cm} (\hspace{-0.3cm}\prod_{j: o_j(\rho)<o_i(\rho)}\hspace{-0.4cm}{(1-G_{j}(\tau_{o_i(\rho)}\hspace{-0.05cm}-\hspace{-0.05cm}\tau_{o_j(\rho)}))})}.
\end{align}

The consumer's utility \eqref{eq:exp-util-cons} is decreasing in terms of the payments $t_i$. Therefore, for each recruitment strategy $\rho$, the consumer prefers to make the minimum possible payment to each provider. Each provider $i$'s received payment $t_i$ must be greater than or equal to its cost $c_i$, as otherwise, the provider finds it unprofitable to perform the task and hence refuses to do so.
If the consumer knew each provider's cost $c_i$, the payment $t_i$ could be simply equal to $c_i$. However, as we mentioned earlier, the cost of each provider is its own private information and the consumer is not aware of it. 
To overcome this lack of knowledge, the consumer can conduct an auction and call for bids. However, since the providers are strategic, there is no guarantee that the submitted bids are equal to the actual costs. 
Receiving incorrect information from the providers may prevent the consumer from taking optimal decisions. Therefore, to avoid such fraudulent behaviors, the consumer would like to design the auction's rules such that they leave no incentive for the providers to be untruthful. 

Based on the discussion above, the consumer's ultimate goal is to design an auction that has the following three properties: 1) it is \textit{incentive compatible}, i.e., it incentivizes providers to reveal the true information; 2) it is \textit{individually rational}, i.e., it motivates providers to participate in the auction; and 3) it is \textit{revenue-maximizer}, i.e., it finds the optimum point of the value-cost tradeoff faced by the consumer. We study the design of an auction with the above three features in the next section.

	\section{Optimal Auction Design Problem} \label{sec:auct-design}
	
In a procurement auction, the set of bids that a service provider may submit is equal to the set of private information it may have. 
In our problem, each provider's private information consists of its cost for performing the task. Therefore, each provider $i$'s bid $b_i$ must belong to the set of possible costs $\mathcal{C}$ and can be interpreted as provider $i$'s reported cost. Notice that each provider $i$'s reported cost is not necessarily equal to its actual cost $c_i$. We denote the vector of all providers' submitted bids by $\boldsymbol{b}=(b_1,\ldots, b_n) \in \mathcal{C}^n$.

A procurement auction $M=(A(.),\boldsymbol{T}(.))$ is defined by an allocation function $A(.)$ and a vector of payment functions $\boldsymbol{T}(.)=(T_1(.),\ldots,T_n(.))$. The allocation function $A: \mathcal{C}^n \rightarrow \mathcal{H}$ determines the recruitment strategy $A(\boldsymbol{b}) \in \mathcal{H}$ that is selected by the consumer based on the received bids $\boldsymbol{b}$, and each payment function $T_i: \mathcal{C}^n \rightarrow \mathbb{R}_{+}$ determines the amount of money $T_i(\boldsymbol{b})$ that will be paid to provider $i$, upon recruitment. As discussed in Section \ref{sec:define-ass-strategy}, the consumer's goal is to design an auction that satisfies incentive compatibility, individual rationality, and revenue maximization.  In the following subsections, we derive conditions on the auction's allocation function $A(.)$ and payment function $\boldsymbol{T}(.)$ that guarantee each of these properties. We then formulate the consumer's auction design problem as an optimization problem.
%As in standard procurement auctions, the payments are assumed to made only to the service providers that are recruited for performing the task \cite{}. These auctions are also referred to as normalized auctions \cite{Dobzinski2011}.

\subsection{Incentive Compatibility (IC)}

Any procurement auction $M=(A(.),\boldsymbol{T}(.))$ induces a non-cooperative game $G=(\mathcal{N},\mathcal{C}^n,U_i(.))$ with incomplete information, where the set of players is the set of service providers $\mathcal{N}$, a bidding strategy $\sigma_i : \mathcal{C} \rightarrow \mathcal{C}$ of each player $i$ is a mapping from its true cost $c_i$ to the bid $b_i$ it submits, and each player $i$'s utility resulting from the action profile $\boldsymbol{b}$ is 
\begin{equation}\label{eq:exp2-util-sp1}
	U_i(A,T_i,c_i,\boldsymbol{b})=P_i(A(\boldsymbol{b}))[-c_i+T_i(\boldsymbol{b})].
\end{equation}
Provider $i$'s utility is the difference between the cost $c_i$ it incurs for performing the task and the payment $T_i(\boldsymbol{b})$ it receives from the consumer if it gets hired (with prob. $P_i(A(\boldsymbol{b}))$), and zero otherwise.

Procurement auction $M=(A(.),\boldsymbol{T}(.))$ is said to be incentive compatible (IC) if truth-telling (i.e., $b_i=c_i, \forall i \in \mathcal{N}$) is a BNE of the induced game $G$ \cite{Borgers2015}, i.e., 
\begin{align}\label{eq:Truth-BNE}
	\int{U_i(A,T_i,c_i,(c_i,\boldsymbol{c}_{-i}))f_{-i}(\boldsymbol{c}_{-i})d\boldsymbol{c}_{-i}} \geq \int{U_i(A,T_i,c_i,(b_i,\boldsymbol{c}_{-i}))f_{-i}(\boldsymbol{c}_{-i})d\boldsymbol{c}_{-i}},
\end{align}
holds for every $b_i$, $c_i$, and $i \in \mathcal{N}$. For ease of notation in the future, we let $U_i(A,T_i,c_i,b_i)$ denote the integral that appears in the right-hand side of \eqref{eq:Truth-BNE}. In fact, $U_i(A,T_i,c_i,b_i)$ is provider $i$'s expected utility in an auction $M=(A(.),\boldsymbol{T}(.))$, when all providers except $i$ submit their costs truthfully, but provider $i$ with cost $c_i$ submits bid $b_i$. 

Using this notation, the incentive compatibility constraint \eqref{eq:Truth-BNE} can be rewritten as 
\begin{align}\label{eq:Truth-BNE-simple}
	U_i(A,T_i,c_i,c_i) \geq U_i(A,T_i,c_i,b_i), \quad \forall c_i, b_i \in \mathcal{C}, i \in \mathcal{N}.
\end{align}

\subsection{Individual Rationality (IR)}
An auction is individually rational (IR) for a provider, if the utility it gains from the auction is non-negative. There are different notions of IR in the literature; the two we focus on here are called \textit{interim IR} and \textit{ex-post IR}. The difference between these two notions is at the time the providers are allowed to drop out of the auction. In interim IR, providers should make their decisions before the auction starts. However, in ex-post IR, the providers are allowed to opt out of the auction even at the very end when the consumer decides to recruit them. 

The interim IR is guaranteed if the expected utility each provider $i$ with cost $c_i$ gains from participating in the auction is non-negative, i.e.
\begin{equation}
	U_i(A,T_i,c_i,c_i) \geq 0, \quad \forall c_i \in \mathcal{C}, i \in \mathcal{N}.\tag{Interim IR}
\end{equation}
However, the ex-post IR requires that no provider has regrets regarding participation even if any bid vector $\boldsymbol{c}_{-i}$ is submitted by other providers, i.e.
\begin{equation}
	T_i(c_i,\boldsymbol{c}_{-i})-c_i \geq 0, \quad \forall c_i \in \mathcal{C}, \boldsymbol{c}_{-i} \in \mathcal{C}^{n-1}, i \in \mathcal{N}.\tag{Ex-post IR}
\end{equation}
Although ex-post IR is stronger and more applicable to our setting, it is also more difficult to deal with in the auction design problem. Therefore, in the design process, we relax this constraint to the interim IR constraint. However, at the end, we will theoretically show that the auction is sufficiently well-designed to satisfy the ex-post IR condition.

\subsection{Revenue Maximization}

In Section \ref{sec:define-ass-strategy}, we derived the consumer's expected revenue $\mathbb{E}\left[ U(\rho,\boldsymbol{t})\right]$ when it employs the recruitment strategy $\rho$ and makes payments $\boldsymbol{t}=(t_i)_{i\in \mathcal{N}}$ to providers upon their recruitment. In the truthful equilibrium of the game induced by auction $M=(A(.),\boldsymbol{T}(.))$, each bid vector $\boldsymbol{c}$ is submitted with probability $f(\boldsymbol{c})$. When bid vector $\boldsymbol{c}$ is received, the consumer employs the recruitment strategy $A(\boldsymbol{c})$ and the payment strategy $\boldsymbol{T}(\boldsymbol{c})$. Therefore, the consumer's expected revenue in an incentive compatible auction $M=(A(.),\boldsymbol{T}(.))$ is
\begin{align}\label{eq:exp2-util-cons}
	&U(A,\boldsymbol{T})=\int{\mathbb{E}\left[ U(A(\boldsymbol{c}),\boldsymbol{T}(\boldsymbol{c}))\right] f(\boldsymbol{c})d\boldsymbol{c}}\nonumber\\
	&=\int{[V P_{succ}(A(\boldsymbol{c}),D)-\sum_{i \in \mathcal{N}}{P_i(A(\boldsymbol{c})) T_i(\boldsymbol{c})}]f(\boldsymbol{c})d\boldsymbol{c}},
\end{align}

The consumer would like to choose the allocation function $A$ and the payment function $\boldsymbol{T}$ so as to maximize the expected revenue $U(A,\boldsymbol{T})$ under the IC and IR constraints. This is equivalent to solving the following optimization problem:
\begin{subequations}\label{eq:consumer-problem}
	\begin{align}
		&\max_{\left\lbrace A,\boldsymbol{T} \right\rbrace} {\quad U(A,\boldsymbol{T})}\\
		\textbf{s.t.} \quad & U_i(A,T_i,c_i,c_i) \geq 0, \quad \forall c_i \in \mathcal{C}, i \in \mathcal{N},\label{eq:IR} \\
		& U_i(A,T_i,c_i,c_i) \geq U_i(A,T_i,c_i,b_i), \quad \forall c_i, b_i \in \mathcal{C}, i \in \mathcal{N}.\label{eq:IC}
	\end{align}
\end{subequations}
	
	\section{Optimal Payment Function} \label{sec:payment-function}
	
Problem \eqref{eq:consumer-problem} is an optimization problem in the space of functions (which are infinite-dimensional vector spaces). Such problems are generally difficult to solve because of their high dimensionality. In this paper, we tackle optimization problem \eqref{eq:consumer-problem} by proceeding as follows. 
In Section \ref{sec:Myerson-eq-conditions}, we derive a necessary and sufficient set of conditions for an auction to be incentive compatible and interim individually rational (i.e., satisfy constraints \eqref{eq:IR}-\eqref{eq:IC}). Replacing \eqref{eq:IR}-\eqref{eq:IC} with this set of constraints, we partially solve problem \eqref{eq:consumer-problem} and derive the optimal payment function $\boldsymbol{T}^*(.)$ in Section \ref{sec:optimal-pay}. The impact of this payment function on reducing the consumer's cost will be discussed in Section \ref{sec:example}. We finally substitute $\boldsymbol{T}^*(.)$ in \eqref{eq:consumer-problem} to express optimization problem \eqref{eq:consumer-problem} solely in terms of the allocation function $A(.)$ in Section \ref{sec:opt-allocation-function}.
Determining the optimal allocation function is the subject of the next section.

\subsection{Simpler Conditions for IC and IR}\label{sec:Myerson-eq-conditions}

We follow Myerson's approach proposed in \cite{Myerson1981} to provide a necessary and sufficient set of conditions for an auction $M=(A(.),\boldsymbol{T}(.))$ to satisfy constraints \eqref{eq:IR}-\eqref{eq:IC}. To this end, we define
\begin{equation}\label{eq:conditional-prob}
	Q_i(A,b_i)=\int{P_i(A(b_i,\boldsymbol{c}_{-i}))f_{-i}(\boldsymbol{c}_{-i})d\boldsymbol{c}_{-i}},
\end{equation}
as the conditional probability that service provider $i$ will be hired in the auction mechanism $M=(A(.),\boldsymbol{T}(.))$ given that it submits bid $b_i$.

\begin{lemma}\label{L-Myerson1}
	An auction $M=(A(.),\boldsymbol{T}(.))$ satisfies the individual rationality and incentive compatibility constraints if and only if the following conditions hold:
	\begin{enumerate}[label=(I\arabic*)]
		\item \label{I1}  The conditional probability function $Q_i(A,b_i)$ defined in \eqref{eq:conditional-prob} is decreasing in its second argument, i.e.,
		\begin{equation}\label{eq:Myerson1}
			(b_i - c_i)( Q_i(A,b_i)- Q_i(A,c_i)) \leq 0;
		\end{equation}
		\item \label{I2} Each provider $i$'s expected utility when it tells the truth satisfies the two following conditions:
		\begin{equation}\label{eq:Myerson2}
			U_i(A,T_i,c_i,c_i)=U_i(A,T_i,c_{max},c_{max})+\int_{c_i}^{c_{max}}{Q_i(A,b_i)db_i},
		\end{equation}
		\begin{equation}\label{eq:Myerson3}
			U_i(A,T_i,c_{max},c_{max}) \geq 0.
		\end{equation}
	\end{enumerate}
\end{lemma}

We present the proofs of all the lemmas and propositions in Appendices \ref{sec:appendix-A}-\ref{sec:appendix-I}.

Condition \ref{I1} states that for a procurement auction to be incentive compatible, the allocation function $A(.)$ must be monotone and give higher chances of selection to providers that report lower costs. This result is intuitively similar to Myerson's monotonicity condition which is well-known in the mechanism design literature \cite{Myerson1981,CARBAJAL2015}. However, there is a slight but important difference between the definition of monotonicity in our setting and that of Myerson. In most of the problems studied in the literature, the selection process is deterministic. That is, based on the received bids $\boldsymbol{b}$, the auctioneer selects a set of service providers and invokes all of them simultaneously. In such settings, monotonicity requires that if service provider $i$ gets invoked when it submits bid $b_i$, it should also be invoked when it submits bid $\hat{b}_i \leq b_i$. However, in our problem, the selection process is gradual and stochastic. That is, each service provider $i$ that submits bid $b_i$ will be invoked with probability $Q_i(A,b_i) \in [0,1]$ defined in \eqref{eq:conditional-prob}. In this setting, monotonicity requires that a service provider that places a lower bid must have a greater probability of getting invoked. This is equivalent to the fact that $Q_i(A,b_i)$ is decreasing in its second argument.

Condition \ref{I2} of Lemma \ref{L-Myerson1} implies that in an incentive compatible and individually rational auction, each provider's expected utility is positive and decreasing in its cost.  This is a consequence of \eqref{eq:Myerson2}-\eqref{eq:Myerson3} and the fact that function $Q_i$ is always positive.

\subsection{Weighted Threshold Payment Scheme}\label{sec:optimal-pay}

Based on Lemma \ref{L-Myerson1}, $M=(A(.),\boldsymbol{T}(.))$ represents an optimal auction if and only if it maximizes $U(A,\boldsymbol{T})$ subject to \eqref{eq:Myerson1}-\eqref{eq:Myerson3}. Using this result, we can derive an explicit closed form expression for the optimal payment function.
\begin{lemma}\label{L-Myerson2}
	The payment function of an optimal auction has the following form:
	\begin{equation}\label{eq:opt-payment}
		T_i^*(\boldsymbol{b})=b_i+\frac{1}{P_i(A(\boldsymbol{b}))}\int_{b_i}^{c_{max}}{P_i(A(\hat{b}_i,\boldsymbol{b}_{-i}))d\hat{b}_i}, \quad \forall i \in \mathcal{N}.
	\end{equation}
\end{lemma}

Lemma \ref{L-Myerson2} gives the lowest amount of money that the consumer should pay to the service providers, based on their bids, to make the auction incentive compatible and individually rational. This payment is different from the well-known threshold payment that is widely used in the literature \cite{Parkes2007,Wang2016}. The traditional threshold payment scheme which is also known as critical payment pays each provider $i$ the maximal cost that $i$ can bid and still gets invoked. This payment rule is designed for settings where the allocation function is deterministic and unconditional. However, in our auction design problem, the allocation function could be conditional and probabilistic. More precisely, based on the received bids, the consumer provides an ordered list $(s_1,\ldots,s_m)$ of candidate providers along with their invocation times $(\tau_1,\ldots,\tau_m)$, and recruits each provider $s_i$ at time $\tau_i$ if the task has not completed by others yet. In this setting, increasing the declared cost may not completely remove a provider from the list of candidate providers, but could delay its turn and hence reduce the chance of being invoked.

The payment function derived in \eqref{eq:opt-payment} is the extension of the threshold payment scheme for auctions with probabilistic allocation functions. In our generalized payment scheme, which we call \textit{weighted threshold payment}, the consumer considers the conditional probability that $i$ got hired if it reported $\hat{b}_i \geq b_i$, given that it is recruited when it submits bid $b_i$, as the relative desirability of cost $\hat{b}_i$ for provider $i$. We denote this relative desirability by $d_{b_i,\boldsymbol{b}_{-i}}^A(\hat{b}_i)$, where 
\begin{align}\label{eq:desirability}
	d_{b_i,\boldsymbol{b}_{-i}}^A(\hat{b}_i)=P_i(A(\hat{b}_i,\boldsymbol{b}_{-i}))/P_i(A(\boldsymbol{b})).
\end{align}
It is clear from \eqref{eq:desirability} that in monotone auctions the relative desirability is always between $0$ and $1$. Based on \eqref{eq:opt-payment}, when the consumer recruits provider $i$, it pays the provider its declared cost $b_i$ plus the integral of the relative desirability over all higher bids. Excess payment is an ``information rent'', that is a payment made by the consumer to service providers in exchange for accurate disclosure of their private information.

\subsection{Potential of Gradual Recruitment in Reducing the Consumer's Cost}\label{sec:example}

The optimal payment function \eqref{eq:opt-payment} provides us a better insight on why gradual recruitment can reduce the consumer's expected cost $C(\rho,\boldsymbol{t})\allowbreak=\allowbreak \sum_{i=1}^n{P_i(\rho)\allowbreak t_i}$ (see Section \ref{sec:cost-strategy}). The reason of this is two-fold: (i) reducing providers' invocation probabilities $P_i(\rho)$ and (ii) reducing the prices $t_i$ the consumer needs to pay to providers upon recruitment. The first reason is obvious from the discussion in Section \ref{sec:define-ass-strategy}. When the consumer employs a gradual recruitment strategy, it can stop the process as soon as one of the hired providers completes the task. This reduces the invocation probabilities of the providers with later turns, and eventually reduces the consumer's expected cost. The second reason is less apparent and originates from the optimal payment function \eqref{eq:opt-payment}. We illustrate this by an example.

\begin{example}\label{Ex0}
	Consider a task with value $V=4$ and deadline $D=1$. Suppose that there are two strategic service providers that can execute the task. The time needed by each of these providers to complete the task is exponentially distributed with mean $1$, i.e., $G_1(t)=G_2(t)=1-e^{-t}$. Each provider $i$'s cost $c_i$, which is only known to itself, is drawn from uniform distribution over $[0,1]$.

	\begin{figure}[t]
		\subfloat[Best simultaneous auction]{\label{fig:ex0-1}
			\includegraphics[height=0.35\textwidth]{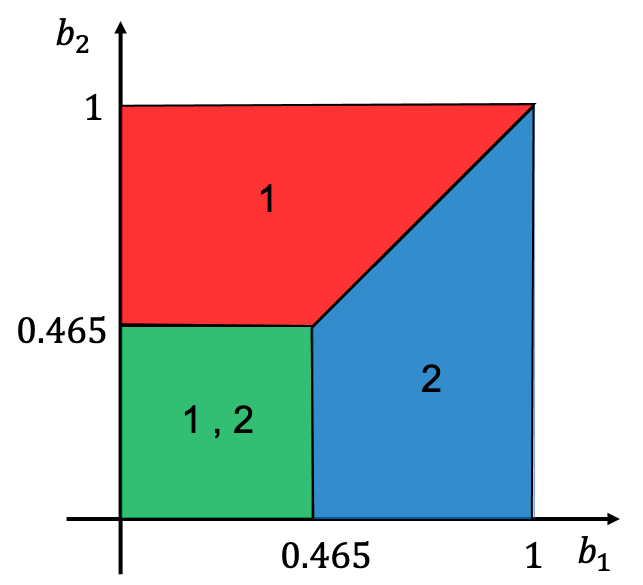}}
		~~~~~
		\subfloat[Best auction with gradual recruitment]{\label{fig:ex0-2}
			\includegraphics[height=0.35\textwidth]{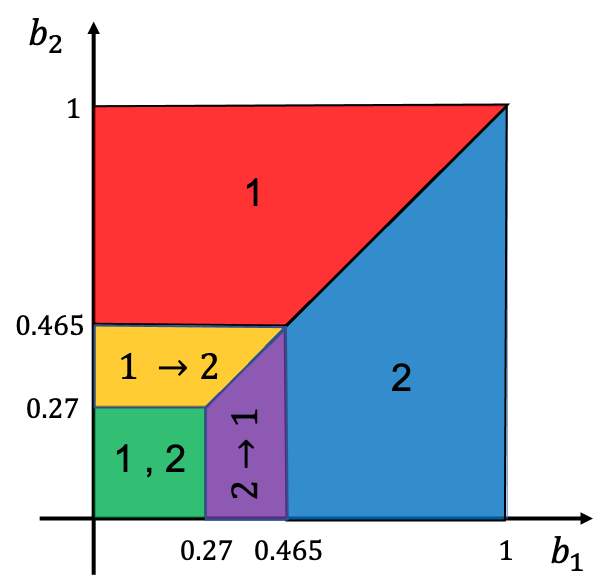}}
		\caption{Two allocation functions for the setting described in Example \ref{Ex0}. To save space, we use short notations to show the recruitment strategy of each region. For example, $1$ is a short notation for $\rho=((1,0))$, $1,2$ is a short notation for $\rho=((1,0),(2,0))$, and $1 \rightarrow 2$ represents any recruitment strategy $\rho=((1,0),(2,t))$ with $t>0$.}\label{fig:heuristic1}
	\end{figure}
	
	In this setting, we denote the best auction in the absence of gradual recruitment by $M_{no}=(A_{no}(.),\boldsymbol{T}_{no}(.))$. The allocation function $A_{no}(.)$, which can be obtained by using the techniques we will provide later in this paper, is shown in Fig. \ref{fig:ex0-1}. This allocation function instructs the consumer to recruit both providers simultaneously at time $0$ if submitted bids $b_1$ and $b_2$ are both below $0.465$. Otherwise, the consumer should only recruit the provider with the minimum submitted bid. 
	The minimum price that the consumer should pay to the providers to make auction $M_{no}$ with allocation function $A_{no}(.)$ incentive compatible and individually rational is derived based on \eqref{eq:opt-payment}, as follows:
	\begin{align}\label{eq:ex0-1}
		\boldsymbol{T}_{no}(b_1,b_2)=
		\begin{cases}
			(0.465,0.465), & \text{if } b_1,b_2 \leq 0.465, \\
			(b_2,0),			& \text{if } b_2 > 0.465, b_1 \leq b_2,\\
			(0,b_1),			& \text{if } b_1 > 0.465, b_2 < b_1.\\
		\end{cases}
	\end{align}
	In each pair returned by payment function $\boldsymbol{T}_{no}(.)$, the first and second elements are the prices paid to providers 1 and 2, respectively, upon recruitment.
	
	Now, let's study what will happen if the consumer takes advantage of gradual recruitment. In this case, it is optimal for the consumer to run auction $M^*=(A^*(.),\boldsymbol{T}^*(.))$ with allocation function $A^*(.)$ shown in Fig. \ref{fig:ex0-2}\footnote{We discuss later in this paper how the optimal allocation function can be computed.}. This allocation function divides the region with $b_1,b_2\leq 0.465$ into three subregions below and provides a different recruitment plan for each subregion:
	\begin{itemize}
		\item \textbf{Subregion 1 ($b_1,b_2 \leq 0.27$):} Both providers get recruited simultaneously at time $0$.
		\item \textbf{Subregion 2 ($0.27< b_2 \leq 0.465$, $b_1 \leq b_2$):} Provider 1 gets recruited at time $0$. Provider 2 is recruited in the future (i.e., at a time $\tau^*(b_2)$), if the task is not completed by then.
		\item \textbf{Subregion 3 ($0.27< b_1 \leq 0.465$, $b_2 < b_1$):} Provider 2 gets recruited at time $0$. Provider 1 is recruited in the future (i.e., at a time $\tau^*(b_1)$), if the task is not completed by then.
	\end{itemize}
	The optimal invocation time $\tau^*(.)$ can be derived by techniques that we will develop in Section \ref{sec:opt-time}. However, for the sake of simplicity and since in this example our goal is just to show how gradual recruitment can reduce the payments, we consider a suboptimal invocation time $\tau(b)=0.3$, for all $0.27< b \leq 0.465$. With this invocation time, the invocation probability of the provider who is second in the queue is $1-G_i(0.3)=0.74$.
	
	For this allocation function, we can derive the optimal payment function according to \eqref{eq:opt-payment}, as follows:
	\begin{align}\label{eq:ex0-2}
		\boldsymbol{T}^*(b_1,b_2)=
		\begin{cases}
			(0.414,0.414), & \text{if } (b_1,b_2) \in \text{Subregion 1}, \\
			(0.3441+0.26 b_2,0.465),			& \text{if } (b_1,b_2) \in \text{Subregion 2},\\
			(0.465,0.3441+0.26 b_1),			& \text{if } (b_1,b_2) \in \text{Subregion 3},\\
			(b_2,0),			& \text{if } b_2 > 0.465, b_1 \leq b_2,\\
			(0,b_1),			& \text{if } b_1 > 0.465, b_2 < b_1.\\
		\end{cases}
	\end{align}
	Comparing \eqref{eq:ex0-2} with \eqref{eq:ex0-1}, we can see that in each Subregion 1-3, the gradual recruitment reduces the total cost of incentivizing providers to submit true bids. This fact is more attractive and apparent in Subregion 1, as in this region, the allocation function $A^*$ is exactly the same as $A_{no}$. Therefore, the gradual recruitment has not changed the providers' invocation probabilities in this region. However, due to the changes it has made to the allocation functions of other regions, the payments of Subregion 1 are reduced as well. 
	
	To see the reason, let's take a closer look at ${T}^*_1(b_1,b_2)$ for Subregion 1. In this subregion, the invocation probability of provider 1 is $1$, i.e., $P_1(A^*(b_1,b_2))=1$. If provider 1 increases its bid $\hat{b}_1$, its invocation probability remains $1$ up until $\hat{b}_1=0.27$, and then drops to $0.74$, for $0.27< \hat{b}_1 \leq 0.465$. Provider 1 will have no chance for being invoked if it submits a bid greater than $0.465$. Based on these values, we can use \eqref{eq:opt-payment} to compute payment ${T}^*_1(b_1,b_2)$ for Subregion 1 as follows:
	\begin{equation}\label{eq:ex0-3}
		{T}^*_1(b_1,b_2)=b_1+\int_{b_1}^{0.27}{1 \hspace{0.05cm} d\hat{b}_1}+\int_{0.27}^{0.465}{0.74 \hspace{0.05cm} d\hat{b}_1}=0.414.
	\end{equation}
	Comparing \eqref{eq:ex0-3} with ${T}_{no,1}(b_1,b_2)=b_1+\int_{b_1}^{0.465}{1 \hspace{0.05cm} d\hat{b}_1}=0.465$ clarifies why payments of Subregion 1 are reduced when gradual recruitment is employed. In auction $M^*$, if providers increase their bids, they will be at the risk of losing their first position in the recruitment  queue and going to the second position. If this happens, the providers' invocation probabilities and hence their expected utility decreases. Therefore, the providers have less incentive to submit higher bids and hence it is easier for the consumer to incentivize them to tell the truth.
	\demo
\end{example}

Example \ref{Ex0} illustrates the impact of gradual recruitment and the weighted threshold payment scheme on lowering the consumer's cost. In the next subsection, we use the result of this section to formulate the optimization problem we need to solve for computing the optimal allocation function.

\subsection{Formulating Problem \eqref{eq:consumer-problem} in terms of the Allocation Function }\label{sec:opt-allocation-function}

Lemma \ref{L-Myerson2} reduces the complexity of designing an optimal auction by expressing the optimal payment function $\boldsymbol{T}(.)$ in terms of the allocation function $A(.)$. This result helps us to focus on optimizing the allocation rule $A(.)$ and then use equation \eqref{eq:opt-payment} to derive the minimum payments required to make the auction with this allocation function incentive compatible and individually rational.

Following this approach, in the next lemma, we state the optimal auction design problem solely in terms of the allocation function $A(.)$.

\begin{lemma}\label{L-Myerson3}
	Suppose that the allocation function $A: \mathcal{C}^n \rightarrow \mathcal{H}$ solves the following optimization problem:
	\begin{subequations}\label{eq:consumer-problem2}
		\begin{align}
			& \max_{ A} {\quad \int{[V P_{succ}(A(\boldsymbol{c}),D)-\sum_{i \in \mathcal{N}}{(c_i+\frac{F_i(c_i)}{f_i(c_i)})P_i(A(\boldsymbol{c}))}]f(\boldsymbol{c})d\boldsymbol{c}}}\label{eq:opt-objective}\\
			\textbf{s.t.} &\quad (b_i - c_i)( Q_i(A,b_i)- Q_i(A,c_i)) \leq 0, \quad \forall i \in \mathcal{N}, c_i,b_i \in \mathcal{C}.\label{eq:monotone-alloc}
		\end{align}
	\end{subequations}
	Then, $M=(A(.),\boldsymbol{T}^*(.))$, where $\boldsymbol{T}^*(.)$ is derived based on \eqref{eq:opt-payment}, represents the optimal revenue-maximizing auction.
\end{lemma}

The objective function of problem \eqref{eq:consumer-problem2} is similar in form to the expected social-welfare of an auction with allocation function $A(.)$, which is defined as
\begin{equation}\label{eq:Social-welfare}
	SW(A)=\int{[V P_{succ}(A(\boldsymbol{c}),D)-\sum_{i \in \mathcal{N}}{c_iP_i(A(\boldsymbol{c}))}]f(\boldsymbol{c})d\boldsymbol{c}}.
\end{equation}
Social-welfare is the sum of the utilities of the consumer and the service providers in an auction. The expected social-welfare $SW(A)$ is derived by taking an expectation over the joint statistics of the providers' costs. We can see that the objective function \eqref{eq:opt-objective} is similar to the expected social-welfare \eqref{eq:Social-welfare}, except that the real cost $c_i$ of each provider $i$ is replaced by a virtual cost $\phi_i(c_i)=c_i+\frac{F_i(c_i)}{f_i(c_i)}$. We see by definition that the providers' virtual costs are always greater than or equal to their actual costs, i.e., $\phi_i(c_i) \geq c_i$. This basically means that the cost that a revenue maximizing consumer evaluates for recruiting each provider $i$ is higher than its actual cost. The reason is simple. According to \eqref{eq:opt-payment}, the consumer should pay each recruited provider $i$ not only its actual cost $c_i$, but also an information rent to induce the provider to disclose its private information. The difference between the virtual costs and the real costs of each provider is the expectation of the information rent that must be paid to the providers to reveal truthful information.

	\section{Optimal Allocation Function} \label{sec:allocation-function}
	
	The purpose of this section is to find an allocation function $A: \mathcal{C}^n \rightarrow \mathcal{H}$ that solves optimization problem \eqref{eq:consumer-problem2}. For each $i, c_i$ and $b_i$, the monotonicity constraint \eqref{eq:monotone-alloc} depends on the recruitment strategies that allocation function $A$ assigns not to a single cost vector, but rather to a set of cost vectors $\{(c_i,\boldsymbol{c}_{-i}),(b_i,\boldsymbol{c}_{-i}),\boldsymbol{c}_{-i} \in \mathcal{C}^{n-1}\}$. This interaction among decision variables makes optimization problem \eqref{eq:consumer-problem2} non-separable across cost vectors and hence difficult to solve.
	However, the next lemma proves that the monotonicity constraint \eqref{eq:monotone-alloc} is satisfied ``for free'' at the optimal allocation.
	
	\begin{lemma} \label{L-monotonicity}
		When the providers' cost distributions $F_1, \ldots, F_n$ are regular, i.e., virtual costs are increasing in costs, the allocation function that maximizes \eqref{eq:opt-objective} is monotone and satisfies constraint \eqref{eq:monotone-alloc}.  
	\end{lemma}
	
	Lemma \ref{L-monotonicity} allows us to relax the monotonicity constraint and be sure that this condition will be automatically satisfied at the final solution. If we drop the monotonicity constraint \eqref{eq:monotone-alloc}, the problem becomes separable and can be decomposed into the following unconstrained subproblems $SP(\boldsymbol{c})$, $\boldsymbol{c} \in \mathcal{C}^n$:
	
	\begin{align}\label{eq:consumer-problem3}
		SP(\boldsymbol{c}): \max_{ A(\boldsymbol{c}) \in \mathcal{H}} {\quad V P_{succ}(A(\boldsymbol{c}),D)-\sum_{i \in \mathcal{N}}{\phi_i(c_i)P_i(A(\boldsymbol{c}))}}. \nonumber
	\end{align}
	
	For each cost vector $\boldsymbol{c}$, problem $SP(\boldsymbol{c})$ aims to find the optimal recruitment strategy $A^*(\boldsymbol{c})=((s_1,\tau_1),\ldots,(s_m,\tau_m))$ that specifies the optimal ordered set of providers $A^*_s(\boldsymbol{c})=(s_1,\ldots,s_m)$ and their optimal invocation times $A^*_{\tau}(\boldsymbol{c})=(\tau_1,\ldots,\tau_m)$. This problem is a mixture of continuous and combinatorial optimization problems. Computing the optimal procurement times $A^*_{\tau}(\boldsymbol{c})$ for each ordered set of providers $A_s(\boldsymbol{c})$ is a continuous optimization problem. However, determining the optimal ordered set $A^*_s(\boldsymbol{c})$ is a combinatorial optimization problem. Combinatorial optimization problems can be viewed as searching for the best element of some set of discrete items. Therefore, in principle, any sort of search algorithm can be used to solve them. However, the number of ordered subsets of $n$ providers can, in practice, become quite large (e.g. 1,956 for $n=6$ providers but 1,302,061,344 for $n=12$ providers). Exploring such a huge search space is computationally prohibitive and needs major accelerations in order to be used practically.
	
	We proceed as follows to solve optimization problem $SP(\boldsymbol{c})$ for each cost vector $\boldsymbol{c}$. In Subsection \ref{sec:opt-time}, we focus on the continuous part of $SP(\boldsymbol{c})$ and discuss how to derive the optimal invocation times $A^*_{\tau}(\boldsymbol{c})=(\tau_1,\ldots,\tau_m)$ for a fixed provider sequence $A_s(\boldsymbol{c})$. We make use of this result in Subsection \ref{sec: BnB-algorithm} to develop a branch-and-bound algorithm for solving the combinatorial part of the problem.
	
	\subsection{Continuous Part: Optimal Invocation Times}\label{sec:opt-time}
	
	In this part, we assume that the optimal subset of providers and their ordering is given. That is, we are given an ordered set of providers $A_s(\boldsymbol{c})=(s_1,\ldots,s_m)$, where $s_i$ is invoked before $s_{i+1}$. To compute the optimal invocation time, we must determine $A_{\tau}(\boldsymbol{c})=(\tau_1,\ldots,\tau_m)$, where $\tau_i$ is the invocation time of $s_i$, such that the objective function of problem $SP(\boldsymbol{c})$ gets maximized. 
	
	Using \eqref{eq:psucc} and \eqref{eq:prob_selection}, we can write the objective function of $SP(\boldsymbol{c})$ in terms of $s_i$ and $\tau_i$, $i=1,\ldots,m$, as follows:
	\begin{equation}
		f(\tau)=V (1-\prod_{i=1}^m{(1-G_{s_i}(D-\tau_i))})-\sum_{i=1}^m{\phi_{s_i}(c_{s_i})\prod_{j=1}^{i-1}{(1-G_{s_j}(\tau_i-\tau_j))}}.
	\end{equation}
	To derive the optimal invocation time, we must solve an optimization problem that maximizes $f(\tau)$ under the following constraints: $\forall i: 0 \leq \tau_i \leq D$ and $\forall i,j: i < j \implies \tau_i \leq \tau_j$. This is a continuous non-convex optimization problem that can be solved numerically by existing methods such as stochastic gradient descent \cite{Robbins1951} and saddle-free Newton \cite{Dauphin2014}.
	We denote the function that returns the optimal invocation time for provider sequence $(s_1,\ldots,s_m)$ by $Times(s_1,\ldots,s_m)$.
	
	\subsection{Combinatorial Part: Optimal Provider Sequence} \label{sec: BnB-algorithm}
	One of the most universally applicable approaches for reducing the search space in combinatorial optimization problems is branch-and-bound algorithm. Such algorithms recursively split the search space into smaller groups; this splitting is called branching. Branching alone would amount to brute-force enumeration of candidate solutions and testing them all. To improve on the performance of brute-force search, a branch-and-bound algorithm keeps track of a lower bound and an upper bound for all solutions in given groups. The algorithm uses these bounds to prune the search space, eliminating groups that it can prove will not contain an optimal solution (i.e., its upper bound is less than the lower bound of another group).
	This systematic partitioning of the search space and removing the groups that cannot contain an optimal solution can potentially reduce the space of solutions that have to be searched. However, the performance of a branch-and-bound algorithm depends crucially upon the appropriate selection of branching and bounding techniques.

	In this paper, due to the similarity of the objective function of problem $SP(\boldsymbol{c})$ to the social-welfare (as discussed in Section \ref{sec:opt-allocation-function}), we adopt the branching and bounding techniques proposed in \cite{STEIN2011} for deriving a social-welfare maximizing recruitment strategy $A(\boldsymbol{c})$. We briefly discuss each of these techniques below.
	
	\textbf{Branching technique:} The main idea is to partition the search space, which is the set of all ordered subsets of providers $\mathcal{N}=\{1,\ldots,n\}$, based on their first $r$ elements. $r$ is a parameter that starts from $1$ and increases gradually over iterations. The pseudo-code of this algorithm is presented in Algorithm \ref{alg2}.
	In the first iteration, the search space is partitioned into $n$ groups $\{\left\langle 1\right\rangle,\left\langle 2\right\rangle,\ldots,\left\langle k\right\rangle\}$ (Line 3), where each $\left\langle i\right\rangle$ represents the set of orderings that have $i$ as their first element. For example, for $n=3$, we have $\left\langle 1\right\rangle=\{(1),(1,2),(1,3),(1,2,3),(1,3,2)\}$. After this partitioning, the algorithm calculates a lower bound and an upper bound for the best consumer's utility within each group (Lines 5-6) and keeps track of the highest lower bound found so far (Lines 7-10). We will discuss how the Lower and Upper functions work later.
	The highest lower bound $U_{low}$ serves as a touchstone for assessing the quality of the groups to be explored in the future. A group will be discarded if its upper bound is lower than or equal to $U_{low}$ (Lines 11-13).
	
	In the second iteration, the algorithm picks the group with the highest lower bound, as this group is most likely to contain the optimal ordering, and partitions it into $n-1$ smaller-sized groups based on the first two elements (Lines 16-18). The function $Expand(o)$ in Line 18 sub-partitions group $o$ based on the next unfixed element. That is, if the orderings within $o$ match in their first $r$ elements, the function $Expand(o)$ partitions them based on their $(r+1)$-$th$ elements. The aim of this function is to create the next generation of groups to be explored more precisely.
	For example, if $\left\langle 1\right\rangle$ is picked at Line 16 when $k=3$, the next generation of groups will be $Expand(\left\langle 1\right\rangle)=\{\left\langle 1,2\right\rangle,\left\langle 1,3\right\rangle\}$.
	These new groups will be evaluated based on the highest potential revenue they can provide for the consumer and will be discarded if their upper bound is below $U_{low}$ (Lines 19-29). The groups that survive from this generation along with the ones that survive from the first generation (Line 30) go for another round of branching and bounding. This process continues until there is no branchable group left.

	\begin{algorithm}[t]
		\begin{multicols}{2}
			\textbf{Input:} Cost vector $\boldsymbol{c}$, Duration functions $\{G_i(.)\}_{i \in \mathcal{N}}$, Deadline $D$\;
			Let $o^*(\boldsymbol{c})=\emptyset, U_{low}=0$\;
			$Q=\{\left\langle 1\right\rangle,\left\langle 2\right\rangle,\ldots,\left\langle n\right\rangle\}$\;
			\For{all $o' \in Q$}{
				$\underbar{u} \gets Lower(o')$\;
				$\bar{u} \gets Upper(o')$\;
				\If{$\underbar{u} > U_{low}$}{
					$o^*(\boldsymbol{c}) \gets o'$\;
					$U_{low} \gets \underbar{u}$\;
				}
				\If{$\bar{u} \leq U_{low}$}{
					$Q \gets Q - \{o'\}$\;
				}
			}
			\While{$Q \neq \emptyset$}{
				$o \gets \argmax_{o \in Q}{Lower(o)}$\;
				$Q \gets Q-\{o\}$\;
				$P \gets Expand(o)$\;
				\For{all $o' \in P$}{
					$\underbar{u} \gets Lower(o')$\;
					$\bar{u} \gets Upper(o')$\;
					\If{$\underbar{u} > U_{low}$}{
						$o^*(\boldsymbol{c}) \gets o'$\;
						$U_{low} \gets \underbar{u}$\;
					}
					\If{$\bar{u} \leq  U_{low}$}{
						$P \gets P - \{o'\}$\;
					}
				}
				$Q \gets \{x \in Q \cup P | Upper(x) > U_{low}\}$\;
			}
			$A^*_s(\boldsymbol{c}) \gets o^*(\boldsymbol{c})$\;
			$A^*_{\tau}(\boldsymbol{c}) \gets Times(o^*(\boldsymbol{c}))$\;
		\end{multicols}
		\textbf{Output: Optimal recruitment strategy $A^*(\boldsymbol{c})=(A^*_s(\boldsymbol{c}),A^*_{\tau}(\boldsymbol{c}))$}
		\caption{Branch-and-Bound Algorithm}\label{alg2}
	\end{algorithm}

	\textbf{Lower bounding technique:} Deriving a lower bound for the highest expected utility the orderings within a group can provide to the consumer is simple. This is because the best expected utility provided by any member of the group can be interpreted as a lower bound for the maximum utility the whole group can provide to the consumer. 
	
	The groups we consider in our branch-and-bound algorithm are of the form $\left\langle {i_1},{i_2},\ldots, {i_j}\right\rangle$, where $j \leq n$, and ${i_s} \in \mathcal{N}$, for all $1 \leq s \leq j$. Each group $o=\left\langle {i_1},{i_2},\ldots, {i_j}\right\rangle$ is the set of all orderings that start with ${i_1},\ldots, {i_j}$. 
	We select ordering $R(o)=({i_1},\ldots, {i_j}) \in o$ as a representative for group $o$. In ordering $R(o)$, the consumer invokes ${i_1},\ldots, {i_j}$ in turn and does not invoke any other providers. The optimal invocation time for ordering $R(o)$ is $(t^*_1,\ldots,t^*_j)=Times(R(o))$, where $Times(.)$ is the function derived in Section \ref{sec:opt-time}. Using this result, we can construct the best recruitment strategy with the ordering $R(o)$ as $\rho^*(o)=(({i_1},t^*_1),\ldots,({i_j},t^*_j))$. We consider the consumer's expected utility when employing recruitment strategy $\rho^*(o)$ as the lower bound for the highest expected utility members of the group $o$ can provide to the consumer. That is, 
	\begin{equation}\label{eq:lower_bnd}
		Lower(o)=V P_{succ}(\rho^*(o),D)-\sum_{i \in \mathcal{N}}{\phi_i(c_i)P_i(\rho^*(o))}.
	\end{equation}

	\textbf{Upper bounding technique:} Calculating an upper bound on the expected utility that members of a group can provide for the consumer is not immediately obvious. Obtaining a tight upper bound requires computing the utility provided by every member of the group, which is too cumbersome and probably impractical. However, we can obtain a looser (but still effective) upper bound by applying the following approach.
	
	For each group $o=\left\langle {i_1},{i_2},\ldots, {i_j}\right\rangle$, we define
	\begin{equation}
		\mathcal{N}'(o)=\{i \in \mathcal{N} | i \notin \{{i_1},{i_2},\ldots, {i_j}\}\}.
	\end{equation}
	If $\mathcal{N}'(o) = \emptyset$, then $({i_1},{i_2},\ldots, {i_j})$ is the only member of group $o$. Therefore, the upper bound is equal to the lower bound discussed above. Otherwise, we create a virtual service provider $s_o$ that dominates all subsets of $\mathcal{N}'(o)$. This service provider has the minimum cost among all providers $\mathcal{N}'(o)$, i.e., $c_{s_o}=\min_{i \in \mathcal{N}'(o)}{c_{i}}$, and can perform the task faster than any combination of providers $\mathcal{N}'(o)$, i.e., $G_{s_o}(x)=1-\prod_{i \in \mathcal{N}'(o)}{(1-G_{i}(x))}$. Therefore, invoking $s_o$ is strictly better than invoking any subset of providers $\mathcal{N}'(o)$. With this reasoning, we obtain a new ordering $\bar{o}=({i_1},{i_2},\ldots, {i_j}, s_o)$ by appending $s_o$ to $o$ and then calculate the upper bound as the maximum revenue obtained by employing recruitment strategy $\rho^*(\bar{o})=(\bar{o},Times(\bar{o}))$, if this revenue is higher than the lower bound $Lower(o)$. Failure to meet this condition indicates that it is not possible to achieve a higher utility by invoking further providers; therefore, we can set the upper bound equal to the lower bound. That is,
	\begin{equation}\label{eq:upper_bnd}
		Upper(o)=\max{(Lower(o),V P_{succ}(\rho^*(\bar{o}),D)-\sum_{i \in \mathcal{N}}{\phi_i(c_i)P_i(\rho^*(\bar{o}))})}.
	\end{equation}
	
	Running Algorithm \ref{alg2} with the lower and upper bounding techniques described above, we can derive the optimal ordering $o^*(\boldsymbol{c})$ for recruiting providers $\mathcal{N}$. Based on the discussions in Section \ref{sec:opt-time}, the optimal time for hiring these providers can be computed as $Times(o^*(\boldsymbol{c}))$. Therefore, we can denote the optimal recruitment strategy as $A^*(\boldsymbol{c})=(o^*(\boldsymbol{c}),Times(o^*(\boldsymbol{c})))$ (Lines 32-34).

	\section{The Weighted Gradual Procurement Auction}\label{sec:properties}
	
\begin{figure}[t]
	\centering
	\includegraphics[height=0.34\textwidth]{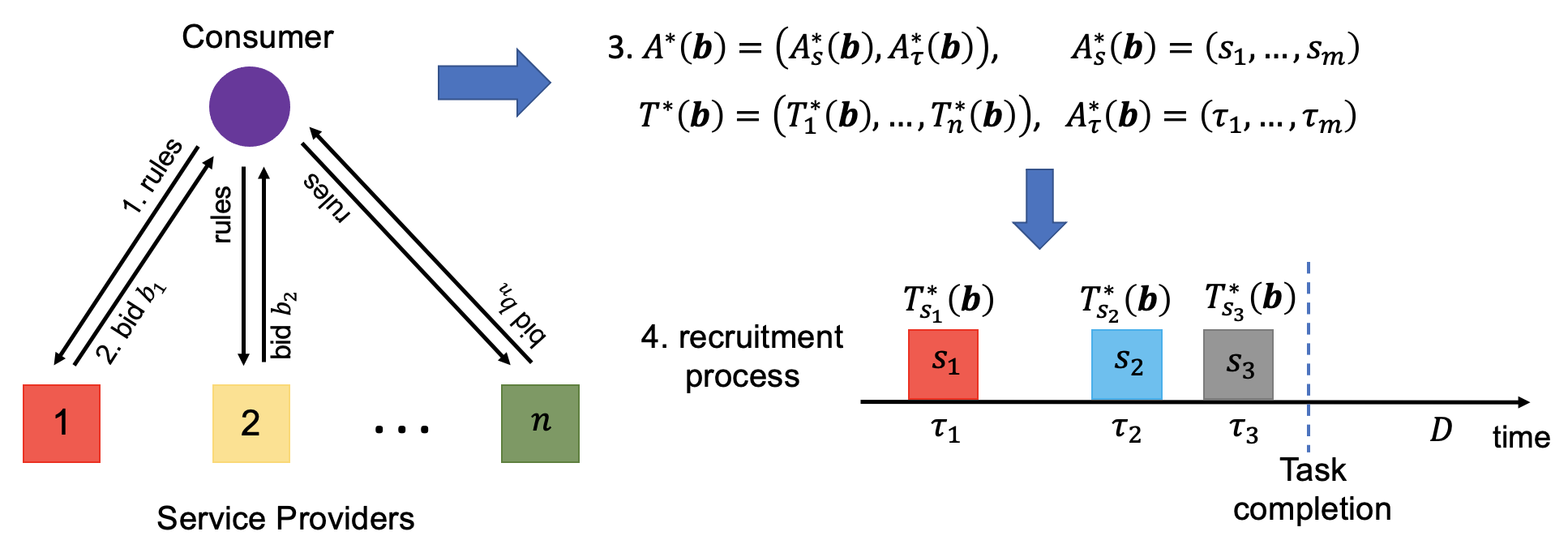}
	\caption{The general structure of the WGPA auction. This auction is composed of the following four steps:  Step 1 (Auction's rules announcement), Step 2 (Bidding), Step 3 (Determining the recruitment and payment strategies), and Step 4 (Recruitment process). These steps are explained thoroughly in Section \ref{sec:properties}.}\label{fig:algorithm-steps}
\end{figure}

The allocation function $A^*(.)$, presented in Section \ref{sec:allocation-function}, along with the weighted threshold payment function $\boldsymbol{T}^*(.)$, derived in Section \ref{sec:optimal-pay}, specifies our proposed auction $M^*=(A^*(.),\boldsymbol{T}^*(.))$. We call this a weighted gradual procurement auction (WGPA) and the general structure of WGPA is as follows (see Fig. \ref{fig:algorithm-steps}):
\begin{enumerate}
	\item \textbf{Auction's rules announcement:} The consumer publicly announces that the allocation function $A^*(.)$ and the payment function $T^*(.)$ are going to be used during the auction. 
	\item \textbf{Bidding:} Each provider $i$ analyzes the auction's rules and submits a bid $b_i$ that best serves its interest;
	\item \textbf{Determining the recruitment and payment strategies:} Based on the submitted bids, the consumer selects the recruitment strategy $A^*(\boldsymbol{b})$ and the payment strategy $\boldsymbol{T}^*(\boldsymbol{b})$;
	\item \textbf{Recruitment process:} The consumer recruits providers according to the ordering $A^*_s(\boldsymbol{b})$ at times $A^*_{\tau}(\boldsymbol{b})$, until either the task is completed  or the deadline is reached. Each provider $i$ receives a payment $T_i^*(\boldsymbol{b})$ once it is recruited.
\end{enumerate}

As we discussed in Section \ref{sec:payment-function}, the weighted threshold payment function $T^*(.)$ is designed to guarantee incentive compatibility and interim individual rationality. Moreover, the allocation function $A^*(.)$ is designed to provide the highest possible revenue to the consumer. We elaborate these arguments in the next subsection to present and prove the fundamental properties of WGPA.

\subsection{Fundamental Properties of WGPA}\label{sec:prop}

In this section, we prove theoretically that our proposed WGPA auction satisfies all three desirable properties we were looking for. That is, the auction is individually rational, incentive compatible, and revenue-optimal. We also prove that WGPA can always guarantee a non-negative expected revenue for the consumer.

\begin{proposition}[Incentive Compatibility]\label{Prop-IC}
	The truth-telling bidding strategy is a BNE of each game induced by WGPA. That is, providers cannot benefit by misreporting their private costs.
\end{proposition}

\begin{proposition}[Interim and Ex-Post IR]\label{Prop-IR}
	WGPA is both interim and ex-post individually rational. That is, 
	\begin{itemize}
		\item (Interim IR) it is rational for each provider $i$ to participate in the auction; 
		\item (Ex-post IR) the recruited providers perform the task upon recruitment and do not regret their decision.
	\end{itemize}
\end{proposition}

\begin{proposition}[Revenue Maximization]\label{Prop-RM}
	WGPA provides the consumer with the maximum expected revenue that any incentive compatible and individually rational auction could provide.
	%	
	%	There is a threshold $B_{th}$ above which the auction $M^*=(A^*(.),\boldsymbol{T}^*(.))$ is revenue maximizer.
\end{proposition}

\begin{proposition}[Non-negative Expected Revenue]\label{Prop-PR}
	The consumer's expected revenue is always non-negative in WGPA. Therefore, it is rational for the consumer to participate in WGPA.
\end{proposition}

\subsection{Computational Complexity}
Propositions \ref{Prop-IC}-\ref{Prop-RM} prove that our proposed WGPA auction achieves the main properties we were looking for. However, this is at the cost of high computational complexity to find the optimal allocation function using Algorithm \ref{alg2}.
%When a bid vector $\boldsymbol{b}$ is submitted, the consumer needs to run Algorithm \ref{alg2} not only for the submitted bid $\boldsymbol{b}$, but also for all bid vectors that appear in the integral of \eqref{eq:opt-payment}, for all $i \in \mathcal{N}$, to compute the payments. The integral of \eqref{eq:opt-payment} can be approximated with a Riemann sum when the interval $[b_i,c_{max}]$ is divided into some subintervals with equal lengths $stepsize$. Using this technique, the number of times the consumer needs to run Algorithm \ref{alg2} to compute all providers' payments is upper bounded by $n c_{max}/stepsize$.
In Algorithm \ref{alg2}, we tried to reduce the computational complexity by using a branch-and-bound algorithm instead of brute-force search. However, this algorithm still searches for the optimal solution and in the worst case, it may need to examine the entire search space. This may be the case, for example, when there are large numbers of highly similar providers and when the value of the task is very large in relation to the service costs. 

To reduce the worst-case complexity, in the next subsection, we present a low-complexity heuristic algorithm (i.e., Algorithm \ref{alg4}) that can be substituted for Algorithm \ref{alg2}. As we will show in Section \ref{sec:numerical}, this algorithm reduces the computational complexity by over $99\%$ with less than $1\%$ performance loss.

\subsection{Heuristic Algorithm for Deriving a Near-Optimal Allocation Function} \label{sec:heuristic}

Algorithm \ref{alg4} is based on a greedy-like search and aims to find a near optimal ordering for recruiting providers. It starts with an empty ordering and then greedily adds, removes or switches providers until a local optimum is reached. Intuitively, this algorithm benefits from selecting providers that offer a good value/cost trade-off. By also allowing providers to be removed or switched, it has some backtracking capabilities — thus an expensive but reliable provider can eventually be replaced by many cheap and unreliable providers that individually do not yield a high expected utility, but in combination result in a better strategy. 

\begin{algorithm}[t]
	\textbf{Input:} Cost vector $\boldsymbol{c}$, Duration functions $\{G_i(.)\}_{i \in \mathcal{N}}$, Deadline $D$\;
	Let $o_{apx}(\boldsymbol{c})=\emptyset, U_{best}=0$\;
	$flag=0$\;
	\While{$flag=0$}{
		$P \gets Neighbors(o_{apx}(\boldsymbol{c}))$\;
		$o' \gets \argmax_{o \in P}{Lower(o)}$\;
		\uIf{$Lower(o') > U_{best}$}{
			$o_{apx}(\boldsymbol{c}) \gets o'$\;
			$U_{best} \gets Lower(o')$\;
		}
		\Else{$flag \gets 1$\;}
	}
	\textbf{Output: Near-optimal ordering $o_{apx}(\boldsymbol{c})$}
	\caption{Heuristic Algorithm (Alternative to Algorithm \ref{alg2})}\label{alg4}
\end{algorithm}

In more detail, the algorithm stores the best ordering found so far and the expected revenue it can provide to the consumer by $o_{apx}(\boldsymbol{c})$ and $U_{best}$, respectively (Line 2). 
%
%to keep the . This set is initialized to the empty set (Line 2). The consumer's expected revenue when it recruits providers based on ordering $o_{apx}(\boldsymbol{c})$ at times $Times(o_{apx}(\boldsymbol{c}))$ is stored by $U_{best}$. 
The algorithm also sets a flag that takes value $1$ once a local optimum is found (Line 3). In each round, the algorithm checks to see if ordering  $o_{apx}(\boldsymbol{c})$ has a neighbor that provides better revenue for the consumer (Lines 5-7). If such neighbors exist, the algorithm replaces $o_{apx}(\boldsymbol{c})$ by its best neighbor and goes to the next round (Lines 7-9). This process continues until the best ordering found so far has no better neighbor and hence is a local optimum.

For the algorithm description to be complete, we need to define the neighborhood of an ordering. The larger the neighborhood is, the better the quality of the returned solution is, but at the same time the longer it takes to search the neighborhood at each iteration (Line 6). As an extreme case, if all orderings are defined to be neighbors, the algorithm will find the exact optimum ordering in just one round, but at the cost of doing an exhaustive search over all orderings.

In this paper, we define the neighborhood of an ordering $o$ as the set of all orderings that can be obtained by any of the three following actions: 1) selecting a provider $i$ which is currently not in $o$ and adding it to $o$ at any possible position, 2) selecting a provider $i \in o$ and removing it, 3) selecting two providers $i, j \in o$ and swapping their turns. We will show in Section \ref{sec:numerical} that this definition of neighborhood is wide enough to ensure the final solution is close to the exact optimum. At the same time, it is narrow enough to enable the algorithm to scale to realistic settings with hundreds of providers. 

We close this section by Proposition \ref{Prop-Heuristic}. This proposition proves that replacing Algorithm \ref{alg2} with Algorithm \ref{alg4} does not ruin the incentive compatibility, individual rationality, or monotonicity of the auction. The only difference is a significant complexity reduction at the expense of less than one percent decrease in the consumer's expected revenue, as will be shown in Section \ref{sec:heuristic-numerical}.

\begin{proposition}[Auction with Heuristic Allocation Function]\label{Prop-Heuristic}
	The auction $M_{apx}=(A_{apx}(.),\boldsymbol{T}^*(.))$, where $A_{apx}(\boldsymbol{c})=(o_{apx}(\boldsymbol{c}),Times(o_{apx}(\boldsymbol{c})))$ is derived by running Algorithm \ref{alg4}, satisfies monotonicity, ex-post and interim individual rationality and incentive compatibility.
\end{proposition}
	
	\section{Numerical Results} \label{sec:numerical}
	
Having proposed a revenue-optimal procurement auction that is able to incentivize providers to reveal their private information truthfully, we now evaluate the performance of this auction in a variety of simulated environments. 
The purpose of this evaluation is four-fold and it should be seen as a complement to the theoretical results of the previous sections. First, we investigate how the redundant allocation and gradual recruitment techniques enable WGPA to adapt its recruitment strategies to different situations (Section \ref{sec:flexibility-numerical}). Second, we demonstrate how much WGPA enhances the performance (in terms of both revenue and social-welfare) compared with those of the available benchmarks (Section \ref{sec:becnhmark}). Third, we evaluate the performance of sub-optimal auction $M_{apx}$ proposed in Section \ref{sec:heuristic} and show that this auction significantly reduces the running time without noticeably reducing the
performance (Section \ref{sec:heuristic-numerical}). Finally, we check the robustness of our results in Section \ref{sec:impact-assumptions}.

\subsection{Flexibility of WGPA}\label{sec:flexibility-numerical}

As discussed in Section \ref{sec:define-ass-strategy}, the consumer's recruitment strategy should be flexible, depending on the task's value and the deadline. In this section, we study how the redundant allocation and gradual recruitment techniques we embedded in WGPA help the consumer to achieve this goal.

To do this, we conduct an experiment with $100$ service providers where each provider $i$'s delivery time has an exponential distribution with cumulative distribution function (cdf) $G_i(t)=1-e^{\lambda_i t}=1-e^{-0.01 i t}$. In this experiment, we call parameter $\lambda_i=0.01 i$ the service rate of provider $i$.
%similar to the setting used in \cite{STEIN2011}, we assume that each provider $i$'s duration distribution is exponential with rate $\lambda_i$, i.e., $G_i(t)=1-e^{-\lambda_i t}$. We consider a continuum of providers with service rates ranging on the interval $[0,1]$ and naturally assume that performing the task faster is more expensive, i.e., $\lambda_i > \lambda_j$ implies $c_i > c_j$. \todo{What happens if this is not the case?} 
We also assume, for ease of demonstration, that the execution cost is an identity function of the service rate, i.e., $c_i=\lambda_i$, for all $i=1,\ldots,100$. However, the nature of the results does not depend on this assumption. 

\begin{figure}[t]
	\subfloat[Setting 1: $V=10$, $D=3$]{\label{fig:visualization-1}
		\includegraphics[height=0.35\textwidth]{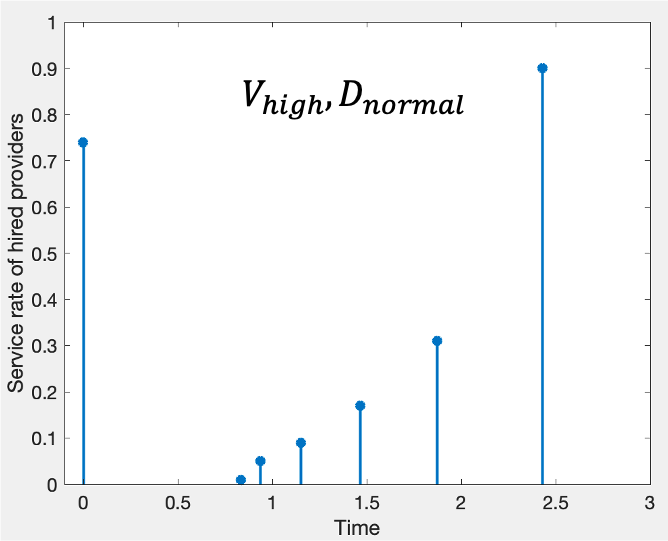}}
	~~~
	\subfloat[Setting 2: $V=4$, $D=3$]{\label{fig:visualization-2}
		\includegraphics[height=0.35\textwidth]{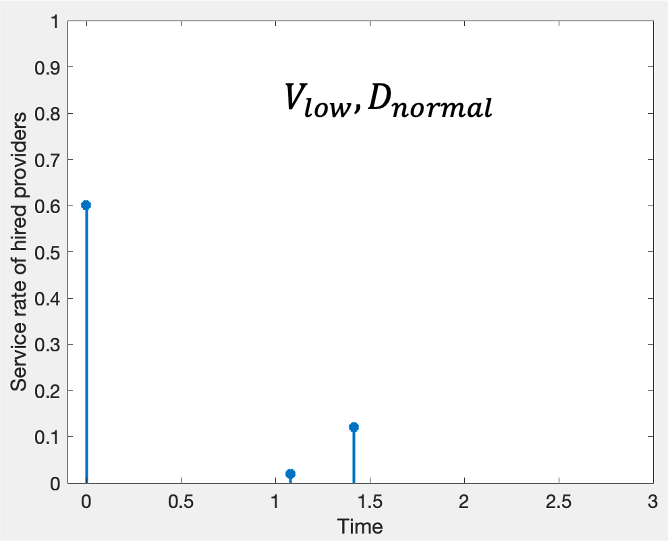}}
	
	\subfloat[Setting 3: $V=10$, $D=1$]{\label{fig:visualization-3}
		\includegraphics[height=0.35\textwidth]{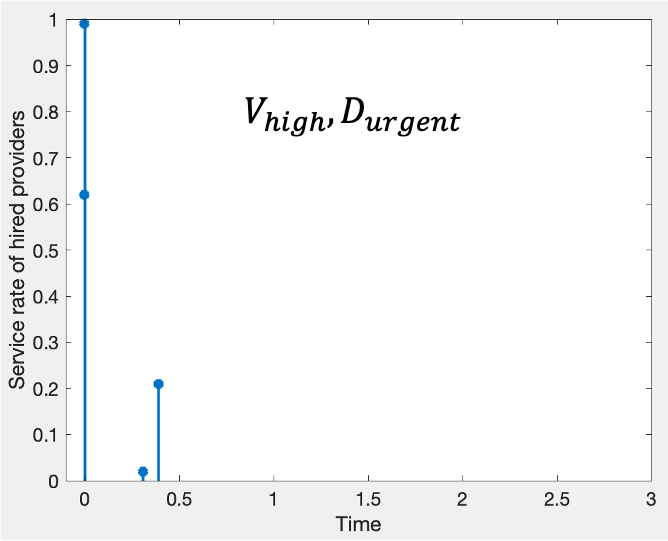}}
	~~~
	\subfloat[Setting 4: $V=4$, $D=1$]{\label{fig:visualization-4}
		\includegraphics[height=0.35\textwidth]{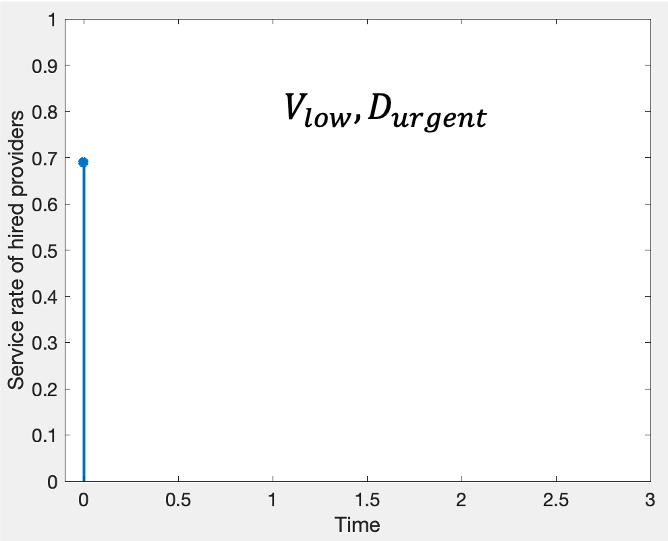}}
	\caption{Revenue-optimal recruitment strategies when $100$ providers with equally-spaced service rates $\lambda \in (0,1]$ are available.}\label{fig:visualization}
\end{figure}

To study a range of environments, we consider four different simulation settings where the task has either a low ($V_{low} =4$) or a high value ($V_{high}=10$) and the deadline is either normal ($D_{normal} =3$) or urgent ($D_{urgent} =1$). In each of these settings, we run Algorithm \ref{alg2} to derive the revenue-optimal recruitment strategy. The results of this experiment are shown in Fig. \ref{fig:visualization}. We can see that in Setting 1 where the task is high-valued and has a normal deadline (i.e., Fig. \ref{fig:visualization-1}), it is optimal for the consumer to start the recruitment process by hiring a relatively fast provider ($\lambda=0.74$) and wait for 0.83 time-units to see if it can perform the task alone. If it cannot, the consumer should recruit some auxiliary providers to increase success likelihood. This phase of recruitment starts with hiring slow but cheap providers and then the closer the consumer is to the deadline, the faster and more expensive providers are hired. Notice that since the completion of the task has a high value for the consumer, it hires a very fast and expensive provider ($\lambda=0.9$) when the deadline is approaching to greatly increase the chance of success.

We can see from Fig. \ref{fig:visualization-2} that in Setting 2 where the task is low-valued and has a normal deadline, the optimal recruitment process has two main differences with that of Setting 1: (i) the first recruited provider is slightly slower ($\lambda=0.6$) and has been given a longer period of time to try the task alone (1.08 time-units); (ii) if the first provider is not successful, the consumer recruits only two cheap auxiliary providers to increase the chance of success. In this case, the task completion is not valuable enough for the consumer to recruit any more expensive providers.

Comparing Figs. \ref{fig:visualization-3} and \ref{fig:visualization-4} with Figs. \ref{fig:visualization-1} and \ref{fig:visualization-2}, respectively, shows how the deadline affects the optimal recruitment strategy. In fact, when the task is urgent, it is better for the consumer to advance the recruitment process by spending more money and recruit faster providers sooner. This basically means that the gradual recruitment technique is of less interest when the task has a tight deadline.

This result motivates us to study the efficiency and necessity of using redundant allocation and gradual recruitment techniques in not just a specific example (i.e., a continuum of providers with $\lambda_i=c_i$), but in more general environments. To this end, for each fixed $V$ and $D$, we run 1000 experiments where in each experiment the number of providers $n$ is chosen uniformly from $[2,20]$ and the costs and service rates are drawn independently and uniformly from $[0,1]$ \footnote{This technique has been previously used in \cite{STEIN2011} for generating random environments.}. To simulate practical settings where faster services are often more costly to procure, in all future experiments (unless otherwise stated), we construct each provider $i$ by assigning both the $i^{th}$ highest cost and the $i^{th}$ highest service rate to it. We will discuss in Section \ref{sec:impact-assumptions}, how the results change if this correlation between the costs and service rates does not exist.

\begin{figure}[t]
	\subfloat[High-valued task ($V=10$)]{\label{fig:behavior-1}
		\includegraphics[height=0.38\textwidth]{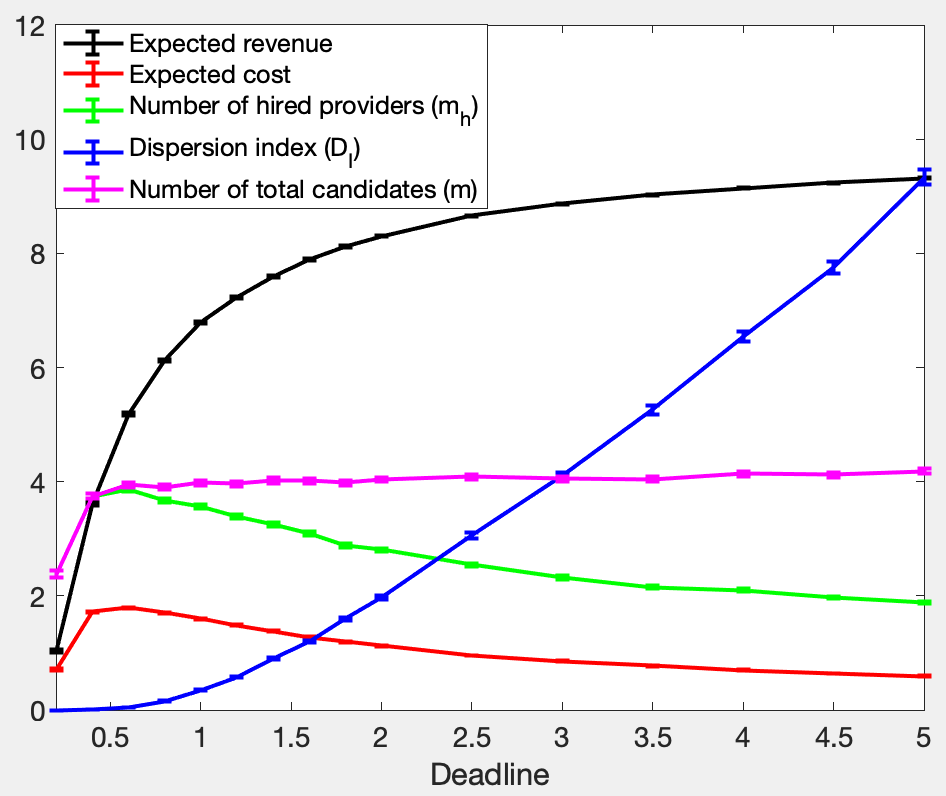}}
	~~~
	\subfloat[Low-valued task ($V=4$)]{\label{fig:behavior-2}
		\includegraphics[height=0.38\textwidth]{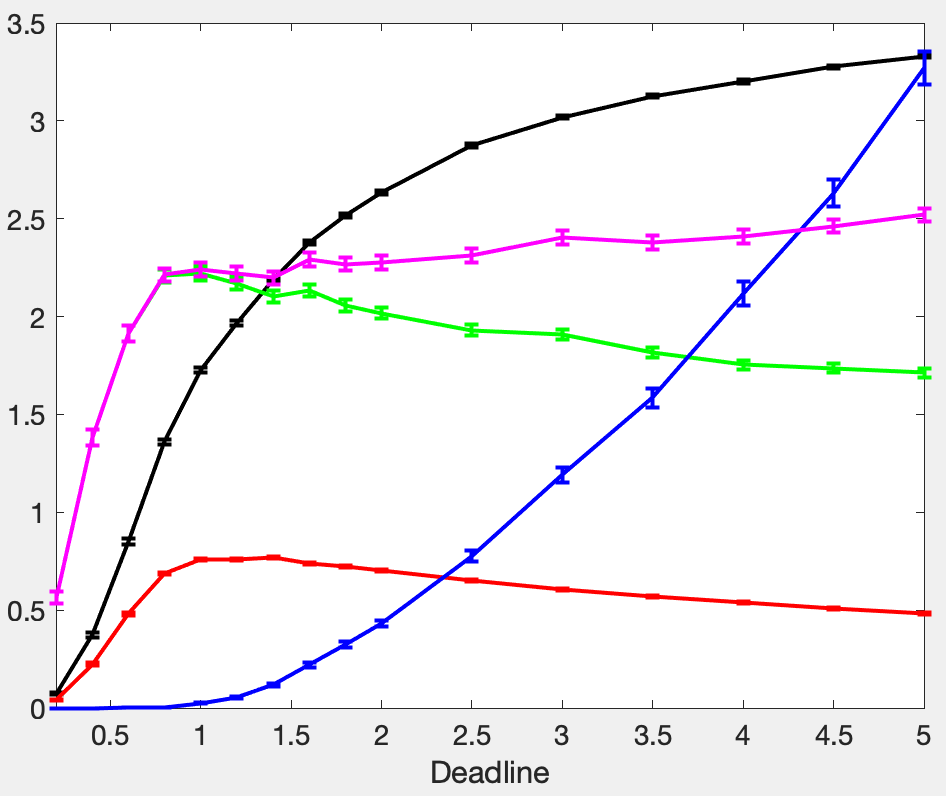}}
	
	\caption{The efficiency metrics for redundant allocation and gradual recruitment techniques}\label{fig:behavior}
\end{figure}

In each experiment, we use the following metrics to measure the efficiency of the redundant allocation and gradual recruitment techniques:
\begin{itemize}
	\item Number of total candidates ($m$): For each recruitment strategy $\rho = ((s_1,\tau_1),\allowbreak \ldots,\allowbreak (s_m,\tau_m))$, $m$ is the number of providers that have the chance to be recruited over time.
	\item Number of hired providers ($m_h$): The recruitment process continues until at least one of the providers completes the task and hence not all of the candidate providers $(s_1,\ldots,s_m)$ are recruited in each experiment. We denote by $m_h$ the number of providers that are actually hired in a single experiment.
	\item Dispersion index ($D_I$): For each recruitment strategy $\rho = ((s_1,\tau_1),\allowbreak \ldots,\allowbreak (s_m,\tau_m))$, we define the dispersion index as $D_I=\tau_1+\ldots + \tau_m$. This index is a measure of the spread of a recruitment process over time. Low values of $D_I$ shows that the recruitment is more centered around time $0$, while high values of $D_I$ shows that the recruitment is distributed over a longer period of time.
\end{itemize}
In Fig. \ref{fig:behavior}, we plot these metrics as well as the expected revenue and the expected cost for the optimal recruitment strategy versus the deadline. We can see that whether the task is high-valued or low-valued, the dispersion index increases when the deadline goes up. This shows that the benefit of using the gradual recruitment technique is highest when the task is less urgent and hence the consumer has more time for recruitment. We can also see that the number of total candidates increases, with a steep change at a certain threshold $th$ ($th=0.4$ when $V=10$ and equals $0.8$ when $V=4$). The number of hired providers, however, increases up to the same threshold $th$ and decreases afterwards. Before threshold $th$, the number of hired candidates is almost the same as the number of candidate providers. This shows that before threshold $th$, redundant allocation is the dominant technique; in this region, the consumer prefers to recruit multiple providers but also prefers to recruit all of them almost simultaneously at time $0$. This fact is also confirmed by small values of $D_I$ before threshold $th$. However, after the threshold,  the importance of gradual recruitment becomes more apparent. In this region, the consumer finds it beneficial to reduce its costs by postponing the hiring of some of the providers.

\subsection{Superiority of WGPA Over the Benchmarks}\label{sec:becnhmark}

In this section, we compare the average revenue and the social-welfare obtained by WGPA to those obtained by the benchmarks (i.e., Bm1-Bm3). We also compare our results to the maximum theoretical revenue that could be achieved if full information was available to the consumer (Bm4).
This theoretical upper bound does not have practical usage in our setting as the cost information is not available to the consumer and must be elicited from self-interested providers.
%
%two benchmark auctions (Bm1-Bm2) that both use threshold payment scheme, but have different allocation functions. We also compare our results with the incentive mechanism proposed in \cite{STEIN2011} for maximizing social-welfare (Bm3). Our last benchmark is the maximum theoretical revenue that could be achieved if full information was available to the consumer (Bm4).
\begin{enumerate}[label=(Bm\arabic*)]
	\item \textbf{Best single auction:} This auction is optimal among non-redundancy-based auctions that assign the task to just one single provider. Such auctions are similar to single-item auctions discussed in Section \ref{sec:literature-optimal} and hence the optimal among them can be derived by using Myerson's idea \cite{Myerson1981}.
	\item \textbf{Best simultaneous auction:} This auction is optimal among redundancy-based auctions that do not employ gradual recruitment technique. Such auctions recruit a set of providers simultaneously at time $0$ to attempt the task in parallel. Auction Bm2 is the solution of a standard homogeneous multi-object auction design problem which can be solved by techniques proposed in \cite{Malakhov2009}.
	\item \textbf{Pairing mechanism \cite{STEIN2011}:} 
	This mechanism uses both redundancy and gradual recruitment, but it is designed to obtain a different objective (i.e., maximizing social-welfare) than maximizing the consumer's expected revenue. The pairing mechanism is currently the state-of-the-art in the class of approximate social-welfare maximizing incentive mechanisms.
	
	%	\item \textbf{Timeout $(p,t)$:} This strategy first orders all providers using a preference ordering given by parameter $p\in \{cost,time, balanced\}$. These choices for $p$, respectively, order all providers by ascending cost, mean duration, or a combination (i.e. cost per mean duration). The strategy then invokes the providers in that order, leaving a waiting time of $t$ between successive invocations. This continues until either the task is completed or the deadline is reached. The Timeout strategy represents approaches that use both redundancy and gradual selection, but not in a smart way. 
	\item \textbf{Best full information mechanism:} This mechanism provides the maximum revenue in full-information environments, where the providers' cost function is fully available to the consumer. Therefore, it serves as a theoretical upper bound for the performance of any mechanism in the incomplete-information setting.
\end{enumerate}

Comparing our optimal auction to these benchmarks allows us to quantify the benefit of using information-contingent strategies in an optimal and principled manner to deal with execution uncertainty. We conduct the comparison by running 1000 experiments with randomly generated providers in each of the four simulation settings described in Section \ref{sec:flexibility-numerical} and use t-tests to ensure statistical significance at the $p < 0.05$ level. In each figure, the error bars show statistical error variations of $68\%$ confidence level.

%Here, our evaluation is guided by the following hypothesis:
%
%\vspace{0.2cm}
%\noindent \textbf{Hypothesis 1.} Our proposed auction has a superior performance in terms of the consumer's expected revenue over the current procurement auctions.

%To test Hypothesis 1, we run 1000 experiments with randomly generated providers in each of the four simulation settings described in Section \ref{sec:flexibility-numerical} and use t-tests to ensure statistical significance at the $p < 0.05$ level. As the associated confidence intervals are generally small, we omit these in all figures for clarity.

%
%To test Hypothesis 1, we adopt the simulation setting in \cite{STEIN2011} and assume that each provider $i$'s duration distribution is exponential with rate $\lambda_i$, i.e. $G_i(t)=1-e^{-\lambda_i t}$. We randomly generate each provider $i$ by drawing its cost $c_i$ and duration rate $\lambda_i$ independently and uniformly at random from $[0,1]$. To consider a range of settings, tasks have either a low ($V_{low} =4$) or a high value ($V_{high}=10$) and their deadline is either normal ($D_{normal} =3$) or urgent ($D_{urgent} =1$). Furthermore, 
%Throughout our evaluation we repeat all experiments 1000 times and use t-tests to ensure statistical significance at the $p < 0.05$ level. As the associated confidence intervals are generally small, we omit these in all figures for clarity.

%\begin{figure}[t]
%	\centering
%	\includegraphics[width=0.85\textwidth,height=0.65\textwidth]{./Figures/Picture1-pair.png}
%	\caption{...}\label{fig:benchmark-no-budget}
%\end{figure}

\subsubsection{Comparison in Terms of Expected Revenue}

\begin{figure}[t]
	\subfloat[Setting 1: $V=10$, $D=3$]{\label{fig:benchmark-no-budget-1}
		\includegraphics[height=0.35\textwidth]{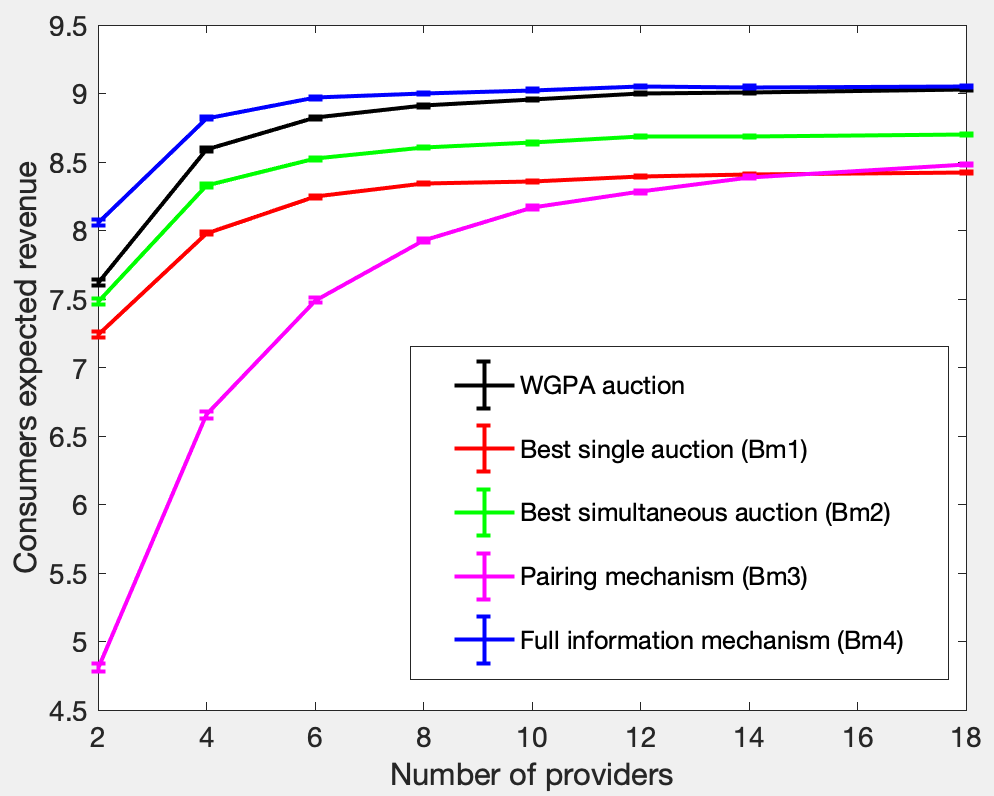}}
	~~~
	\subfloat[Setting 2: $V=4$, $D=3$]{\label{fig:benchmark-no-budget-2}
		\includegraphics[height=0.35\textwidth]{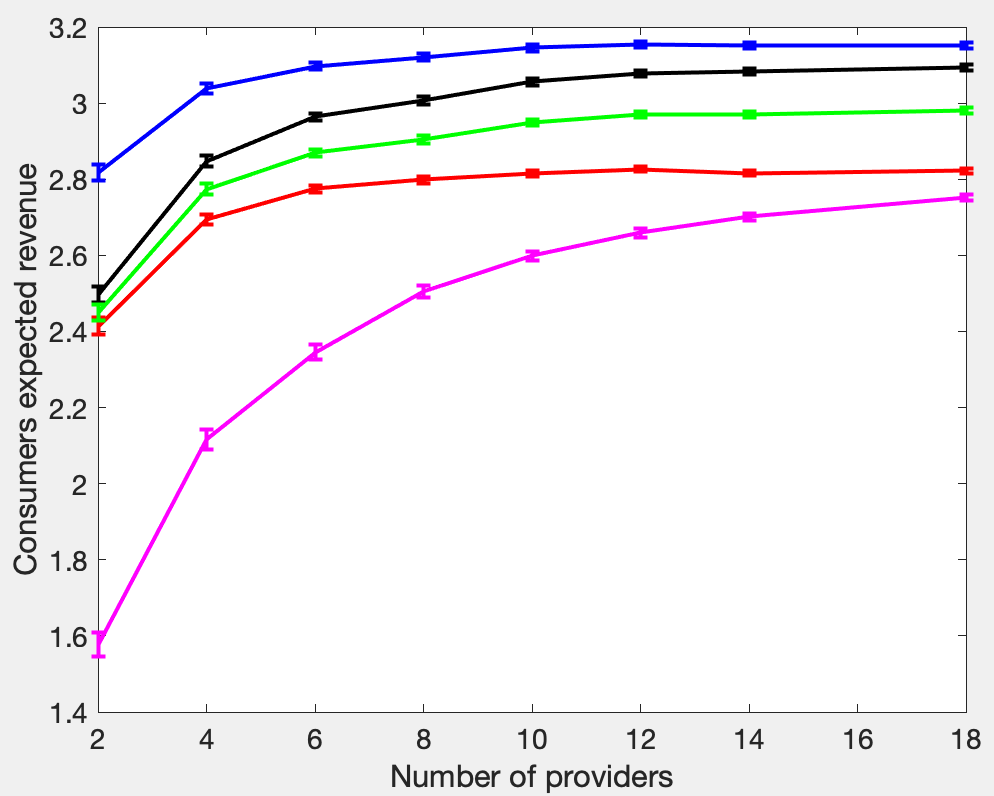}}
	
	\subfloat[Setting 3: $V=10$, $D=1$]{\label{fig:benchmark-no-budget-3}
		\includegraphics[height=0.35\textwidth]{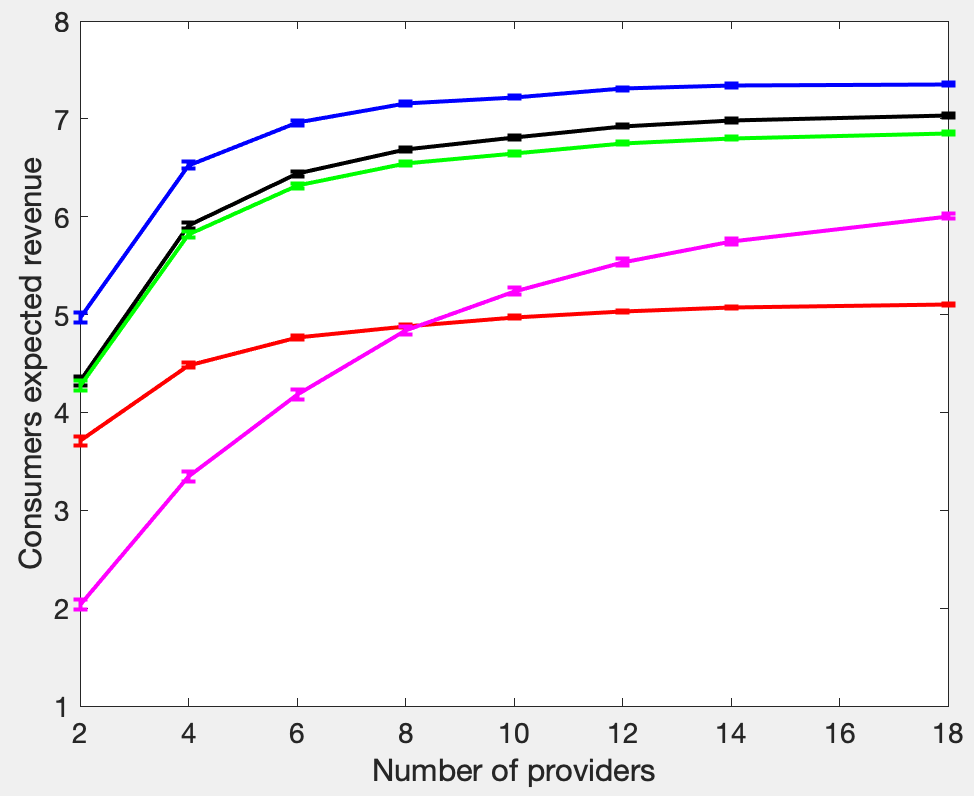}}
	~~~
	\subfloat[Setting 4: $V=4$, $D=1$]{\label{fig:benchmark-no-budget-4}
		\includegraphics[height=0.35\textwidth]{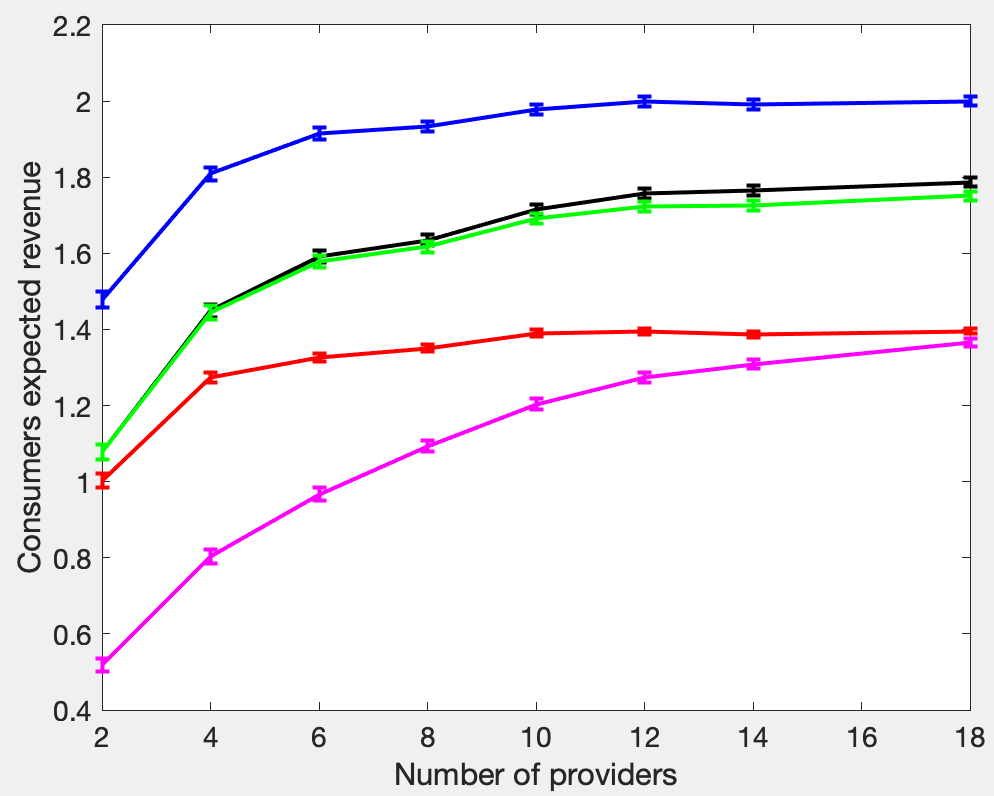}}
	\caption{Comparison of the performance of the WGPA auction with benchmarks introduced in Section \ref{sec:becnhmark} in terms of revenue }\label{fig:benchmark-no-budget}
\end{figure}
In Fig. \ref{fig:benchmark-no-budget}, we plot the consumer's expected revenue versus the number of providers in the above-described settings. The results show that the redundant allocation has a huge impact on the consumer's expected revenue over a wide range of environments. This can be seen by comparing the expected revenue of our auction with that of Bm1 where only a single provider is recruited. We can see that the optimal auction proposed in this paper results in $6.5\%$, $8.1\%$, $36.5\%$, and $26.9\%$ higher revenues compared to those obtained by Bm1 in Settings 1-4, respectively. The percentage of improvement goes up to $120\%$ when the value of the task and the deadline approach infinity and zero, respectively.

We can see from Fig. \ref{fig:benchmark-no-budget} that the advantage of gradual recruitment shows itself more apparently when the task is not urgent (i.e., Figs. \ref{fig:benchmark-no-budget-1}-\ref{fig:benchmark-no-budget-2}).
\footnote{This result is consistent with our discussion in Section \ref{sec:flexibility-numerical}.} When the task is urgent (see Figs. \ref{fig:benchmark-no-budget-3}-\ref{fig:benchmark-no-budget-4}), the optimal auction is at most $2\%$ better than Bm2. However, when the task has a normal deadline, this superiority goes up to $5\%$. 

We can also see from Fig. \ref{fig:benchmark-no-budget} that our proposed WGPA auction significantly outperforms the pairing mechanism proposed in \cite{STEIN2011}, in terms of the consumer's expected revenue. In fact, the expected revenues obtained by WGPA are up to $59.5\%$, $61.8\%$, $122.4\%$, and $111\%$ higher than those obtained by Bm3, in Settings 1-4, respectively. This superiority comes from three aspects:
\begin{enumerate}
	\item Ignoring half of the providers in the pairing mechanism: As discussed previously, the pairing mechanism pairs the service providers randomly and disposes of the providers with the maximum cost at each pair. Some of the providers that are ignored by the mechanism could be the ones that are optimal to be recruited. The effect of this non-optimal decision making is more noticeable when the number of providers is low (e.g. $n \leq 10$), as in this case, the variety of services is low and hence there might not be a good alternative to the ignored providers.
	\item Different payment functions: WGPA employs the weighted threshold payment scheme which guarantees IC and IR at the minimum cost. However, the pairing mechanism uses a heuristic payment function, which based on the results provided in Fig. \ref{fig:benchmark-no-budget}, pays too high prices  to providers to incentivize them to tell the truth.
	\item Different objectives: The pairing mechanism has the goal of maximizing social-welfare (not the consumer's expected revenue). Therefore, it might be the case that lower consumer's expected revenue is the penalty the mechanism pays to increase the social-welfare. However, as we will show in Section \ref{sec:compare-SW}, this turns out not to be the case.
\end{enumerate}

\subsubsection{Comparison in Terms of Social-Welfare}\label{sec:compare-SW}

%\vspace{0.2cm}
%\noindent \textbf{Hypothesis 2.} Our proposed auction provides a greater social welfare than the pairing mechanism proposed in \cite{STEIN2011} for maximizing social welfare.

\begin{figure}[t]
	\centering
	\includegraphics[height=0.35\textwidth]{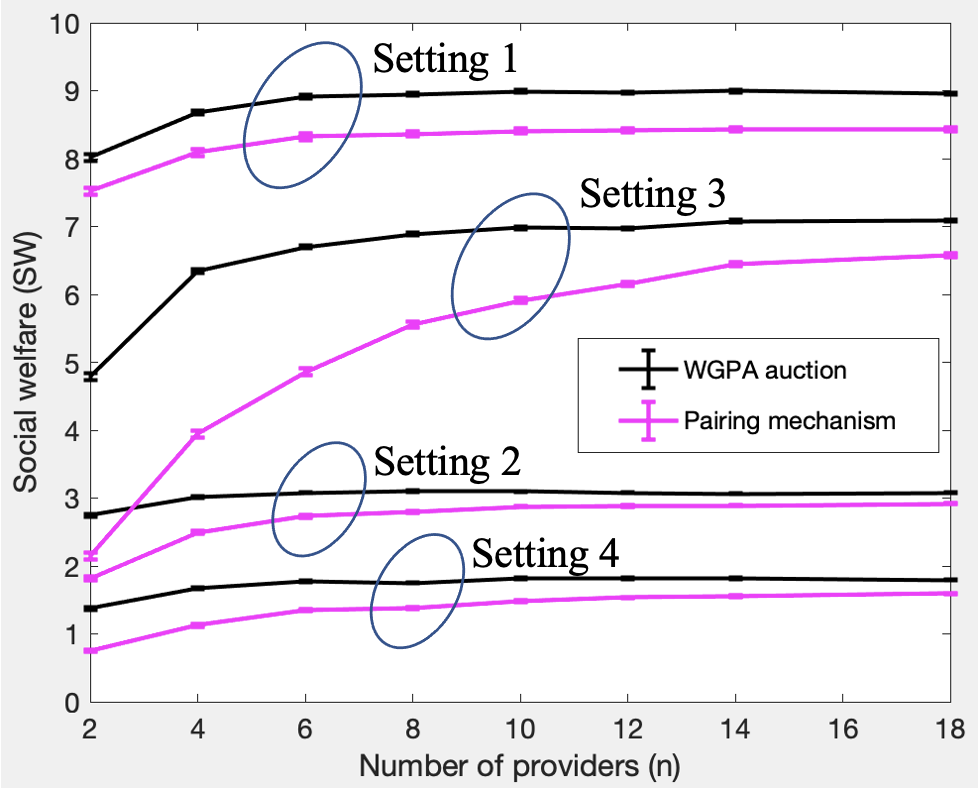}
	\caption{Comparison of the performance of the WGPA auction with benchmarks introduced in Section \ref{sec:becnhmark} in terms of social-welfare}\label{fig:Stein-no-budget}
\end{figure}

The pairing mechanism \cite{STEIN2011} is the best social-welfare maximizing incentive mechanism known so far. The purpose of this section is to show that although the main focus of WGPA is to maximize the consumer's expected revenue, it outperforms the pairing mechanism in terms of the social-welfare. This claim is supported by Fig. \ref{fig:Stein-no-budget}. 

In more detail, in Fig. \ref{fig:Stein-no-budget}, we consider the four simulation settings described previously and plot the social-welfare obtained by WGPA and the pairing mechanism, versus the number of providers. We can see that in all settings, WGPA provides a significantly higher social-welfare than the pairing mechanism \cite{STEIN2011}. The superiority is up to $7.3\%, 51.5\%, 122.7\%$, and $83.5\%$, in Settings 1-4, respectively. This is a significant and somewhat surprising result as it demonstrates that although WGPA is designed for maximizing the consumer's expected revenue, it is so well-designed that it also outperforms the available mechanisms in terms of the social-welfare. Therefore, WGPA can also be seen as the best social-welfare maximizing auction available so far and can be used by consumers whose objective is to maximize the social-welfare.
%
%The difference is much more evident when the number is providers is low. The reason is that the pairing mechanism throws away half of the providers.

\subsection{Heuristics to Reduce the Computational Complexity} \label{sec:heuristic-numerical}

In Section \ref{sec:heuristic}, we proposed auction $M_{apx}$ with the low-complexity allocation function $A_{apx}$ and claimed that this auction significantly reduces the running time without significantly reducing the
performance when compared with the optimal auction WGPA. 
The purpose of this section is to demonstrate that claim. 

To this end, for each fixed $V$ and $D$, we conduct several experiments with different number of providers $n$ and different cost vectors $\boldsymbol{c}$ and compare the running times and the provided revenues of WGPA and $M_{apx}$. In terms of time complexity, experiments show that auction $M_{apx}$ reduces the running time by several orders of magnitude (in the average range of $64\%-99.8\%$), and hence gives the ability to the auction to handle large number of providers (e.g. 100 providers runs in 4.4 seconds).

\begin{figure}[t]
	\subfloat[Maximum percentage of inefficiency caused by running auction $M_{apx}$]{\label{fig:heuristic-no-budget-1}
		\includegraphics[height=0.35\textwidth]{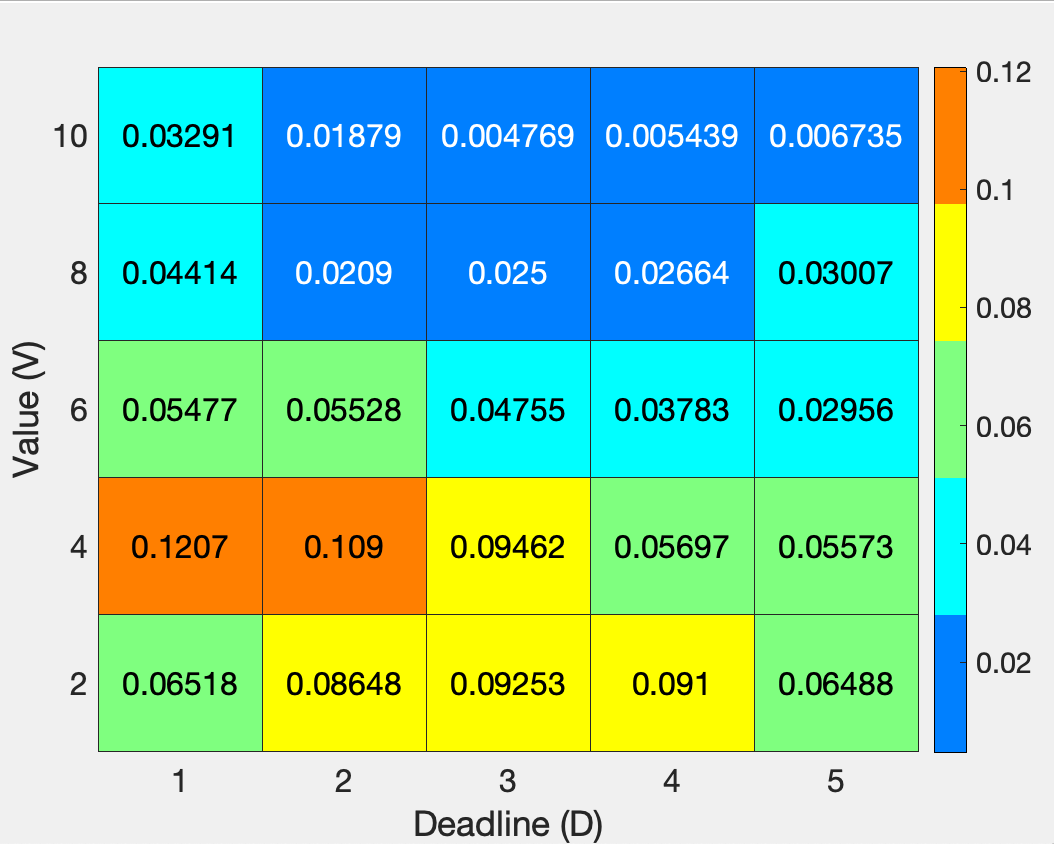}}
	~~~
	\subfloat[Average percentage of inefficiency caused by running auction $M_{apx}$]{\label{fig:heuristic-no-budget-2}
		\includegraphics[height=0.35\textwidth]{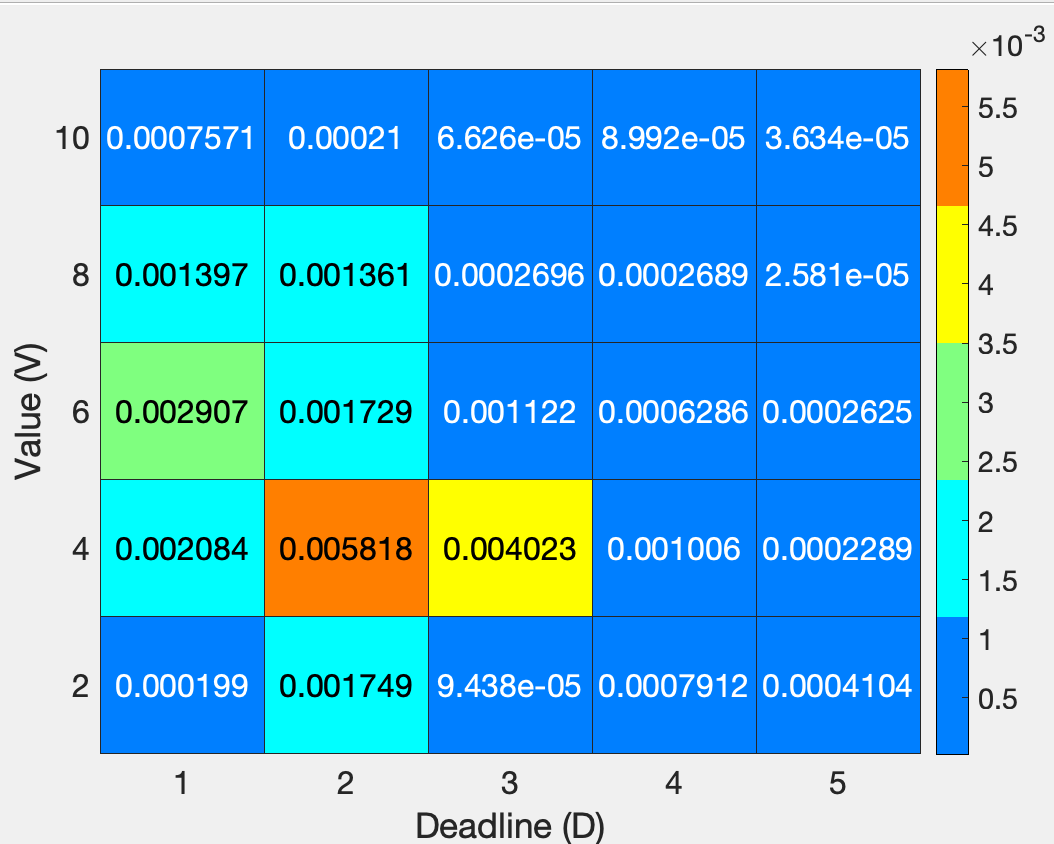}}
	\caption{Accuracy of the suboptimal auction $M_{apx}$ introduced in Section \ref{sec:heuristic}}\label{fig:heuristic-no-budget}
\end{figure}

In terms of performance, we report the maximum and the average percentage of inefficiency caused by running the suboptimal auction $M_{apx}$ in Figs. \ref{fig:heuristic-no-budget-1} and \ref{fig:heuristic-no-budget-2}, respectively.
We can see that in each single experiment, the revenue provided by $M_{apx}$ is at most $12.07\%$ lower than the optimal revenue. This means that the worst-case approximation ratio of auction $M_{apx}$ is $0.8793$. However, based on Fig. \ref{fig:heuristic-no-budget-2}, auction $M_{apx}$ has a very close-to-optimal performance in terms of the ``expected'' revenue. In fact, the expected revenue the consumer expects to achieve from $M_{apx}$ is at least $0.9942$ of the optimal revenue. This shows that the worst-case instances are very rare and do not appear in practice very frequently. Therefore, auction $M_{apx}$ with the average-case approximation ratio of $0.9942$ is a good choice for a consumer with limited computational power that aims to maximize its expected revenue.

\subsection{Relaxing Assumptions}\label{sec:impact-assumptions}

We made the following assumptions in all our previous simulations (except Fig. \ref{fig:visualization}):
\begin{enumerate}[label=A\arabic*]
	\item The consumer has precise knowledge about the providers' duration functions;\label{Ass1}
	\item The providers' costs and service rates are correlated, i.e., providing a faster service incurs a higher cost;\label{Ass2}
	\item The providers' population is homogeneous, i.e. providers' costs and service rates are chosen from the same distribution;\label{Ass3}
	\item The providers' duration distributions are exponential. \label{Ass4}
\end{enumerate}
To check the robustness of the results in wide range of circumstances, in the following subsections, we study how the results change when each of these assumptions is relaxed.

\subsubsection{WGPA's Robustness to Incorrect Information}\label{sec:robustness}

Our goal in this section is to assess the robustness of the WGPA auction regarding the consumer's incorrect information about providers' duration functions. To this end, for each percentage of inaccuracy $\delta \in (0,40)$, we conduct several experiments in which the consumer's belief about the providers' service rates $\boldsymbol{\lambda}=(\lambda_1,\ldots,\lambda_n)$ has at most $\delta$ percent error (i.e., the consumer's belief is in a sphere with center $\boldsymbol{\lambda}$ and radius $0.01\delta |\boldsymbol{\lambda}|$). In each of these experiments, we compare the consumer's expected revenue when it employs different available mechanisms.

%This experiment is guided by two following hypotheses:
% 
%\vspace{0.2cm}
%\noindent \textbf{Hypothesis 3.} The optimal auction outperforms the current state of the art even if the consumer's knowledge about the providers is inaccurate. 
%
%\vspace{0.2cm}
%\noindent \textbf{Hypothesis 4.} Compared to the available benchmarks, auction  $M^*$ is more robust to the consumer's inaccurate information.
%
%To test Hypotheses 3-4, we run the following experiment in each of Settings 1-4. We assume that the consumer's belief about the providers' service rates has at most $\delta$ percent error, i.e. the consumer's belief about each provider $i$'s service rate $\lambda_i$ is drawn uniformly over $[(1-\delta) \lambda_i, (1+\delta) \lambda_i,]$. In this situation, we first compare the consumer's expected revenue when it employs different available mechanisms. We then assess the performance loss occurred due to belief inaccuracy.

\begin{figure}[t]
	\subfloat[Setting 1: $V=10$, $D=3$]{\label{fig:robust-nobudget-1}
		\includegraphics[height=0.35\textwidth]{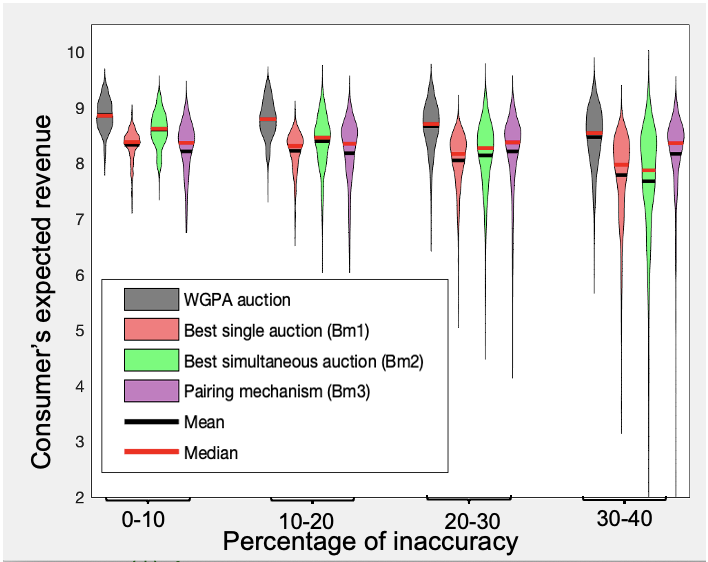}}
	~~~
	\subfloat[Setting 2: $V=4$, $D=3$]{\label{fig:robust-nobudget-2}
		\includegraphics[height=0.35\textwidth]{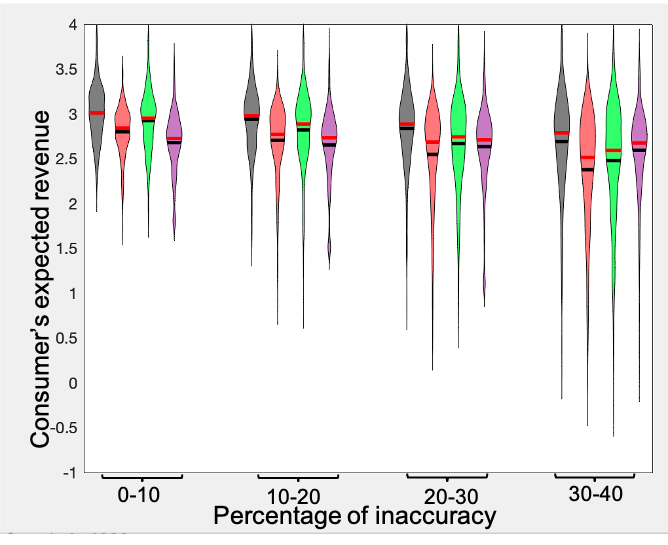}}
	
	\subfloat[Setting 3: $V=10$, $D=1$]{\label{fig:robust-nobudget-3}
		\includegraphics[height=0.35\textwidth]{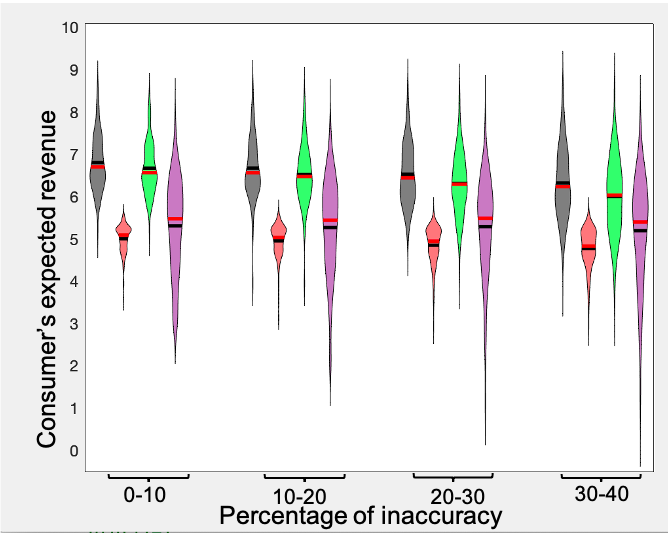}}
	~~~
	\subfloat[Setting 4: $V=4$, $D=1$]{\label{fig:robust-nobudget-4}
		\includegraphics[height=0.35\textwidth]{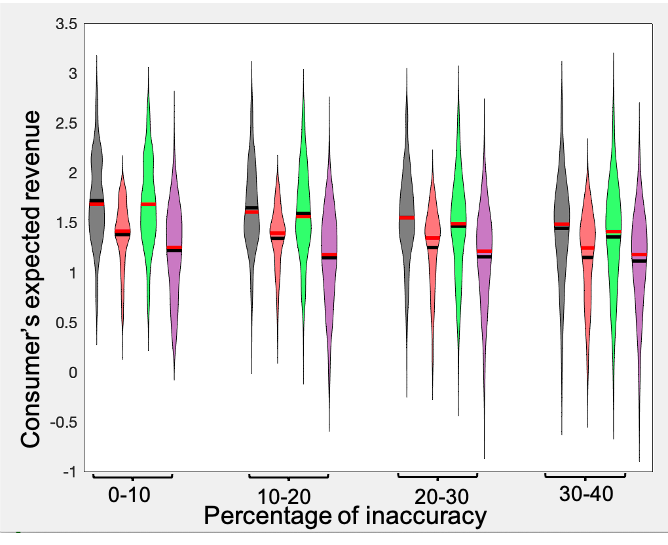}}
	\caption{Robustness of WGPA as well as the other available benchmarks to information inaccuracy}\label{fig:robust-nobudget}
\end{figure}

The results are shown in Figs. \ref{fig:robust-nobudget-1}-\ref{fig:robust-nobudget-4} for Settings 1-4, respectively.  In Figs. \ref{fig:robust-nobudget-1}-\ref{fig:robust-nobudget-4}, we plot the distribution of consumer's revenue for different ranges of inaccuracy. It can be seen that in all cases, the WGPA auction keeps its superiority over the benchmarks. We can also see that the main impact of the information inaccuracy is not on the expected value but on the variance of the revenue gained by different auctions. This is due to the nature of the problem. In a procurement auction, the decisions are often made based on the relative order and not absolute values of providers' service rates. The information inaccuracy, however, often changes the providers' absolute service rates but less frequently their relative order. This feature allows the auctions to bypass the inaccuracy and still choose a good recruitment strategy even if the information is inaccurate.

%
%
%consumer's expected utility vs the percentage of inaccuracy $\delta$. It can be seen from this figure that the superiority of the optimal auction $M^*$ over Benchmarks Bm1-Bm3 increases when the rate of inaccuracy goes up. Therefore, even if the consumer's belief is inaccurate, auction $M^*$ is the best mechanism it can use to maximize its expected revenue. To explore the robustness of different mechanisms, we compare the consumer's expected revenue with the revenue it would obtain if it had accurate information. In Fig. \ref{fig:robust-nobudget2}, we plot the distribution of this performance loss. We can see that when the consumer employs auction $M^*$, the performance loss has both a lower mean and a lower variance compared to the other methods. 

%\begin{figure}[t]
%	\subfloat[]{\label{fig:robust-nobudget1}
%		\includegraphics[width=0.45\textwidth,height=0.31\textwidth]{./Figures/Robust-nobudget1.png}}
%	~
%	\subfloat[]{\label{fig:robust-nobudget2}
%		\includegraphics[width=0.45\textwidth,height=0.31\textwidth]{./Figures/Robust-nobudget2.png}}
%
%	\caption{Robustness Analysis}\label{fig:robust-nobudget}
%\end{figure}

\subsubsection{Independent Costs and Service Rates}

%\begin{figure}[t]
%	\centering
%	\includegraphics[width=0.94\textwidth,height=0.37\textwidth]{./Figures/Sorted-Unsorted.png}
%	\caption{...}\label{fig:sorted-unsorted}
%\end{figure}

\begin{figure}[t]
	\subfloat[Correlated environment (with Assumption \ref{Ass2})]{\label{fig:sorted-unsorted-1}
		\includegraphics[height=0.35\textwidth]{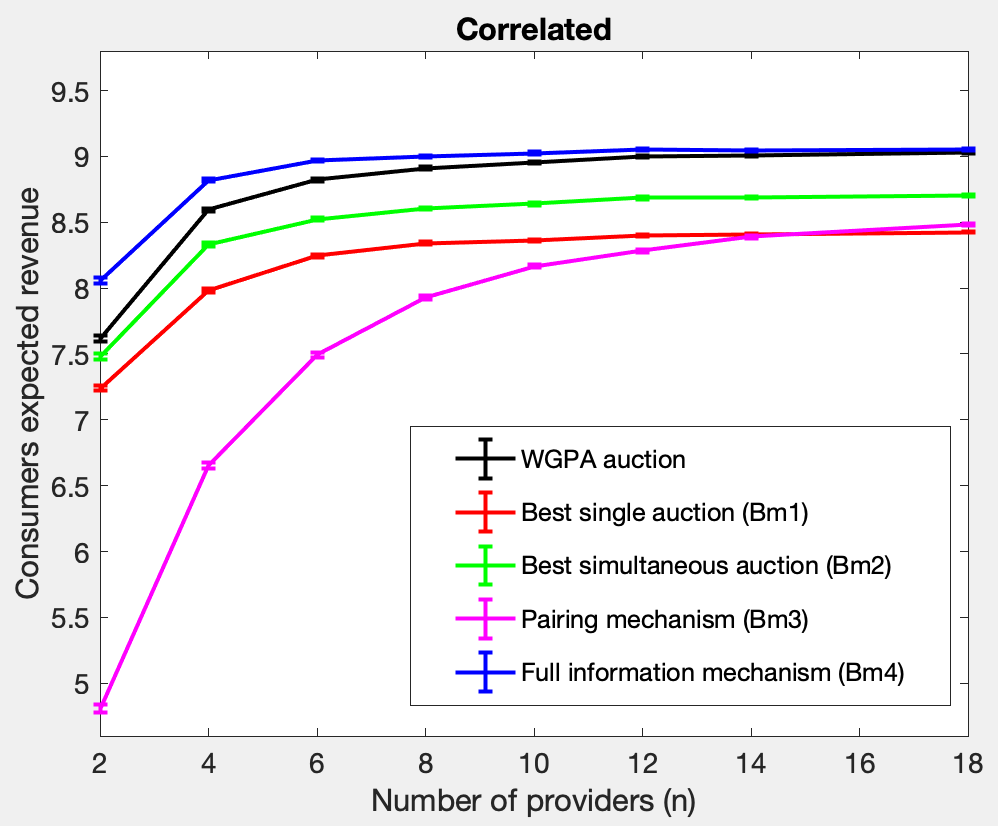}}
	~~~
	\subfloat[Independent environment (without Assumption \ref{Ass2})]{\label{fig:sorted-unsorted-2}
		\includegraphics[height=0.35\textwidth]{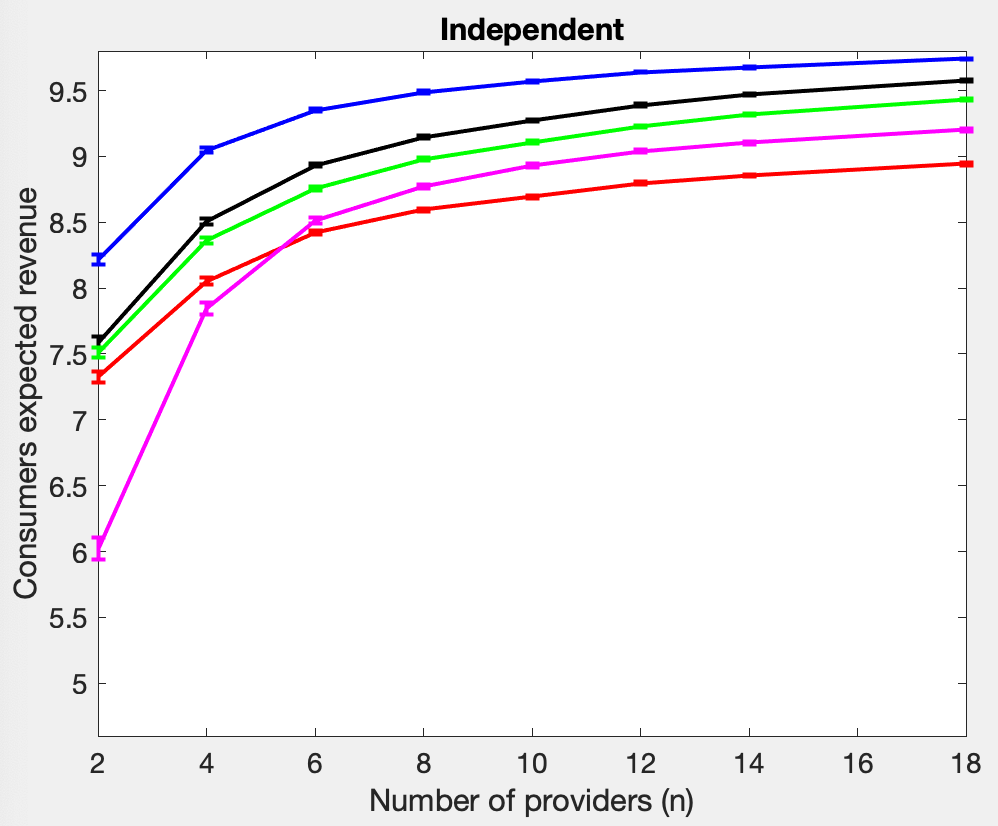}}
	\caption{Impact of removing Assumption \ref{Ass2} on the relative performance of WGPA and the available benchmarks}\label{fig:sorted-unsorted}
\end{figure}

In this section, we study how the results change if Assumption \ref{Ass2} is lifted. To this end, we repeated all the experiments performed in Sections \ref{sec:becnhmark} and \ref{sec:heuristic-numerical} in a new environment where providers' costs and service rates are chosen independently and uniformly from $[0,1]$.\footnote{In this environment, unlike the previous one, there may exist a provider $i$ that provides a faster service than $j$ at a lower cost. In such cases, provider $i$ is said to dominate provider $j$.} Such experiments show that Assumption \ref{Ass2} has no substantial effect on the results and hence all the trends discussed in Sections \ref{sec:becnhmark} and \ref{sec:heuristic-numerical} still exist when this assumption is lifted. However, to show the nature of the changes caused by removing Assumption \ref{Ass2}, we compare the results of  the same study as that in Fig. \ref{fig:benchmark-no-budget-1} for correlated and independent environments in Fig. \ref{fig:sorted-unsorted}. 

We can see from Fig. \ref{fig:sorted-unsorted} that the trends of the curves in both environments are quite similar. However, there are two differences in detail:
\begin{enumerate}
	\item Each mechanism provides a higher expected revenue in the independent environment compared to the correlated one.
	\item The curves get flat after certain points in the correlated environment, however, the same cannot be observed in the independent setting.
\end{enumerate}
The reason for the first difference is that in the correlated setting, the consumer has to spend more money if it wants to recruit a faster provider. However, in the independent environment, there is a chance that the consumer finds a fast provider at a low cost. This feature reduces the cost and increases the revenue of any mechanism in the independent environment. The second difference also comes from a similar reason. In the independent environment, the chance of there existing a high-speed provider with a low cost increases as the number of providers goes up. Therefore, all the available mechanisms provide higher revenues when more service providers are available.

\subsubsection{Heterogeneous Population}

Consider an environment where the population is not homogeneous, but consists of two groups of providers with different properties. Providers of group 1 (G1) are cheap to hire but slow in delivering the task (i.e. $0 < c,\lambda <0.5$), while providers of the second group (G2) are high-speed but expensive to hire (i.e. $0.5 < c,\lambda <1$). We are interested to find the answers to the following three questions: 1) How does the optimal recruitment strategy behave in heterogeneous environments? 2) How much does the performance of WGPA depend on the heterogeneity of the environment? 3) How does WGPA compare to the available benchmarks in heterogeneous environments?

\begin{figure}[t]
	\centering
	\includegraphics[height=0.5\textwidth]{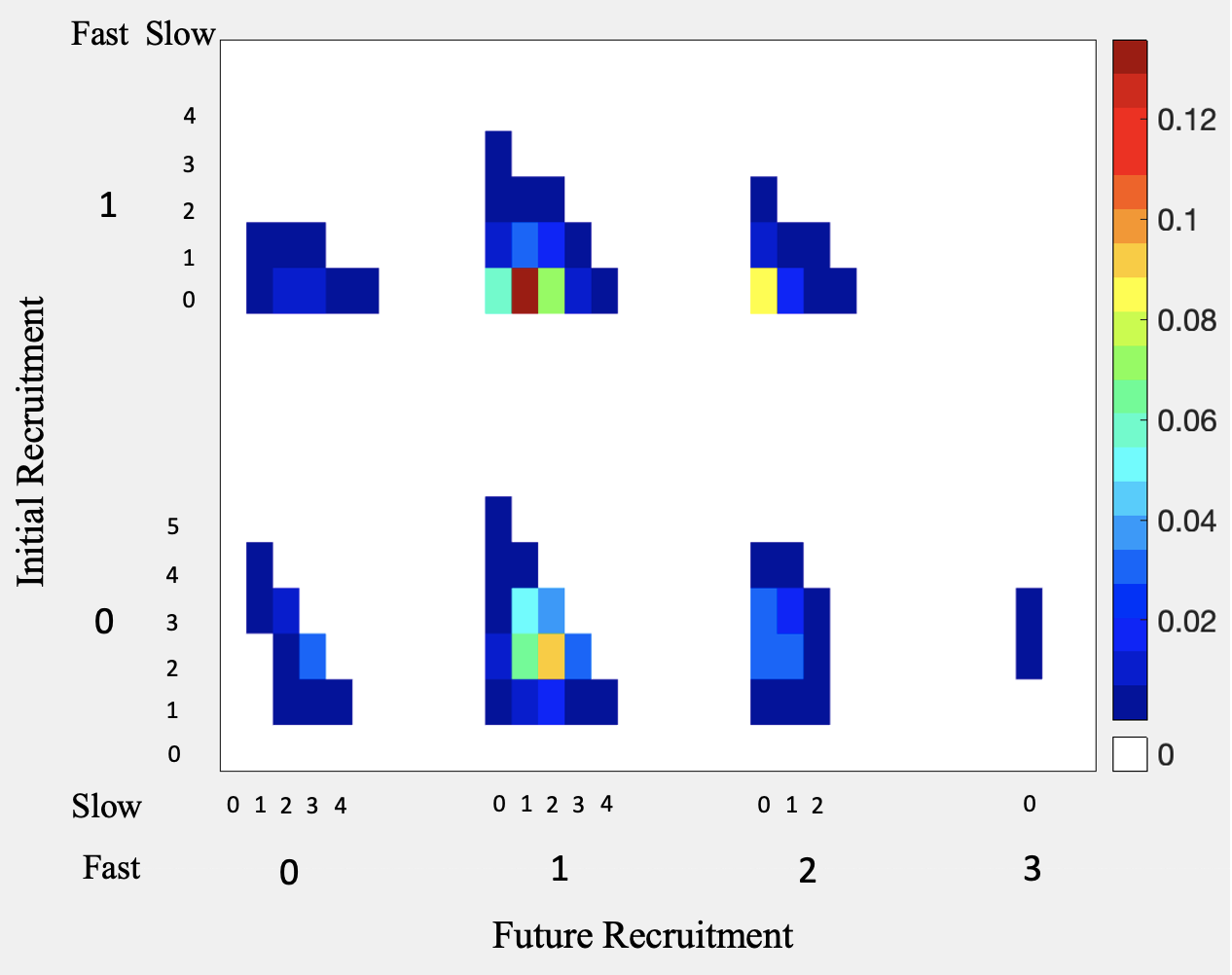}
	\caption{The distribution of different recruitment strategies in the revenue-optimal auction for a heterogeneous environment with $n=10$, $\theta=0.5$, $V=10$, and $D=3$. The y-axis shows the numbers of fast and slow providers that are recruited at time $0$. The x-axis shows how many fast and slow providers should be recruited gradually at $t>0$.}\label{fig:uneven-behavior}
\end{figure}

To answer the first question, we made several studies over environments with different percentages $\theta$ of cheap-slow providers, different task values $V$, and different deadlines $D$. However, due to a lack of space, we bring one representative sample of the results in Fig. \ref{fig:uneven-behavior}, where $n=10$, $\theta=0.5, V=10$, and $D=3$. This figure shows the distribution of different recruitment strategies in the revenue-optimal auction. It can be seen that in the most popular recruitment strategy, which we call FSF, the consumer recruits one fast provider at time $0$ and then recruits one slow and one fast provider gradually over time. In the second popular strategy, however, the consumer recruits two slow providers at time $0$, recruits two more slow providers gradually over time, and then recruits a fast provider if none of the slow providers was successful in delivering the task. Popular recruitment strategies vary by $\theta$, $V$, and $D$, however FSF is often among the most
popular ones.

\begin{figure}[t]
	\subfloat[Setting 1: $V=10$, $D=3$]{\label{fig:uneven-nim-nim-1}
		\includegraphics[height=0.35\textwidth]{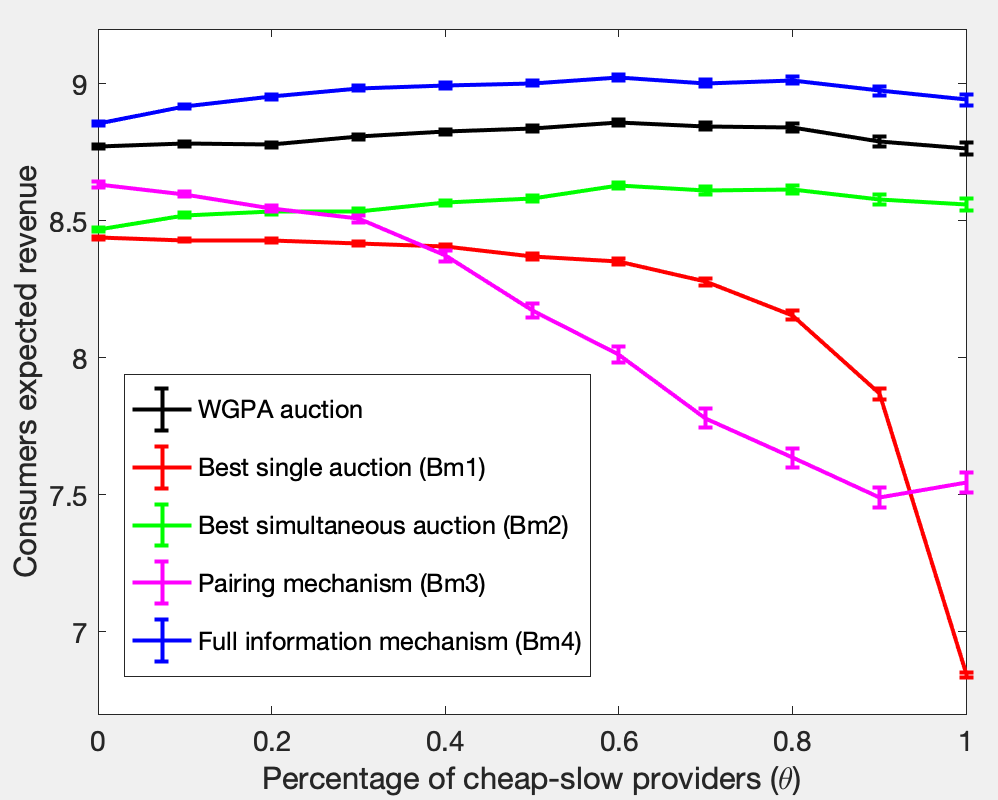}}
	~~~
	\subfloat[Setting 2: $V=4$, $D=3$]{\label{fig:uneven-nim-nim-2}
		\includegraphics[height=0.35\textwidth]{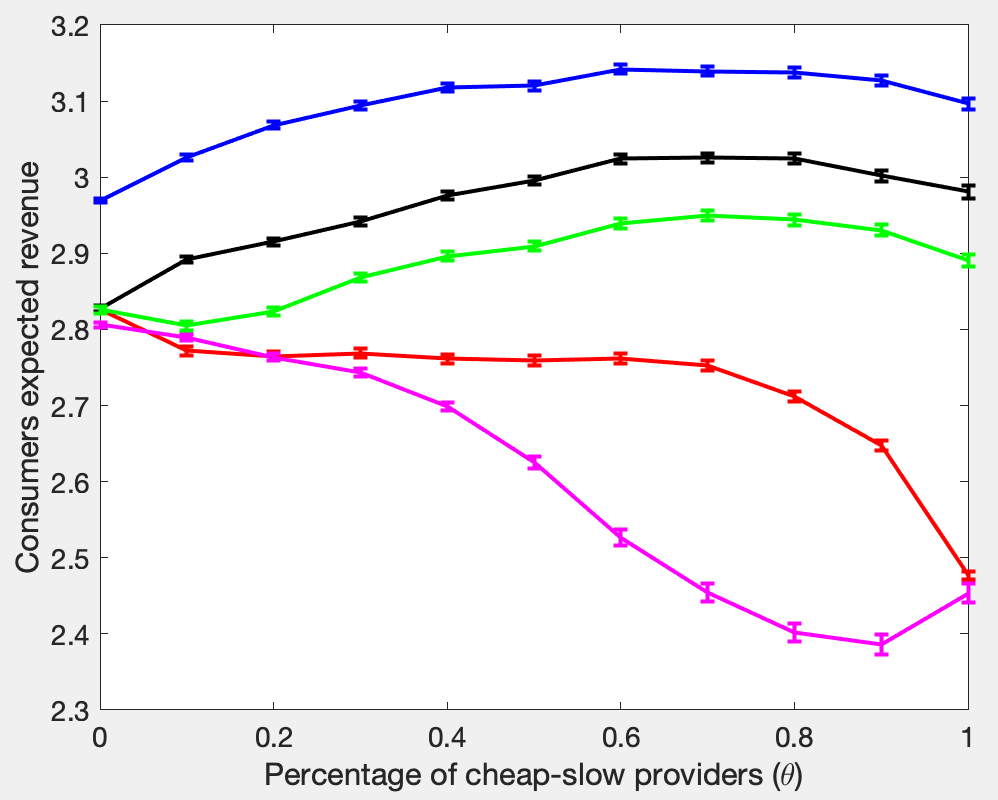}}
	
	\subfloat[Setting 3: $V=10$, $D=1$]{\label{fig:uneven-nim-nim-3}
		\includegraphics[height=0.35\textwidth]{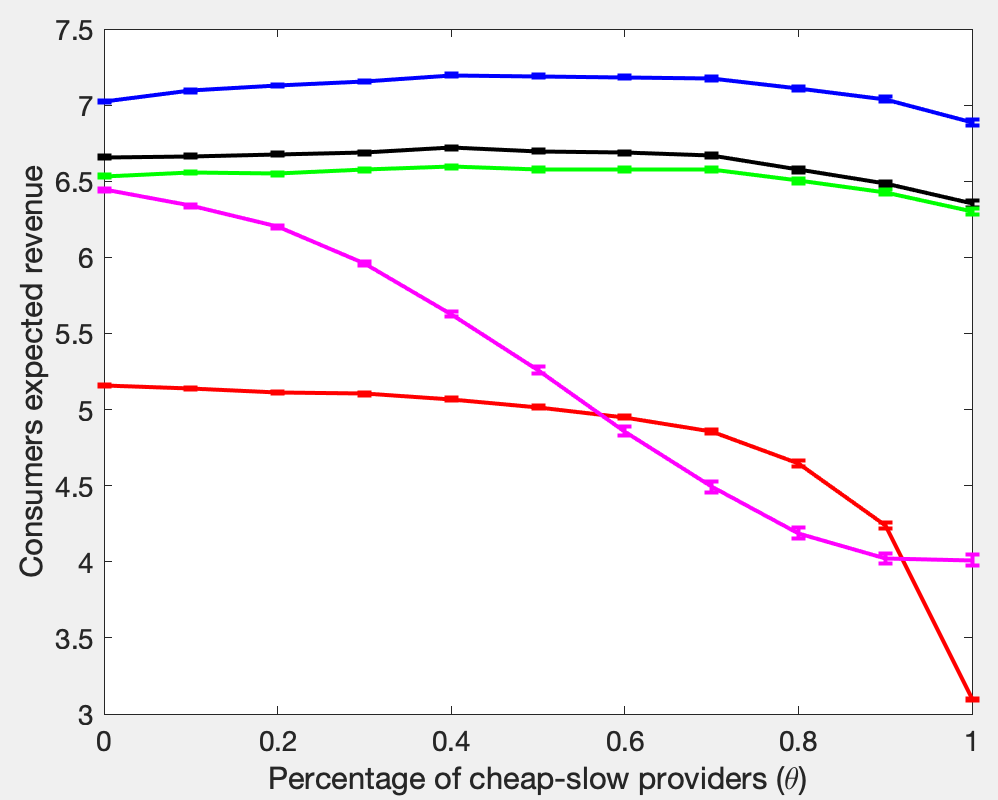}}
	~~~
	\subfloat[Setting 4: $V=4$, $D=1$]{\label{fig:uneven-nim-nim-4}
		\includegraphics[height=0.35\textwidth]{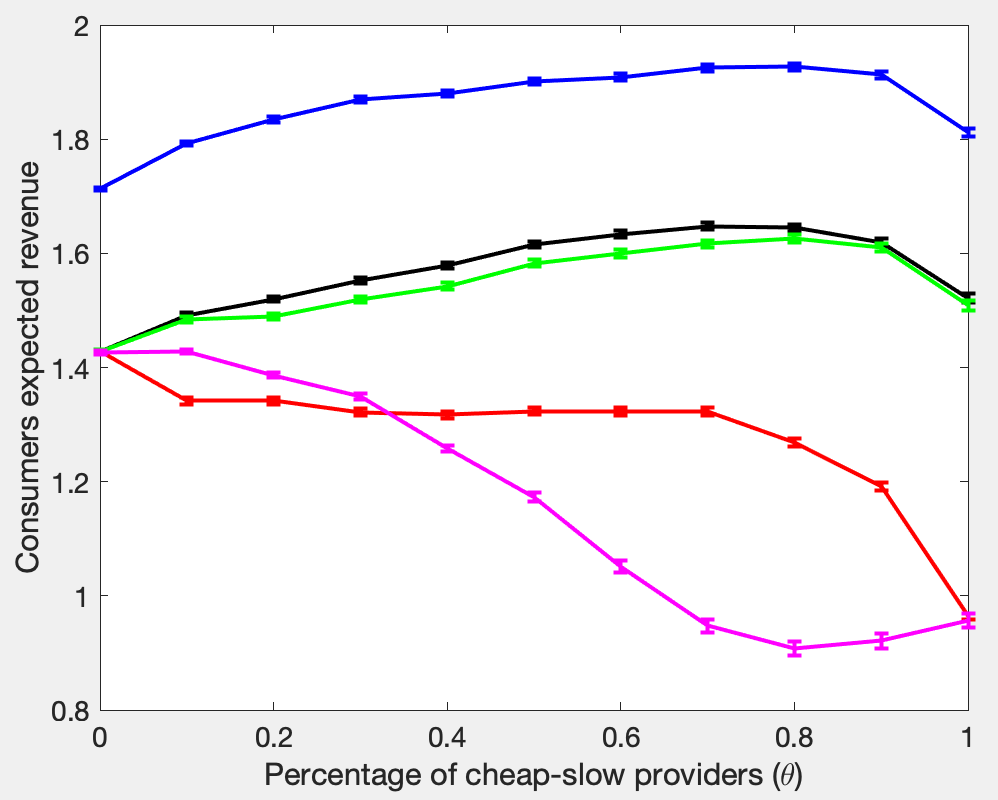}}
	\caption{Comparison of the performance of the WGPA auction with benchmarks in terms of revenue in heterogeneous setting.}\label{fig:uneven-nim-nim}
\end{figure}

To answer the second and third questions, we compare the performance of WGPA with those of other benchmarks when the percentage $\theta$ of cheap-slow providers varies between $0$ to $1$. The results of this comparison in Settings 1-4 are shown in Figs. \ref{fig:uneven-nim-nim-1}-\ref{fig:uneven-nim-nim-4}, respectively. Fig. \ref{fig:uneven-nim-nim} confirms that WGPA keeps its superiority over the available benchmarks in the heterogeneous setting. Another interesting point that can be observed from Fig. \ref{fig:uneven-nim-nim} is that WGPA as well as Bm2 and Bm4, can adapt themselves to the situations and provide a very stable revenue in the whole region. In fact, the revenue provided by WGPA has less than $12\%$ variation, while the revenues provided by Bm1 and Bm3 have over $40\%$ and $36\%$ variations respectively.

\subsubsection{Other Distributions} \label{sec:other-dist}

As the last experiment, we explored different duration distributions and observed that the results are very similar to what reported in previous sections. The only meaningful difference is the percentage of superiority of WGPA over Bm2. Experiments reveal that WGPA shows its superiority over the best simultaneous auction more significantly when delivery times have multi-modal (mixture) distributions \cite{Razali2013}. 
A multi-modal distribution is mainly used when the provider has two or more operation modes with different average delivery times. In such distributions, when the time passes the first mean and the task is not delivered yet, the consumer may believe that the provider is not in its best mode and hence update its estimation of the provider's delivery time accordingly. This phenomenon reveals some valuable information to the consumer over time and often gives it extra incentives to hire providers gradually.

\begin{figure}[t]
	\subfloat[Consumer's expected revenue versus the number of providers]{\label{fig:multi-modal-1}
		\includegraphics[height=0.35\textwidth]{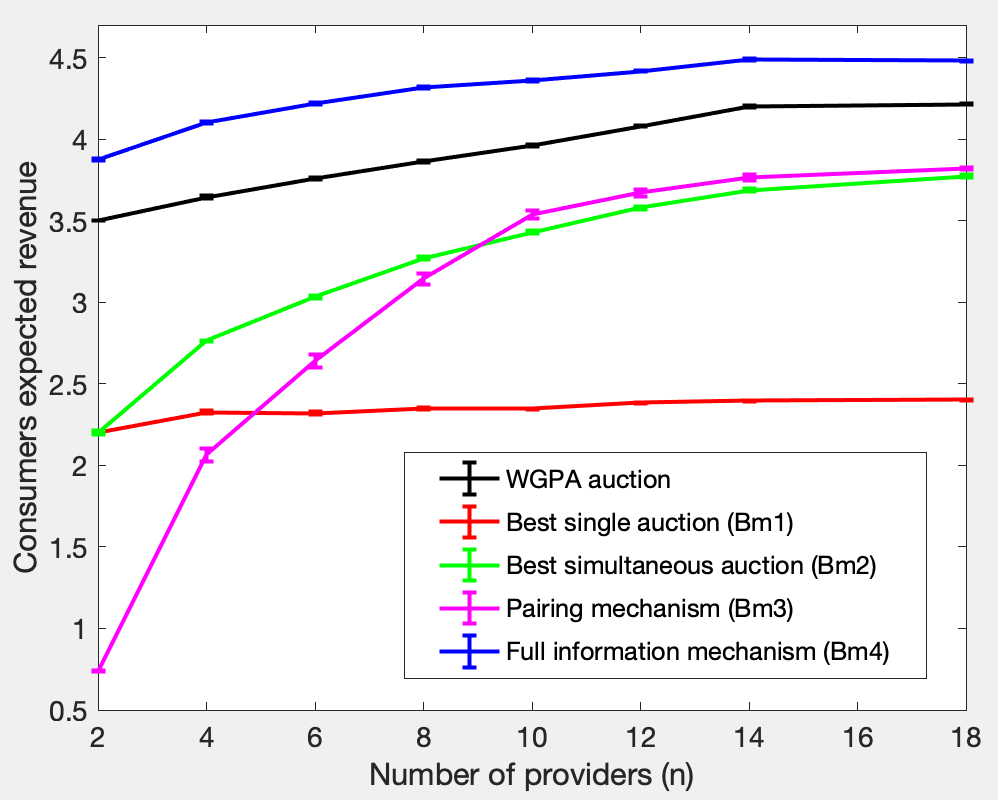}}
	~~~
	\subfloat[Percentage of increase in revenue, success probability, and total cost for WGPA compared to Bm2]{\label{fig:multi-modal-2}
		\includegraphics[height=0.35\textwidth]{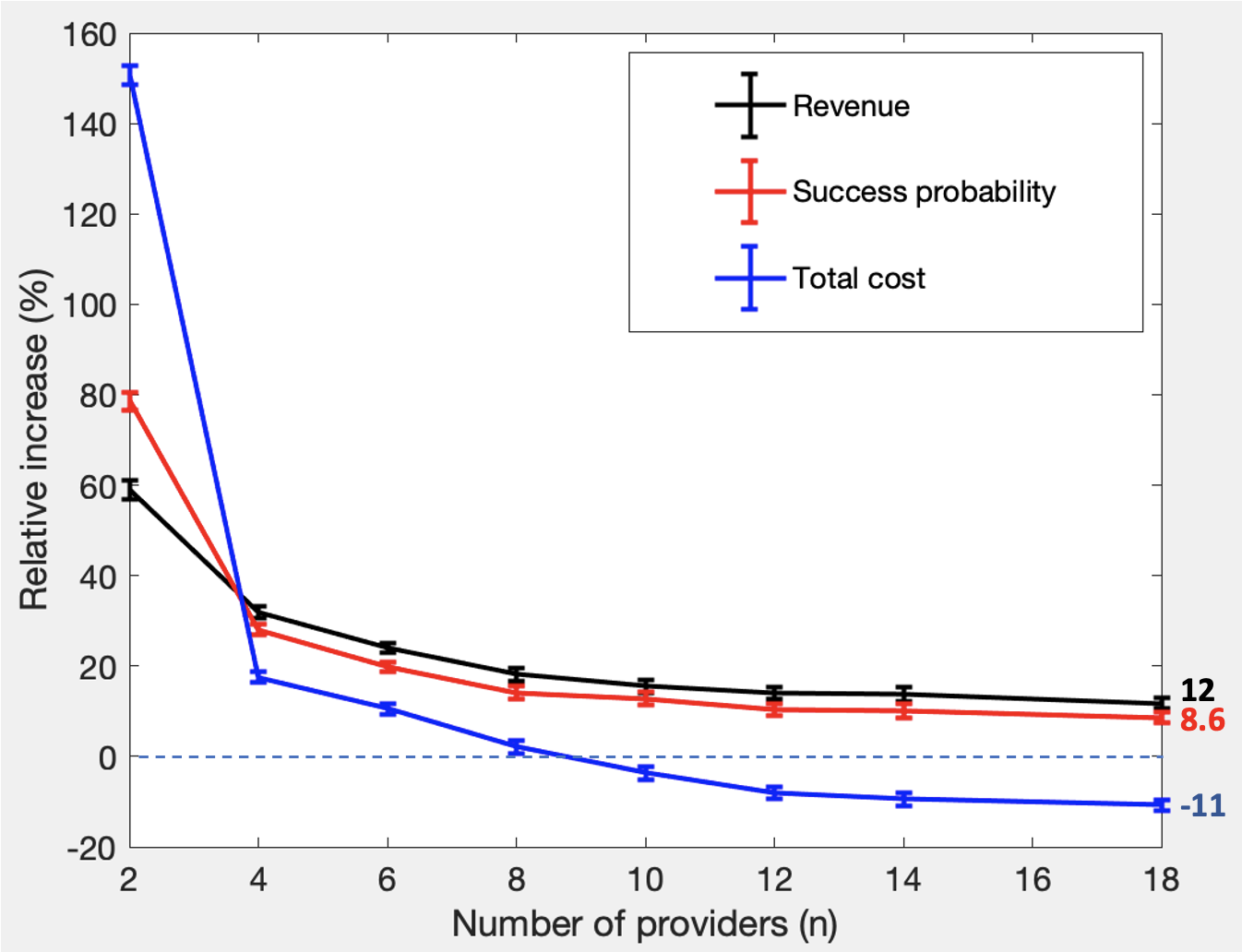}}
	\caption{Performance comparison between WGPA and the benchmarks when the delivery times of a group of providers have multi-modal distributions.}\label{fig:multi-modal}
\end{figure}

In Fig. \ref{fig:multi-modal}, we bring one simple example to support this claim.\footnote{This example is similar to Example \ref{Ex-gradual-recruitment} in Section \ref{sec:gradual-recruitment}.} In this example, we consider two groups of providers, where each provider of the first group has a deterministic delivery time $D/2$, and the delivery time of the second-group providers has a multi-modal distribution with two modes; in the first mode, the provider needs $D/2$ units of time to deliver the task, however, the delivery time of the second mode is $2D$.
The second-group providers can be in any of these modes with an equal probability of $0.5$. In such settings, we compare the performance of WGPA with other benchmarks and report the results in Fig. \ref{fig:multi-modal}. In Fig. \ref{fig:multi-modal-1}, we present the consumer's expected revenue provided by different mechanisms. Fig. \ref{fig:multi-modal-2} shows the percentage of increase in revenue, success probability, and total cost for WGPA compared to the best simultaneous auction Bm2. We can see that WGPA achieves $59\%$ improvement in revenue and $79\%$ in success probability over Bm2 when the number of providers is low, i.e. $n=2$. In this case, the consumer needs to spend $150\%$ more money compared to Bm2 to provide these advantages. When the number of providers goes up ($n \geq 15$), the WGPA auction provides $12\%$ more revenue compared to Bm2 with over $8.6\%$ increase in success probability and $11\%$ reduction in total costs.

%\begin{figure}[t]
%	\centering
%	\includegraphics[width=0.5\textwidth,height=0.31\textwidth]{./Figures/Multi-modal.png}
%	\caption{An example of settings where WGPA outperforms the best simultaneous auction (i.e., Bm2) significantly.}\label{fig:multi-modal}
%\end{figure}

	\section{Conclusions and Future Work} \label{sec:conclusion}
We have designed a revenue-optimal service procurement auction, called WGPA, for situations where the cost information is privately and asymmetrically distributed among self-interested providers. WGPA provides an optimal combination of parallel and serial redundancies, found by our proposed branch-and-bound algorithm, to provide the highest possible revenue to the service consumer. The incentive compatibility of WGPA is also guaranteed by our novel weighted threshold payment scheme proposed. This payment scheme opens a path towards designing optimal auctions for stochastic planning problems. Besides revenue maximization and incentive compatibility, WGPA satisfies both interim and ex-post individual rationality, i.e. it motivates providers to participate in the auction and perform the task upon recruitment, with no possibility of experiencing regret in the future.

We showed by numerical results that the revenue of WGPA significantly surpasses those of the best available procurement technologies by up to $120\%$. The results also reveal that WGPA outperforms the current state-of-the-art in terms of \textit{social-welfare}. This is a surprising fact as optimality (revenue maximization) and efficiency (social-welfare maximization) are normally conflicting requirements and not met simultaneously in one mechanism. However, WGPA reconciles these two apparently conflicting goals and should now be considered as the state-of-the-art in both areas. To further enhance the applicability of WGPA, we devised a low-complexity heuristic algorithm that can reduce the computational complexity of the branch-and-bound algorithm used in WGPA by over $99\%$ with less than $1\%$ performance loss.

We plan to extend our work in several ways in the future. First, we would like to relax the assumption that the providers' duration functions are known to the consumer, and consider settings where both the duration functions and the cost must be elicited from self-interested providers. Optimal auction design when agents have multi-dimensional private information is more challenging and hence less studied as such agents have more opportunities to strategic behavior. Second, we will deal with consumers with limited budget, as this setting often happens in practical service-oriented applications. A limited budget makes the problem more complex as the recruitment strategy optimizations become non-separable across the cost vectors. Therefore, an integrated approach, which determines the optimal recruitment strategies for all cost vectors simultaneously, is needed.

Finally, we intend to investigate settings with multiple interdependent tasks. To address such settings, we will extend our branch-and-bound algorithm and also consider more general utility functions that depend not on the completion time of each individual task but on the workflow completion time.

	\vspace{0.3cm}
	
	\appendix
	\section*{Appendix}
	
	%\begin{appendices}
	%%
	\section{Proof of Lemma  \ref{L-Myerson1}}\label{sec:appendix-A}
	
	Using \eqref{eq:exp2-util-sp1}, we get
	\begin{align}\label{eq:proof-L1-1}
		U_i(A,T_i,c_i,b_i)&=\int{P_i(A(b_i,\boldsymbol{c}_{-i}))[-c_i+T_i(b_i,\boldsymbol{c}_{-i})] f_{-i}(\boldsymbol{c}_{-i})d\boldsymbol{c}_{-i}}\nonumber\\
		&=\int{P_i(A(b_i,\boldsymbol{c}_{-i}))[-b_i+(b_i-c_i)+T_i(b_i,\boldsymbol{c}_{-i})] f_{-i}(\boldsymbol{c}_{-i})d\boldsymbol{c}_{-i}}\nonumber\\
		&=U_i(A,T_i,b_i,b_i)+(b_i-c_i)Q_i(A,b_i).
	\end{align}
	Thus, the incentive compatibility constraint \eqref{eq:IC} is equivalent to
	\begin{equation}\label{eq:proof-L1-2}
		U_i(A,T_i,c_i,c_i) \geq U_i(A,T_i,b_i,b_i)+(b_i-c_i)Q_i(A,b_i), \quad \forall c_i, b_i \in \mathcal{C}, i \in \mathcal{N}.
	\end{equation}
	Now, we want to show that Conditions \ref{I1}-\ref{I2} are necessary and sufficient conditions for an auction to be incentive compatible and individually rational. 
	\vspace{0.15cm}
	
	\noindent \textbf{Necessity of Conditions \ref{I1}-\ref{I2}:}  Using \eqref{eq:proof-L1-2} twice (once with the roles of $c_i$ and $b_i$ switched), we get
	\begin{equation}\label{eq:proof-L1-3}
		(b_i-c_i)Q_i(A,b_i)	\leq U_i(A,T_i,c_i,c_i) - U_i(A,T_i,b_i,b_i) \leq (b_i-c_i)Q_i(A,c_i).
	\end{equation}
	Condition \ref{I1} is directly concluded from \eqref{eq:proof-L1-3}.
	
	Equation \eqref{eq:proof-L1-3} can be written for any $b_i$ and $c_i$. Let's consider $c_i=b_i-\delta$, for $\delta>0$. Then, we have
	\begin{equation}\label{eq:proof-L1-4}
		Q_i(A,b_i)	\leq \frac{U_i(A,T_i,b_i-\delta,b_i-\delta) - U_i(A,T_i,b_i,b_i)}{\delta} \leq Q_i(A,b_i-\delta).
	\end{equation}
	Since $Q_i(A,b_i)$ is decreasing in $b_i$, it is Riemann integrable. So, 
	\begin{equation}\label{eq:proof-L1-5}
		\int_{c_i}^{c_{max}}{Q_i(A,b_i)db_i}=U_i(A,T_i,c_i,c_i) - U_i(A,T_i,c_{max},c_{max}),
	\end{equation}
	which gives us \eqref{eq:Myerson2}. The second part of Condition \ref{I2} follows directly from the individual rationality constraint \eqref{eq:IR}. So, both Conditions \ref{I1}-\ref{I2} follow from \eqref{eq:IC} and \eqref{eq:IR}.
	\vspace{0.15cm}
	
	\noindent \textbf{Sufficiency of Conditions \ref{I1}-\ref{I2}:} Individual rationality constraint \eqref{eq:IR} is deduced from Condition \ref{I2} and the fact that function $Q(.)$ always returns positive values.
	
	To show incentive compatibility \eqref{eq:IC} (or equivalently \eqref{eq:proof-L1-2}), we use \eqref{eq:Myerson2} once for $b_i$ and once for $c_i$ and then subtract the ensuing equations to get
	\begin{equation}\label{eq:proof-L1-6}
		U_i(A,T_i,b_i,b_i)=U_i(A,T_i,c_i,c_i)+\int_{b_i}^{c_i}{Q_i(A,\hat{b}_i)d\hat{b}_i}.
	\end{equation}
	Now, we divide the proof into two cases:
	\begin{enumerate}
		\item $b_i \leq c_i$: We can conclude from Condition \ref{I1} that $Q_i(A,\hat{b}_i) \leq Q_i(A,b_i)$, for all $\hat{b} \in [b_i,c_i]$. Therefore, we have
		\begin{equation}\label{eq:proof-L1-7}
			U_i(A,T_i,b_i,b_i) \leq U_i(A,T_i,c_i,c_i)+(c_i-b_i)Q_i(A,b_i),
		\end{equation}
		which is identical to the equivalence of incentive compatibility constraint \eqref{eq:proof-L1-2}.
		\item $c_i \leq b_i$: In this case, we first rewrite  \eqref{eq:proof-L1-6} as
		\begin{equation}\label{eq:proof-L1-8}
			U_i(A,T_i,b_i,b_i)=U_i(A,T_i,c_i,c_i)-\int_{c_i}^{b_i}{Q_i(A,\hat{b}_i)d\hat{b}_i}.
		\end{equation}
		Based on Condition \ref{I1}, we have $Q_i(A,\hat{b}_i) \geq Q_i(A,b_i)$, for all $\hat{b} \in [c_i,b_i]$. Substituting this inequality into \eqref{eq:proof-L1-8}, we again get \eqref{eq:proof-L1-7}, which is equivalent to the incentive compatibility constraint. This shows that in both cases, Conditions \ref{I1}-\ref{I2} are sufficient to guarantee incentive compatibility.
	\end{enumerate}
	
	\section{Proof of Lemma  \ref{L-Myerson2}}
	The consumer's expected revenue \eqref{eq:exp2-util-cons} can be rewritten as
	\begin{align}\label{eq:proof-L2-1}
		&U(A,\boldsymbol{T})=\int{[V P_{succ}(A(\boldsymbol{c}),D)-\sum_{i \in \mathcal{N}}{c_i P_i(A(\boldsymbol{c}))}]f(\boldsymbol{c})d\boldsymbol{c}}-\sum_{i \in \mathcal{N}}{\int{P_i(A(\boldsymbol{c}))[-c_i+T_i(\boldsymbol{c})]}f(\boldsymbol{c})d\boldsymbol{c}}.
	\end{align}
	Using \eqref{eq:exp2-util-sp1}, \eqref{eq:Myerson2}, and the definition of $U_i(A,T_i,c_i,c_i)$, we get
	\begin{align}\label{eq:proof-L2-2}
		&\int{P_i(A(\boldsymbol{c}))[-c_i+T_i(\boldsymbol{c})]}f(\boldsymbol{c})d\boldsymbol{c}=\int_{0}^{c_{max}}{U_i(A,T_i,c_i,c_i)f_i(c_i)dc_i}\nonumber\\
		& \qquad =\int_{0}^{c_{max}}{[U_i(A,T_i,c_{max},c_{max})+\int_{c_i}^{c_{max}}{Q_i(A,b_i)db_i}]f_i(c_i)dc_i}\nonumber\\
		& \qquad = U_i(A,T_i,c_{max},c_{max}) + \int_{0}^{c_{max}}{Q_i(A,b_i) \int_{0}^{b_i}{f_i(c_i)dc_i}db_i}\nonumber\\
		& \qquad = U_i(A,T_i,c_{max},c_{max}) + \int_{0}^{c_{max}}{Q_i(A,c_i) F_i(c_i) dc_i} \nonumber\\
		& \qquad = U_i(A,T_i,c_{max},c_{max}) + \int_{0}^{c_{max}}{F_i(c_i) \Big[\int{P_i(A(\boldsymbol{c}))f_{-i}(\boldsymbol{c}_{-i})d\boldsymbol{c}_{-i}}\Big] dc_i}\nonumber\\
		& \qquad = U_i(A,T_i,c_{max},c_{max}) + \int{\frac{F_i(c_i)}{f_i(c_i)}P_i(A(\boldsymbol{c}))f(\boldsymbol{c})d\boldsymbol{c}}.
	\end{align}
	Substituting \eqref{eq:proof-L2-2} into \eqref{eq:proof-L2-1} gives us
	\begin{align}\label{eq:proof-L2-3}
		&U(A,\boldsymbol{T})\hspace{-0.1cm}=\hspace{-0.1cm}\int{\hspace{-0.1cm}[V P_{succ}(A(\boldsymbol{c}),D)\hspace{-0.06cm}-\hspace{-0.1cm}\sum_{i \in \mathcal{N}}{(c_i\hspace{-0.06cm}+\hspace{-0.06cm}\frac{F_i(c_i)}{f_i(c_i)}) P_i(A(\boldsymbol{c}))}]f(\boldsymbol{c})d\boldsymbol{c}}-\hspace{-0.1cm}\sum_{i \in \mathcal{N}}{U_i(A,T_i,c_{max},c_{max})}.
	\end{align}
	The consumer's problem is to maximize \eqref{eq:proof-L2-3} subject to constraints \eqref{eq:Myerson1}-\eqref{eq:Myerson3}. In this formulation, $T_i$ appears only in the last term of the objective function and in the constraints \eqref{eq:Myerson2}-\eqref{eq:Myerson3}. The objective function 
	\eqref{eq:proof-L2-3} is decreasing in terms of each $U_i(A,T_i,c_{max},c_{max})$. Based on \eqref{eq:Myerson3}, the minimum value for $U_i(A,T_i,c_{max},c_{max})$ in an individually rational auction is zero. Therefore, a payment function is optimal if it guarantees  $U_i(A,T_i,c_{max},c_{max})=0$, for all $i$, while it satisfies constraint \eqref{eq:Myerson2}. This requirement is equivalent to the following condition:
	\begin{equation}\label{eq:proof-L2-4}
		U_i(A,T_i,c_i,c_i)-\int_{c_i}^{c_{max}}{Q_i(A,b_i)db_i} = 0,
	\end{equation}
	which can be rewritten as
	%\begin{align}\label{eq:proof-L2-5}
	%	&\int{P_i(A(\boldsymbol{c}))[-c_i+T_i(\boldsymbol{c})]f_{-i}(\boldsymbol{c}_{-i})d\boldsymbol{c}_{-i}}-\int_{c_i}^{c_{max}}{\Big[ \int{P_i(A(b_i,\boldsymbol{c}_{-i}))f_{-i}(\boldsymbol{c}_{-i})d\boldsymbol{c}_{-i}}\Big]db_i}  \nonumber \\
	%	& \qquad = \int{\Big[P_i(A(\boldsymbol{c}))[-c_i+T_i(\boldsymbol{c})]-\int_{c_i}^{c_{max}}{P_i(A(b_i,\boldsymbol{c}_{-i}))db_i}\Big]f_{-i}(\boldsymbol{c}_{-i})d\boldsymbol{c}_{-i}} = 0.
	%\end{align}
	\begin{equation}\label{eq:proof-L2-5}
		\int{\Big[P_i(A(\boldsymbol{c}))[-c_i+T_i(\boldsymbol{c})]-\int_{c_i}^{c_{max}}{P_i(A(b_i,\boldsymbol{c}_{-i}))db_i}\Big]f_{-i}(\boldsymbol{c}_{-i})d\boldsymbol{c}_{-i}}  = 0.
	\end{equation}
	To satisfy this constraint, we must have 
	\begin{equation}
		P_i(A(\boldsymbol{c}))[-c_i+T_i(\boldsymbol{c})]-\int_{c_i}^{c_{max}}{P_i(A(b_i,\boldsymbol{c}_{-i}))db_i}=0,
	\end{equation}
	and hence
	\begin{equation}
		T_i(\boldsymbol{c})=c_i+\frac{1}{P_i(A(\boldsymbol{c}))}\int_{c_i}^{c_{max}}{P_i(A(b_i,\boldsymbol{c}_{-i}))db_i},
	\end{equation}
	for all $i \in \mathcal{N}$. This completes the proof of Lemma \ref{L-Myerson2}.
	
	\section{Proof of Lemma  \ref{L-Myerson3}}

In the proof of Lemma \ref{L-Myerson2}, we have shown that designing the optimal revenue-maximizing auction is equivalent to maximizing \eqref{eq:proof-L2-3} subject to constraints \eqref{eq:Myerson1}-\eqref{eq:Myerson3}. We then derived optimal payment function \eqref{eq:opt-payment} which guarantees satisfaction of constraints \eqref{eq:Myerson2}-\eqref{eq:Myerson3}. Using this result, our optimal auction design problem can be re-written as
\begin{align}\label{eq:proof-L3-1}
	& \max_{ A} {\quad \int{[V P_{succ}(A(\boldsymbol{c}),D)-\sum_{i \in \mathcal{N}}{(c_i+\frac{F_i(c_i)}{f_i(c_i)})P_i(A(\boldsymbol{c}))}]f(\boldsymbol{c})d\boldsymbol{c}}}\nonumber\\
	&	\textbf{s.t.} \hspace{0.5cm} \text{ Constraint \eqref{eq:Myerson1}},
\end{align}
which is identical to problem \eqref{eq:consumer-problem2}.
	
	\section{Proof of Lemma  \ref{L-monotonicity}}

Consider optimization problem \eqref{eq:consumer-problem2} when the monotonicity constraint is relaxed. This problem can be written as follows
\begin{align}\label{eq:proof-L4-1}
	& \max_{ A} {\quad \int{[V P_{succ}(A(\boldsymbol{c}),D)-\sum_{i \in \mathcal{N}}{\phi_i(c_i)P_i(A(\boldsymbol{c}))}]f(\boldsymbol{c})d\boldsymbol{c}}}.
\end{align}
We can see that in this problem, each provider $i$'s virtual cost $\phi_i(c_i)$ works to the disadvantage of its hiring. Therefore, if provider $i$'s virtual cost $\phi_i(c_i)$ goes up when everything else is unchanged, provider $i$'s invocation probability $P_i(A(\boldsymbol{c}))$ cannot be increased in the optimal solution. When the providers' cost distributions are regular, each provider $i$'s virtual cost $\phi_i(c_i)$ is an increasing function of its cost $c_i$. In this case, each provider $i$'s invocation probability is a decreasing function of its cost as well. That is,
\begin{equation}\label{eq:proof-L4-2}
	b_i \geq c_i \Rightarrow P_i(A(b_i,\boldsymbol{c}_{-i})) \leq  P_i(A(c_i,\boldsymbol{c}_{-i})), \quad \forall \boldsymbol{c}_{-i}.
\end{equation}
Substituting \eqref{eq:proof-L4-2} into \eqref{eq:conditional-prob} results in
\begin{equation}\label{eq:proof-L4-3}
	b_i \geq c_i \Rightarrow Q_i(A,b_i) \leq  Q_i(A,c_i),
\end{equation}
which is equivalent to the monotonicity constraint \eqref{eq:monotone-alloc}. This shows that the solution of problem \eqref{eq:proof-L4-1} is monotone and hence, completes the proof of Lemma  \ref{L-monotonicity}.
	
	\section{Proof of Proposition  \ref{Prop-IC}}

	Using \eqref{eq:exp2-util-sp1} and \eqref{eq:opt-payment}, we get
\begin{align}\label{eq:proof-P1-1}
	& U_i(A^*,T^*_i,c_i,c_i) \hspace{-0.06cm} - \hspace{-0.06cm} U_i(A^*,T^*_i,c_i,b_i) \hspace{-0.06cm}=\hspace{-0.06cm}\int{\hspace{-0.1cm}[U_i(A^*,\hspace{-0.05cm}T^*_i,\hspace{-0.05cm}c_i,\hspace{-0.06cm}(c_i,\boldsymbol{c}_{-i}))\hspace{-0.06cm}-\hspace{-0.06cm}U_i(A^*,\hspace{-0.05cm}T^*_i,\hspace{-0.05cm}c_i,\hspace{-0.06cm}(b_i,\boldsymbol{c}_{-i}))]f_{-i}(\boldsymbol{c}_{-i})d\boldsymbol{c}_{-i}}  \nonumber\\
	& \qquad  =\hspace{-0.05cm} \int{\Big[P_i(A^*(c_i,\boldsymbol{c}_{-i}))[-c_i+T^*_i(c_i,\boldsymbol{c}_{-i})]\hspace{-0.06cm}-\hspace{-0.06cm}P_i(A^*(b_i,\boldsymbol{c}_{-i}))[-c_i+T^*_i(b_i,\boldsymbol{c}_{-i})]\Big]f_{-i}(\boldsymbol{c}_{-i})d\boldsymbol{c}_{-i}} \nonumber\\
	& \qquad  =\hspace{-0.05cm} \int{\Big[P_i(A^*(b_i,\boldsymbol{c}_{-i}))(c_i-b_i)+
		\int_{c_i}^{b_i}{P_i(A^*(\hat{b}_i,\boldsymbol{b}_{-i}))d\hat{b}_i}\Big]f_{-i}(\boldsymbol{c}_{-i})d\boldsymbol{c}_{-i}}\nonumber\\
	& \qquad  \hspace{-0.05cm}\geq \int{\Big[P_i(A^*(b_i,\boldsymbol{c}_{-i}))(c_i-b_i)+
		P_i(A^*(b_i,\boldsymbol{c}_{-i}))(b_i-c_i)\Big]f_{-i}(\boldsymbol{c}_{-i})d\boldsymbol{c}_{-i}} =0,
\end{align}
where the last inequality follows from the monotonicity of $P_i$ shown in the proof of Lemma \ref{L-monotonicity}. Therefore, we have $U_i(A^*,T^*_i,c_i,c_i) \geq U_i(A^*,T^*_i,c_i,b_i)$, for all $i, c_i, b_i$ which proves incentive compatibility of WGPA.
	
	\section{Proof of Proposition  \ref{Prop-IR}}
	
\textbf{Proof of Interim IR:}  We have shown in the proof of Lemma \ref{L-Myerson2} that the weighted payment scheme is designed such that
	\begin{equation}\label{eq:proof-P2-1}
		U_i(A^*,T^*_i,c_i,c_i)=\int_{c_i}^{c_{max}}{Q_i(A^*,b_i)db_i},
	\end{equation}
for all $i,c_i$ (see \eqref{eq:proof-L2-4}). Function $Q(.)$ is positive by definition. Therefore, we have $U_i(A^*,T^*_i,c_i,c_i) \allowbreak \geq 0$, for all $i,c_i$, which guarantees interim individual rationality.
	
\noindent\textbf{Proof of Ex-post IR:} Based on Proposition \ref{Prop-IC},  all providers report their costs truthfully at the equilibrium. When a provider $i$ reports its cost truthfully, it  receives payment
	\begin{equation} T_i^*(\boldsymbol{c})=c_i+\frac{1}{P_i(A^*(\boldsymbol{c}))}\int_{c_i}^{c_{max}}{P_i(A^*(\hat{b}_i,\boldsymbol{c}_{-i}))d\hat{b}_i},
	\end{equation}
upon recruitment, which is clearly greater than its own cost $c_i$. This fact is irrespective of the other providers' bids and hence proves the ex-post individual rationality of WGPA.
	
	\section{Proof of Proposition  \ref{Prop-RM}}
	We designed WGPA's allocation function as the solution of optimization problem \eqref{eq:consumer-problem2}. The WGPA's payment function is also designed based on the weighted threshold payment scheme \eqref{eq:opt-payment}. Therefore, the revenue-optimality of WGPA follows directly from Lemma \ref{L-Myerson3}.
	
	\section{Proof of Proposition  \ref{Prop-PR}}
		Based on Proposition \ref{Prop-RM}, WGPA provides the consumer with the maximum expected revenue that any incentive compatible and individually rational auction could provide. Consider the auction $\tilde{M}=(\tilde{A}(.),\boldsymbol{\tilde{T}}(.))$ with
	\begin{equation}
		\tilde{A}(\boldsymbol{c})=\emptyset,  \quad \tilde{T}_i(\boldsymbol{c})=0, \quad \forall \boldsymbol{c} \in \mathcal{C}^n.
	\end{equation}
	This auction neither recruits any provider nor makes any payment. This auction is both incentive compatible and interim individually rational as the providers' expected utilities are $0$ regardless of the cost they declare. Therefore, WGPA should outperform $\tilde{M}$ in terms of the consumer's expected revenue. That is,
	\begin{equation}
		U(A^*,\boldsymbol{T}^*) \geq U(\tilde{A},\tilde{M})=0.
	\end{equation}
	
	\section{Proof of Proposition  \ref{Prop-Heuristic}}\label{sec:appendix-I}
	\textbf{Monotonicity:} To prove monotonicity, we show that a provider $i$'s invocation probability $P_i(.)$ increases if it declares a lower cost $c_i$. This increase occurs for two reasons:
	\begin{enumerate}
		\item Increasing the chance of $i$ being among the candidate providers $o_{apx}$: In Algorithm \ref{alg4}, each ordering $o$ is valued according to the expected utility $Lower(o)$ it can provide to the consumer. For regular distributions, this expected utility is decreasing in terms of $c_i$ for all $i \in o$ (see \eqref{eq:lower_bnd}). Therefore, declaring a lower cost by provider $i$ increases the values of all orderings that include $i$, while keeps the value of all other orderings fixed. This provides a higher chance for orderings containing $i$ to beat other orderings and get selected as the final ordering $o_{apx}$, and ultimately increases the invocation probability $P_i(.)$.
		\item Advancing provider $i$'s invocation: As discussed in Section \ref{sec:opt-time}, for each ordering $o=(1,\ldots,m)$, the providers' optimal invocation times $\tau^*_1,\ldots,\tau^*_m$ are derived by solving an optimization problem that maximizes $f(\tau)$ under some constraints. It can be shown, by some algebra, that each $\tau^*_i$ is an increasing function of $\phi_{i}(c_{i})$. Therefore, when distributions are regular, reporting a lower cost $c_{i}$ decreases $\phi_{i}(c_{i})$ and hence $\tau^*_i$, and ultimately increases $P_{i}(.)$.
	\end{enumerate}
	\noindent \textbf{Incentive compatibility and individual rationality:} Given the monotonicity of allocation function $A_{apx}$, incentive compatibility and ex-post and interim individual rationality of auction $M_{apx}=(A_{apx}(.),\boldsymbol{T}^*(.))$ can be proved by following the same procedure as in the proofs of Propositions \ref{Prop-IC} and \ref{Prop-IR}.

	\bibliography{ilbib.bib}
	\bibliographystyle{theapa}
	
\end{document}